\documentclass{aa}    

\usepackage{txfonts}
\usepackage{longtable}  
\usepackage{natbib}
\usepackage{graphicx}

\bibpunct[, ]{(}{)}{;}{a}{}{,}
\usepackage{supertabular}

\usepackage{color}

\newcommand{\LyudmilasStar}{HE~2327$-$5642}        
\newcommand{\tefft}{$T_{\mbox{\scriptsize eff}}$}  
\newcommand{\teffm}{T_{\mbox{\scriptsize eff}}}    

\newcommand{\eps}[1]{\log\varepsilon_{\rm #1}}
\newcommand{\kms}{km\,s$^{-1}$}
\newcommand{\Vmac}{V_{\rm mac}}
\newcommand{\iso}[2]{\mbox{$^{#1}{\rm #2}$}}

\begin{document}

\title{The Hamburg/ESO R-process Enhanced Star survey (HERES) \thanks{Based on
    observations collected at the European Southern Observatory, Paranal,
    Chile (Proposal numbers 170.D-0010, and 280.D-5011).}}
\subtitle{V. Detailed abundance analysis of the r-process enhanced 
          star HE~2327$-$5642}

\author{
  L. Mashonkina\inst{1,2}
  N. Christlieb\inst{3} \and
  P.S. Barklem\inst{4} \and
  V. Hill\inst{5} \and
  T.C. Beers\inst{6} \and
  A. Velichko\inst{2} 
}

\offprints{L. Mashonkina; \email{lima@inasan.ru}}
\institute{
     Universit\"ats-Sternwarte M\"unchen, Scheinerstr. 1, D-81679 M\"unchen, 
     Germany \\ \email{lyuda@usm.lmu.de}
\and Institute of Astronomy, Russian Academy of Sciences, RU-119017 Moscow, 
     Russia \\ \email{lima@inasan.ru}
\and Zentrum f\"ur Astronomie der Universit\"at Heidelberg, Landessternwarte,
     K\"onigstuhl 12, D-69117 Heidelberg, Germany \\
     \email{N.Christlieb@lsw.uni-heidelberg.de}
\and Department of Astronomy and Space Physics, Uppsala University, Box 515, 
     75120 Uppsala, Sweden
\and Observatoire de Paris, GEPI and URA 8111 du CNRS, 92195 Meudon Cedex,
     France
\and Department of Physics and Astronomy, Michigan State University, East
     Lansing, MI 48824, USA
}

\date{Received  / Accepted }

\abstract{}{We report on a detailed abundance analysis of the strongly
  r-process enhanced giant star newly discovered in the HERES project,
  HE~2327$-$5642 ([Fe/H] = --2.78, [r/Fe] = +0.99).}  
{Determination of
  stellar parameters and element abundances was based on analysis of
  high-quality VLT/UVES spectra. The surface gravity was calculated from the
  non-local thermodynamic equilibrium (NLTE) ionization balance between
  \ion{Fe}{i} and \ion{Fe}{ii}, and \ion{Ca}{i} and \ion{Ca}{ii}.} 
{Accurate
  abundances for a total of 40 elements and for 23 neutron-capture elements
  beyond Sr and up to Th were determined in {\LyudmilasStar}.  
 For every chemical species, the dispersion of the single line measurements around the mean does not exceed 0.11~dex.
The heavy
  element abundance pattern of HE~2327$-$5642 is in excellent agreement with
  those previously derived for other strongly r-process enhanced stars such as
  CS\,22892-052, CS\,31082-001, and HE\,1219-0312.  Elements in the range from
  Ba to Hf match the scaled Solar $r$-process pattern very well. No firm
  conclusion can be drawn with respect to a relationship between the fisrt
  neutron-capture peak elements, Sr to Pd, in {\LyudmilasStar} and the Solar
  $r$-process, due to the uncertainty of the latter.  A clear distinction in
  Sr/Eu abundance ratios was found between the halo stars with different
  europium enhancement. The strongly r-process enhanced stars reveal a low
  Sr/Eu abundance ratio at [Sr/Eu] = $-0.92\pm0.13$, while the stars with $0
  <$ [Eu/Fe] $< 1$ and [Eu/Fe] $< 0$ have 0.36\,dex and 0.93\,dex larger Sr/Eu
  values, respectively. Radioactive dating for {\LyudmilasStar} with the
  observed thorium and rare-earth element abundance pairs results in an
  average age of 13.3~Gyr, when based on the high-entropy wind calculations, and 5.9~Gyr,
  when using the Solar r-residuals. {\LyudmilasStar} is suspected to be radial-velocity variable based on our high-resolution spectra, covering $\sim 4.3$ years.}
{} 


\titlerunning{Abundance analysis of HE~2327$-$5642}
\authorrunning{Mashonkina et al.}

\maketitle
%
\section{Introduction}\label{Sect:intro}

The detailed chemical abundances of Galactic halo stars contain unique
information on the history and nature of nucleosynthesis in our Galaxy. A
number of observational and theoretical studies have established that in the
early Galaxy the rapid ($r$) process of neutron captures was primarily
responsible for formation of heavy elements beyond the iron group \citep[we
cite only the pioneering papers of][]{Spite,Truran81}. The onset of the slow
($s$) process of neutron captures occurred at later Galactic times (and higher
metallicities) with the injection of nucleosynthetic material from long-lived
low- and intermediate-mass stars into the interstellar medium \citep[see][and
references therein]{Travaglio99}. Since 1994, a few rare stars have been found that
exhibit large enhancements of the $r$-process elements, as compared to Solar
ratios, suggesting that their observed abundances are dominated by the
influence of a single, or at most very few nucleosynthesis events. The
$r$-process is associated with explosive conditions of massive-star
core-collapse supernovae \citep{Woosley1994}, although the astrophysical
site(s) of the $r$-process has yet to be identified. Observations of the
stars with strongly enhanced $r$-process elements have placed important
constraints on the astrophysical site(s) of their synthesis.

\citet{sne1994} found that the extremely metal-poor
([Fe/H]\footnote{In the classical notation, where [X/H] = $\log(N_{\rm
    X}/N_{\rm H})_{star} - \log(N_{\rm X}/N_{\rm H})_{Sun}$.} $\sim -3.1$)
giant CS\,22892-052 is neutron-capture-rich, [Eu/Fe]$\simeq +1.6$ (following
the suggestion of \citealt{Beers2005}, we hereafter refer to stars having
$\mathrm{[Eu/Fe]} > +1$ and $\mathrm{[Ba/Eu]} < 0$ as r-II stars), and that
the relative abundances of nine elements in the range from Ba to Dy are
consistent with a scaled Solar system $r$-process abundance distribution.
Later studies of CS\,31082-001 \citep{hill02}, BD+17$^\circ$\,3248
\citep{cowan2002}, CS\,22892-052 \citep{sne03}, HD\,221170 \citep{ivans2006},
CS\,22953-003 \citep{francois2007}, HE\,1219-0312 and CS\,29491-069
\citep{hayek09} provided strong evidence for a universal production ratio of
the second $r$-process peak elements from Ba to Hf during the Galaxy history.
CS\,31082-001 \citep{hill02} provided the first solid evidence that
variations in progenitor mass, explosion energy, and/or other intrinsic and
environmental factors may produce significantly different $r$-process yields
in the actinide region ($Z \ge 90$). The third $r$-process peak ($76 \le Z \le
83$) is so far not well constrained, because, in most r-II stars, it is only sampled by abundance
measurements of two elements, Os and Ir. The abundances of platinum and gold were obtained for CS\,22892-052 \citep{sne03}. The only detection of lead in a r-II star so far is in CS31082-001 \citep{plez04}.

\citet{sne03} reported an underabundance of elements in the range of $40 < Z <
56$ relative to the scaled Solar $r$-process, which prompted a discussion
of multiple $r$-process sites \citep[see, for example,][]{Travaglio04, QW08,
  far09}. 
The detection of the radioactive elements thorium and uranium provided new
opportunities to derive the ages of the oldest stars and hence to determine a
lower limit for the age of the Universe \citep[see the pioneering papers
of][]{sne1996,cayr01}. It appears that all the r-II stars with measured Th (and
U) can be devided into two groups: (a) stars exhibiting an actinide
boost (e.g., CS\,31082-001, HE\,1219-0312), and (b) stars with no obvious
enhancement of thorium with respect to the scaled Solar r-process pattern
(e.g., CS\,22892-052, CS\,29497-004; for a full list of stars, see
\citealt{Roederer2009}). For the actinide boost stars, ages cannot be derived
when only a single radioactive element, either Th or U, is detected. 

To make clear an origin of the heavy elements beyond the iron group
in the oldest stars of the Galaxy, more and better measurements of additional elements are required. Currently, there are about ten r-II stars reported in the
literature
\citep{hill02,sne03,HERESpaperI,honda2004,HERESpaperII,francois2007,frebel2007,Lai2008,hayek09}.
Abundance pattern for a broad range of nuclei,
based on high-resolution spectroscopic studies, have been
reported for only six of these stars.

Continuing our series of papers on the Hamburg/ESO R-process-Enhanced Star
survey (HERES), we aim at extending our knowledge about synthesis of heavy
elements in the early Galaxy by means of a detailed abundance analysis of the
strongly $r$-process enhanced star {\LyudmilasStar}. In our study we also
investigate the reliability of multiple Th/$X$ chronometers for
{\LyudmilasStar}, where $X$ is an element in the Ba--Hf range.

{\LyudmilasStar} was identified as a candidate metal-poor
star in the Hamburg/ESO Survey (HES; see \citet{Christliebetal:2008a} for
details of the candidate selection procedures). Moderate-resolution
($\Delta\lambda = 2$\,{\AA}) spectroscopy obtained at the Siding Spring
Observatory (SSO) 2.3\,m-telescope with the Double Beam Spectrograph (DBS)
confirmed its metal-poor nature. Therefore, it was included in the target list
of the HERES project. A detailed description of the project and its aims can
be found in \citet[][hereafter Paper~I]{HERESpaperI}, and methods of automated
abundance analysis of high-resolution ``snapshot'' spectra have been described
in \citet[][hereafter Paper~II]{HERESpaperII}.  ``Snapshot'' spectra having
$R\sim 20,000$ and $S/N\sim 50$ per pixel at 4100\,{\AA} revealed that
{\LyudmilasStar} exhibits strong overabundances of the $r$-process elements,
with [Eu/Fe] = $+1.22$ and $\mathrm{[Ba/Eu]}=-0.56$ (Paper\,II).  

This paper is structured as follows. After describing the observations
(Section~\ref{Sect:Observations}), we report on the abundance analysis of
{\LyudmilasStar} in Sections~\ref{Sect:AbundanceAnalysis} and
\ref{Sect:Results}, based on high-quality VLT/UVES spectra and MAFAGS model
atmospheres \citep{Fuhr1}. The heavy element abundance pattern of {\LyudmilasStar} is discussed in
Section~\ref{Sect:ssr}. Section~\ref{Sect:ages} reports on the radioactive decay age determination. Conclusions are given in Section~\ref{Sect:DiscussionConclusions}.

\section{Observations}\label{Sect:Observations}

For the convenience of the reader, we list the coordinates and photometry of {\LyudmilasStar} in Table~\ref{Tab:CoordsPhotometry}. The photometry was taken from \citet{Beersetal:2007}. High-quality spectra of
this star was acquired during May--November 2005 with the VLT and UVES in
dichroic mode. The BLUE390$+$RED580 (4\,h total integration time) and
BLUE437$+$RED860 (10\,h) standard settings were employed to ensure a large
wavelength coverage. The slit width of $0.8''$ in both arms yielded a
resolving power of $R=60,000$. A $1\times 1$ pixel binning ensured a proper
sampling of the spectra. The observations are summarized in
Table~\ref{Tab:Observations}.

The pipeline-reduced spectra were shifted to the stellar rest frame and then co-added
in an iterative procedure in which pixels in the individual spectra affected
by cosmic ray hits not fully removed during the data reduction, affected by
CCD defects, or other artifacts, were identified. These pixels were flagged
and ignored in the final iteration of the co-addition. Both co-added blue
spectra have a signal-to-noise ratio ($S/N$) of at least 50 per pixel at
$\lambda > 3800$\,{\AA}.  At the shortest wavelengths, the $S/N$ of the
BLUE390 and BLUE437 is $10$ (at 3330\,{\AA}) and 70 (at 3756\,{\AA}),
respectively. The red arm spectra have $S/N > 100$ per pixel in most of the
covered spectral range.

%
%
\begin{table}[htbp]
 \centering
 \caption{\label{Tab:CoordsPhotometry} Coordinates and photometry of 
   {\LyudmilasStar}. }
  \begin{tabular}{lr}
   \hline\hline
   R.A.(2000.0) & 23:30:37.2\\
   dec.(2000.0) & $-$56:26:14\\
   $V$          & $13.881\pm 0.003$\\
   $B-V$        & $ 0.709\pm 0.005$\\
   $V-R$        & $ 0.456\pm 0.004$\\
   $V-I$        & $ 0.933\pm 0.005$\\\hline
  \end{tabular}
\end{table}

%
%
\begin{table}[htbp]
 \centering
 \caption{\label{Tab:Observations} VLT/UVES observations of
   {\LyudmilasStar}. }
  \begin{tabular}{ccrr}\hline\hline
   Setting  & $\lambda^1$  & \multicolumn{1}{c}{$t_{\mbox{\scriptsize exp}}$} & \multicolumn{1}{c}{$S/N^2$} \\\hline
   BLUE390  & $3330$--$4510$\,{\AA} & $ 4.0$\,h &  $10$--$65$ \\ 
   BLUE437  & $3756$--$4978$\,{\AA} & $10.0$\,h &  $70$--$120$\\ 
   REDL580  & $4785$--$5745$\,{\AA} & $ 4.0$\,h & $100$--$120$\\ 
   REDU580  & $5830$--$6795$\,{\AA} & $ 4.0$\,h &  $60$--$120$\\ 
   REDL860  & $6700$--$8508$\,{\AA} & $10.0$\,h & $100$--$180$ \\\hline 
\multicolumn{4}{l}{$ ^1$ \ $\lambda$ refers to rest frame wavelengths,} \\
\multicolumn{4}{l}{$ ^2$ \ $S/N$ refers to the signal-to-noise ratio per pixel.} \\
\end{tabular}
\end{table}

\begin{figure}[htbp]
  \centering
  \includegraphics[clip=true,bb=62 420 541 714,width=8.8cm]{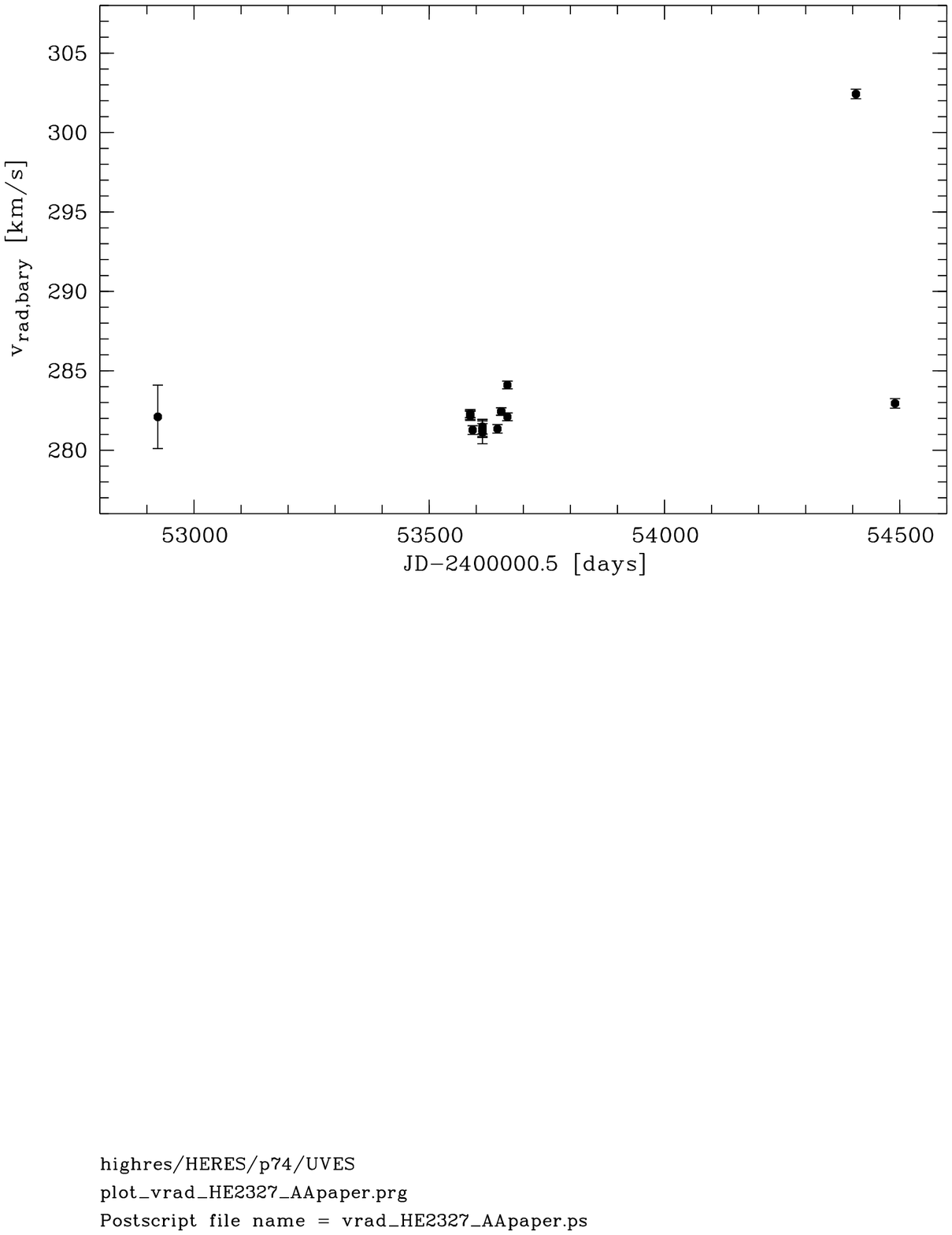}
  \caption{\label{Fig:vrad} Radial velocity measurements of {\LyudmilasStar}.}
\end{figure}

 Barycentric radial velocities of {\LyudmilasStar} as measured with
   Gaussian fits of selected absorption lines in our high-resolution spectra,   
covering $\sim 4.3$ years, indicate that the star 
is radial-velocity variable, although no signatures of a 
double-lined spectroscopic binary star were found.
The data taken during the
Modified Julian Date (MJD) period $53587.2$--$53666.2$ (Table~\ref{Tab:MJDvrad}) suggests that the
radial velocity varies on timescales of $\sim 10$\,days, and that the radial
velocity curve went through a minimum approximately at MJD 53620 (see
Fig.~\ref{Fig:vrad}). Furthermore, the measurement at MJD $54407.019$ deviates
by $\sim 20$\,km/s from the average of the radial velocities measured at the
other epochs, and by a similar amount from the measurement taken only about
three months later. The available data cannot be fitted satisfactorily by a
sinusoidal curve, and therefore we suspect that the orbit of the system is
highly elliptical. Additional observations are needed to confirm the
variability, and for determining the period as well as the nature of the
orbit.

\begin{table}[htbp]
 \centering
 \caption{\label{Tab:MJDvrad} Barycentric radial velocities
   $v_{\mbox{\scriptsize rad,bary}}$ of {\LyudmilasStar}.  
}
  \begin{tabular}{lcc}\hline\hline
    MJD$^*$    & ~~~$v_{\mbox{\scriptsize rad,bary}}$~~~ & $\sigma^{**}$\\
    $\mathrm{[days]}$ & [km/s]                 & [km/s]  \\\hline
    $52923.137$ & $282  $ & $2  $\\
    $53587.172$ & $282.2$ & $0.3$\\
    $53587.215$ & $282.3$ & $0.2$\\
    $53587.257$ & $282.3$ & $0.3$\\
    $53587.321$ & $282.2$ & $0.3$\\
    $53592.289$ & $281.3$ & $0.3$\\
    $53613.092$ & $281.3$ & $0.4$\\
    $53613.135$ & $281.5$ & $0.5$\\
    $53613.178$ & $281.1$ & $0.7$\\
    $53613.221$ & $281.4$ & $0.6$\\
    $53645.081$ & $281.3$ & $0.3$\\
    $53653.044$ & $282.4$ & $0.2$\\
    $53653.087$ & $282.4$ & $0.2$\\
    $53666.112$ & $284.1$ & $0.2$\\
    $53666.157$ & $284.0$ & $0.2$\\
    $54407.020$ & $302.4$ & $0.3$\\
    $54490.027$ & $282.9$ & $0.3$\\\hline
\multicolumn{3}{l}{$ ^*$ The date refers to the start of the exposure.} \\
\multicolumn{3}{l}{$ ^{**}$ The error bars are the 1\,$\sigma$ scatter of the} \\
\multicolumn{3}{l}{ \ \ \ \ measurements based on different lines.} \\
\end{tabular}
\end{table}

\section{Analysis method}\label{Sect:AbundanceAnalysis}

Our determinations of the stellar parameters and the elemental abundances are
based on line profile and equivalent width analyses. We ignored any lines with equivalent widths larger than 100\,m{\AA}. An exception is the elements, such as strontium, where only strong lines can be detected in {\LyudmilasStar}. For a number of chemical
species, namely, \ion{H}{i}, \ion{Na}{i}, \ion{Mg}{i}, \ion{Al}{i},
\ion{Ca}{i-ii}, and \ion{Fe}{i-ii}, non-local thermodynamic equilibrium
(NLTE) line formation was considered. The theoretical spectra of the remaining
elements were calculated assuming LTE. The coupled radiative transfer and
statistical equilibrium equations were solved with the code NONLTE3
\citep{code-sakh,code-nonlte3} for \ion{H}{i}\, and \ion{Na}{i}, and an
updated version of the DETAIL code \citep{code-detail} for the remaining NLTE
species. The departure coefficients were then used to calculate the synthetic
line profiles with the code SIU \citep{Reetz}. The metal line list has been
extracted from the {\sc VALD} database \citep{vald}. For molecular lines, we
applied the data compiled by \citet{Kur94a}. In order to compare with
observations, computed synthetic profiles are convolved with a profile that
combines instrumental broadening with a Gaussian profile of 3.6\,\kms\ and
broadening by macroturbulence. We employ the macroturbulence parameter $\Vmac$ in the radial-tangential form as prescribed in \citet{Gray1992}. From analysis of many line profiles in the spectrum of {\LyudmilasStar}, $\Vmac$ = 3.3\,\kms\ was empirically found with some allowance to vary by $\pm$0.3\,\kms\ (1$\sigma$).

The abundance analysis based on equivalent widths was performed with the code
WIDTH9\footnote{\tt http://kurucz.harvard.edu/programs/WIDTH/} \citep{width9}.
The SIU and WIDTH9 codes both treat continuum scattering correctly; i.e.,
scattering is taken into account not only in the absorption coefficient, but
also in the source function.

Both in SIU and WIDTH9, we used the updated partition functions from the
latest release of the MOOG code\footnote{\tt
  http://verdi.as.utexas.edu/moog.html} \citep{MOOG}, with exception of
\ion{Ho}{ii} and \ion{Ir}{ii}. For \ion{Ho}{ii}, we adopted the partition
function calculated by \citet{ho2-pf}. The \ion{Ir}{ii} partition function was
revised based on the measured energy levels from \citet{levels-Ir2}. For the
temperature range we are concerned with this makes a difference of
 +0.08/ +0.2\,dex in the Ho/Ir abundance determined from the
\ion{Ho}{ii}/\ion{Ir}{i} lines. 

\subsection{Stellar parameters and atmospheric models}\label{Sect:stellarParameters}

In Paper~II, an effective temperature of $5048\pm 100$\,K was derived from
photometry when adopting the reddening derived from the maps of
\citet{Schlegel1998}.  A subsequent analysis of the snapshot spectrum gave $
\log g = 2.22\pm0.25$ and [Fe/H] = $-2.95\pm 0.12$. For the stellar parameter
determination and abundance analysis, we use plane-parallel, LTE,
and line-blanketed MAFAGS model atmospheres \citep{Fuhr1}. Enhancements of the
$\alpha$-elements Mg, Si, and Ca by the amounts determined in a close-to-final
iteration of our analysis were taken into account when computing these model
atmospheres. Since suitable lines of oxygen are not covered by our spectra, we
could not determine the oxygen abundance, hence we adopted
$\mathrm{[O/Fe]} = +0.5$, which is a typical value of other stars at the
metallicity of {\LyudmilasStar}. It is worth noting that oxygen in cool
stellar atmospheres plays a minor role as a donator of free electrons and as an 
opacity source, hence an uncertainty in the oxygen abundance does not significantly
affect the calculated atmospheric structure.

\citet{Heiter2006} investigated the effect of geometry on atmospheric
structure and line formation for Solar abundance models, and concluded that
plane-parallel models can be applied in abundance analysis for the stars with
$\log g > 2$ and $\teffm > 4000$\,K. Therefore, {\LyudmilasStar} lies in the
stellar parameter range where the usage of plane-parallel models is
appropriate.  This is confirmed by flux and abundance comparisons between a
MAFAGS plane-parallel and MARCS \citep{Gustafssonetal:2008}\footnote{\tt
  http://marcs.astro.uu.se} spherical models with stellar parameters close to
those of {\LyudmilasStar}; i.e., $\teffm/\log g/\mathrm{[M/H]} = 5000/2.0/-3$.
Synthetic spectra were computed in the wavelength range 3500--16,000\,{\AA}
with the code SIU, which solves the equation of radiative transfer in only one
depth variable. For the absolute flux, we compared three different
combinations of model atmosphere and spectrum synthesis geometries:
consistently plane-parallel (MAFAGS + SIU, $p-p$), inconsistent (MARCS + SIU,
$s-p$), and consistently spherical (MARCS model atmosphere library, $s-s$). The
difference in absolute flux between these three models does not exceed
0.001\,dex for wavelengths longer than 6600\,{\AA}. For $\lambda <
6600$\,{\AA}, the $p-p$ and $s-p$ fluxes are lower than that from the $s-s$
model with a maximum difference of 0.01\,dex and 0.02\,dex, respectively, at
wavelengths around 3500\,\AA.

\begin{figure}[htbp]
  \resizebox{85mm}{!}{\includegraphics{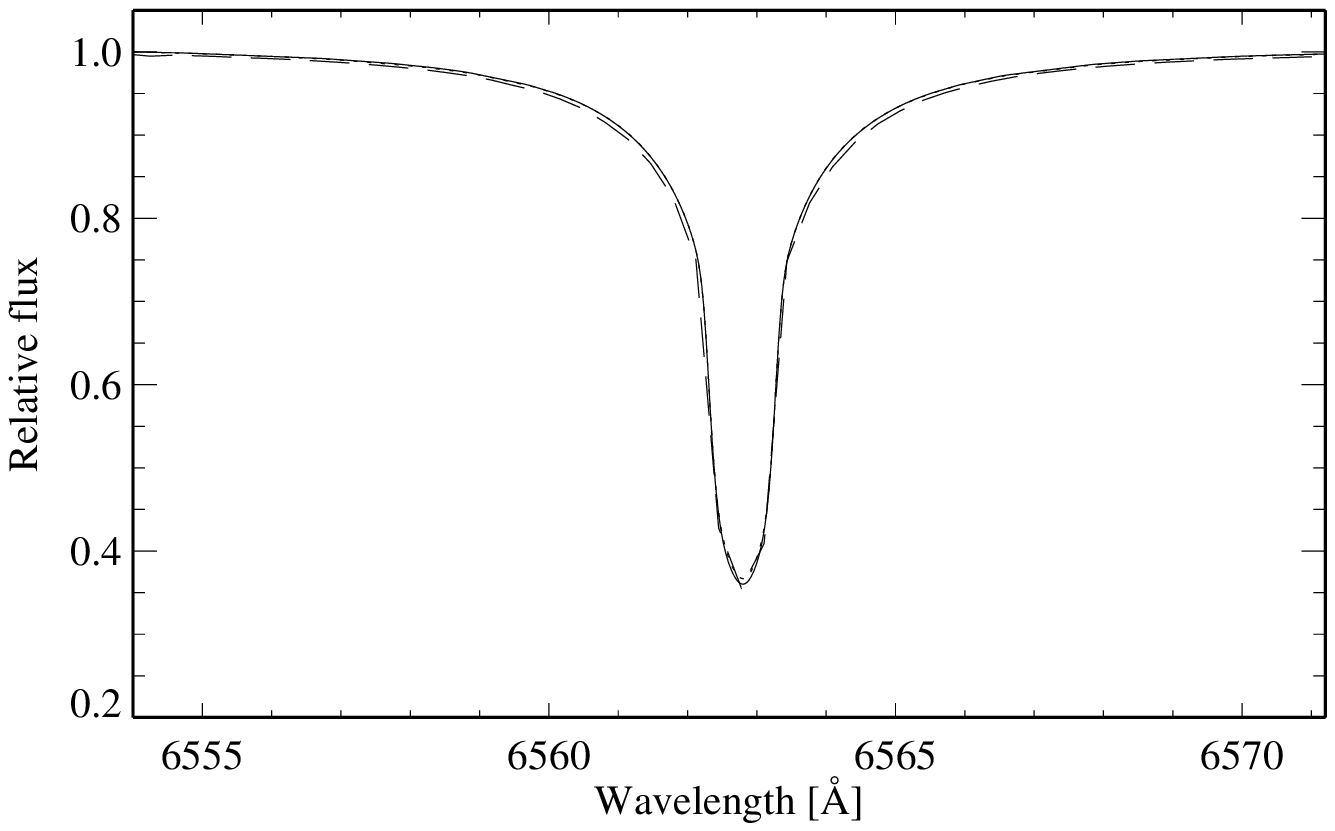}}

  \vspace{-8mm}
  \resizebox{85mm}{!}{\includegraphics{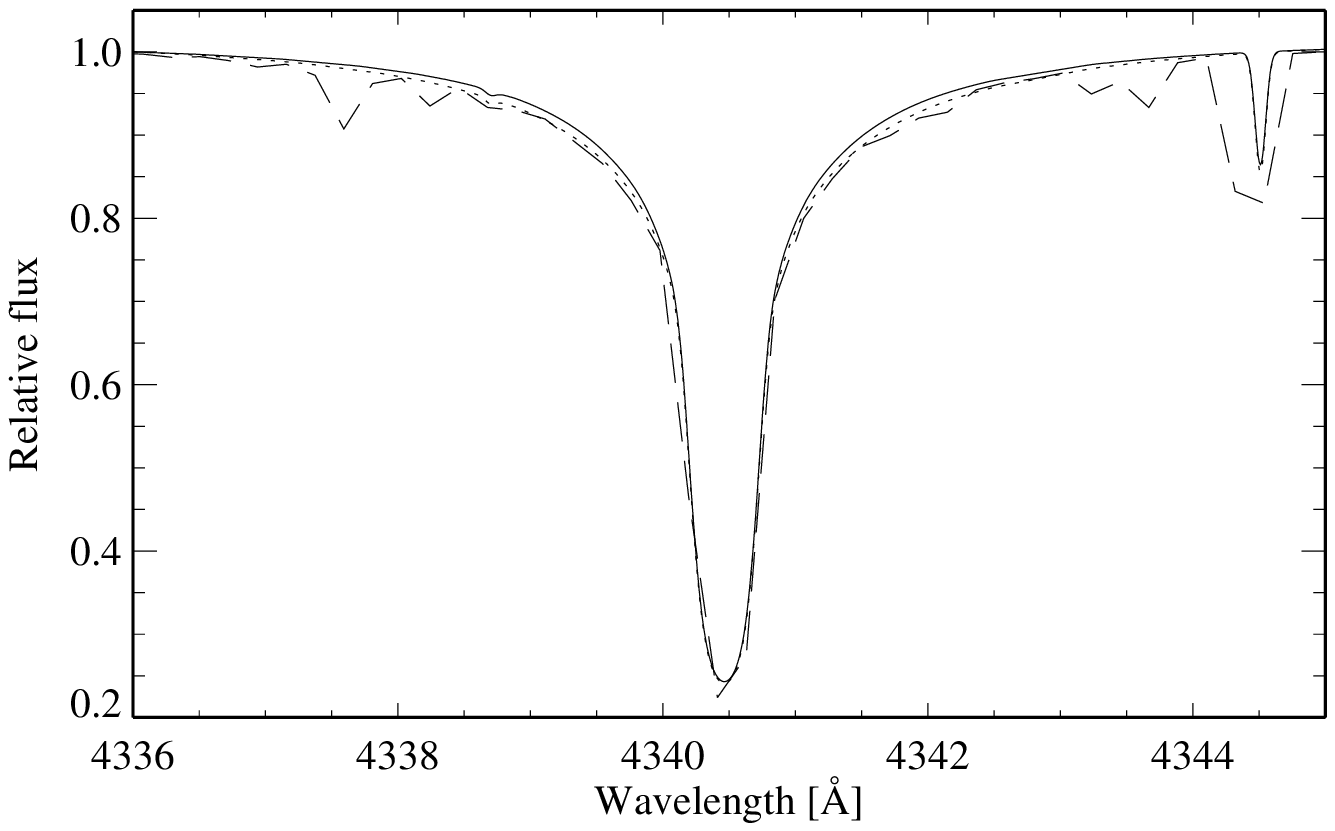}}
  \caption{\label{Fig:Hlines} Synthetic profiles of H$\alpha$ (top
    panel) and H$\gamma$ (bottom panel) from the $s-s$ (dashed curve), $s-p$
    (continuous curve), and $p-p$ (dotted curve) models. The calculations for
    the $s-p$ and $p-p$ models were made for pure hydrogen lines.}
\end{figure}

In Fig.~\ref{Fig:Hlines}, we show the line profiles of H$\alpha$ and H$\gamma$
for all three models. The H$\gamma$ profile from the MAFAGS model is
consistent with that from the $s-s$ model. The difference in H$\alpha$
relative fluxes between the MAFAGS and $s-s$ models translates to an effective
temperature difference of 60\,K. The abundance differences for the selected
spectral lines were obtained between the $p-p$ and $s-p$ models by fitting the
calculated synthetic spectra to the observed ones. The difference in absolute
abundances, $\Delta\eps{}((p-p)-(s-p))$, is always negative, but does not exceed
0.01 and 0.02\,dex for the lines of neutral and ionized species, respectively.
The differences in abundance ratios are also negligible; i.e., smaller than
0.01\,dex.

\begin{figure}[htbp]
  \resizebox{88mm}{!}{\includegraphics{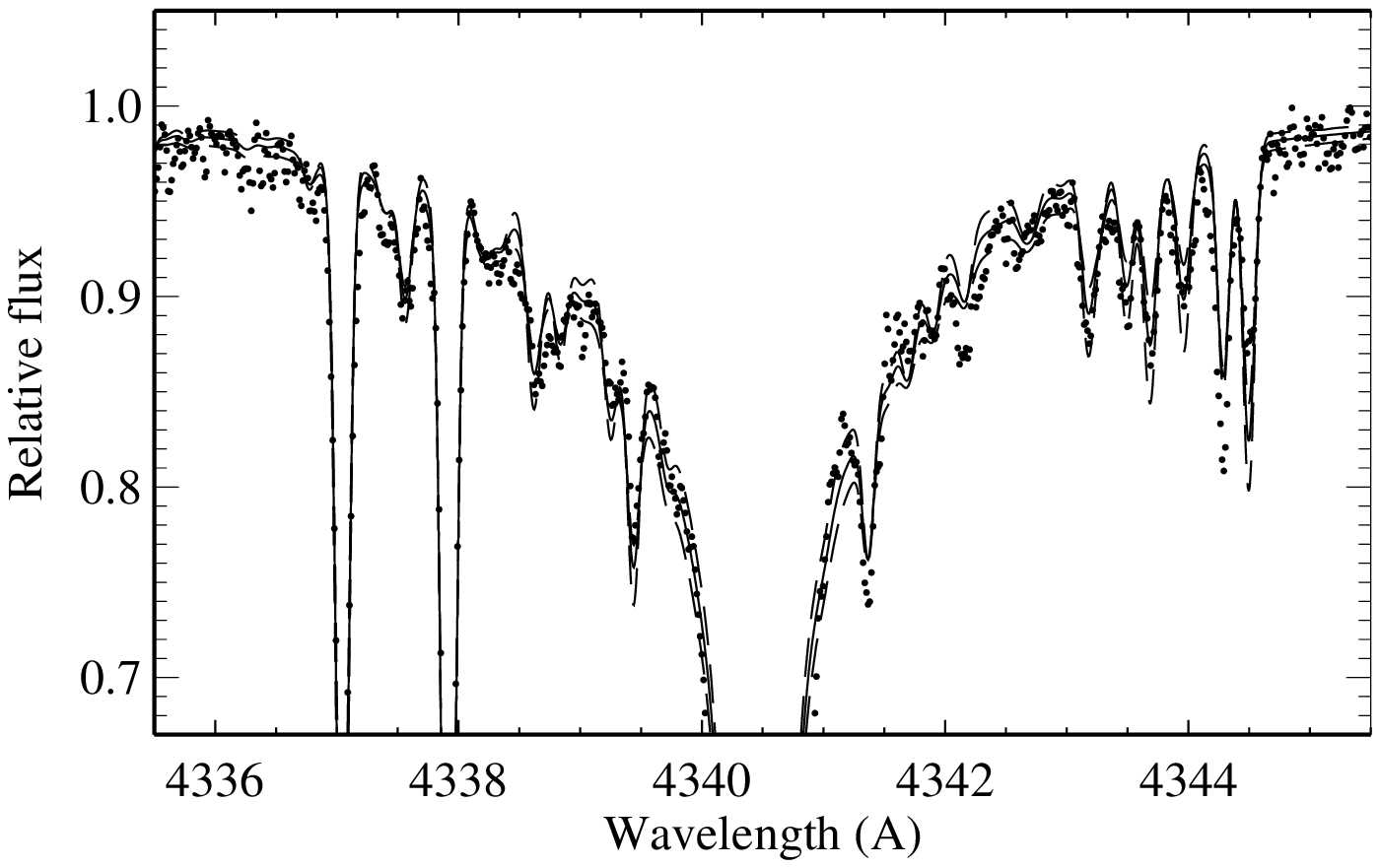}}

  \vspace{-5mm}
  \resizebox{88mm}{!}{\includegraphics{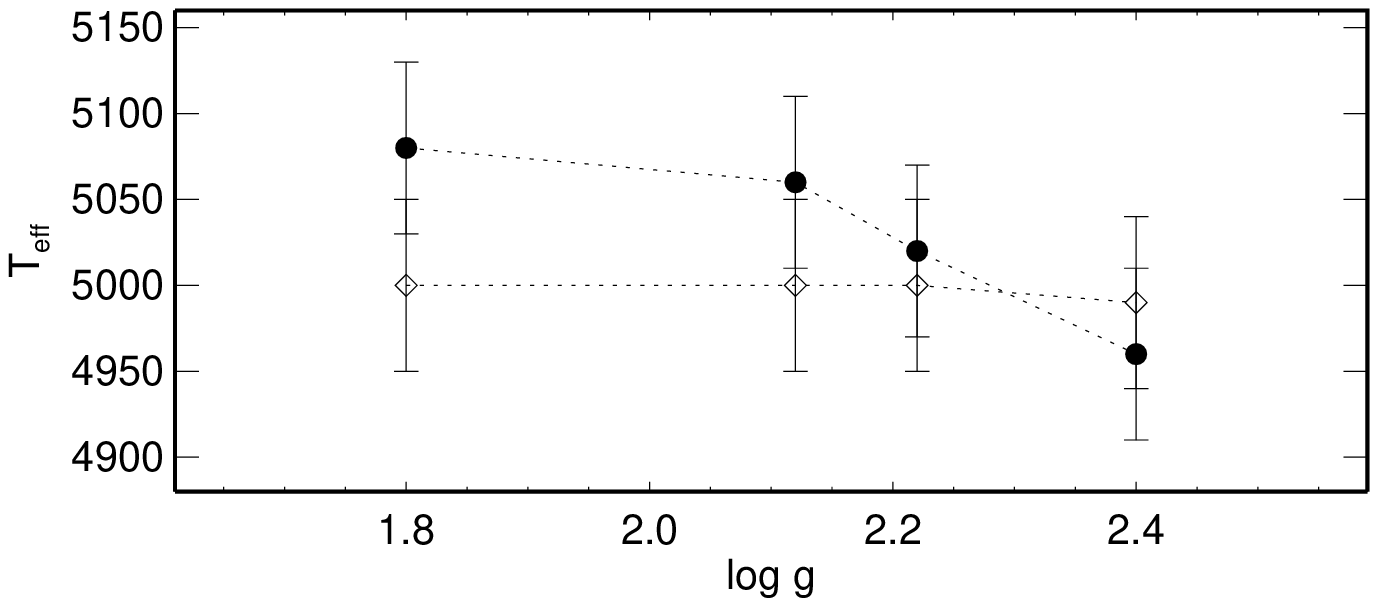}}
  \caption{\label{Fig:teff} Top panel: synthetic flux profile of H$\gamma$ 
    computed for $\teffm = 5000$\,K (continuous curve) compared to the
    observed spectrum of {\LyudmilasStar} (bold dots). The dashed curves show
    the effect of a 80\,K variation in the effective temperature on the
    synthetic spectrum. In all calculations, we assumed $\log g = 2.3$,
    $\mathrm{[Fe/H]} = -2.95$, and $\xi = 1.7$\,\kms. Bottom panel: effective
    temperature derived from the H$\alpha$ (filled circles) and H$\gamma$
    (open diamonds) line wings in {\LyudmilasStar} as a function of surface
    gravity. The error bars show the uncertainty of {\tefft} arising from
    profile fitting.}
\end{figure}

The effective temperature of {\LyudmilasStar} was also determined from a profile
analysis of H$\alpha$ and H$\gamma$ based on NLTE line formation calculations
of \ion{H}{i} using the method described by \citet{ml-nlte-H}.  Only these two
lines were employed because an accurate continuum rectification was not
possible in the spectral regions covering other Balmer lines. The metallicity
and microturbulence velocity from Paper\,II were adopted during the analysis
of these Balmer lines, while the gravity was varied between $\log g = 1.8$ and
2.4. The theoretical profiles of H$\alpha$ and H$\gamma$ were computed by
convolving the profiles resulting from the thermal, natural, and Stark
broadening \citep{vcs70,vcs73}, as well as self-broadening. For the latter, we
use the self-broadening formalism of \citet{BPO}.

We obtained weak NLTE effects for the H$\gamma$ profile beyond
the core, such that the difference in {\tefft} derived from this line
between NLTE and LTE does not exceed 20\,K. It was also found that the H$\gamma$
line wings are insensitive to a variation in surface gravity in the stellar
parameter range we are concerned with. The best fit is achieved at $\teffm
= 5000$\,K. Figure\,\ref{Fig:teff} (top panel) illustrates the quality of the
fits. 

Based on the $S/N$ ratio of the observed spectrum and the sensitivity of the
Balmer lines to variations of {\tefft}, we estimate the uncertainty of
{\tefft} arising from profile fitting as 50\,K for each line. For H$\alpha$,
NLTE leads to a weakening of the core-to-wing transition compared to the LTE
case, resulting in a 80--100\,K higher {\tefft}, depending on surface gravity.
The effective temperature of {\LyudmilasStar} obtained from H$\alpha$ is also
dependent on $\log g$, as shown in bottom panel of Fig.~\ref{Fig:teff}. The
temperature deduced from combining the analyses of H$\alpha$ and H$\gamma$ is
$\teffm = 5000\pm70$\,K, and a favorable range of $\log g$ is between 1.95 and
2.40.

The surface gravity and microturbulence velocity were redetermined from Ca and
Fe lines based on NLTE line formation for \ion{Ca}{i-ii} and \ion{Fe}{i-ii},
using the methods of \citet{ml-nlte-ca,ml-nlte-fe}. For \ion{Ca}{i-ii}, we
employ the lines listed in \citet{ml-nlte-ca} along with the atomic data 
on $gf-$values and  van der Waals damping constant. In total, 8 lines of
\ion{Ca}{i} and the only suitable \ion{Ca}{ii} line covered by our spectra, at
8498\,{\AA}, were used. For \ion{Fe}{i-ii}, 49 lines of \ion{Fe}{i} and 8
lines of \ion{Fe}{ii} were selected from the linelists of \citet{ml-nlte-fe},
Paper\,II, \citet{jonsell}, and \citet{ivans2006}. 
Van der Waals broadening of the Fe lines was accounted for
with the most accurate data available from 
calculations of \citet{omara_sp,omara_pd,omara_ion,omara_df,omara-fe2}. 
Hereafter, these collected papers by  Anstee, Barklem, and 
O'Mara are referred to as the $ABO$ theory. The lines used are
listed in Table~\ref{linelist} (Online material) together with transition
information, references for the adopted $gf-$values, and final element
abundances.

For Ca and Fe, we apply a line-by-line differential NLTE approach, in the
sense that stellar line abundances are compared with individual abundances of
their Solar counterparts. It is worth noting that, with the adopted atomic
parameters, the absolute Solar NLTE abundances obtained from the two
ionization stages, \ion{Ca}{i} and \ion{Ca}{ii}, \ion{Fe}{i} and \ion{Fe}{ii},
were consistent within the error bars: $\eps{\odot}$(\ion{Ca}{i}) =
$6.36\pm0.06$, $\eps{\odot}$(\ion{Ca}{ii}\,8498\,\AA) = $6.29$,
$\eps{\odot}$(\ion{Fe}{i}) = $7.47\pm0.10$, and $\eps{\odot}$(\ion{Fe}{ii}) =
$7.46\pm0.07$ (we refer to abundances on the usual scale, where $\eps{H} =
12$).

NLTE computations were performed for a small grid of model atmospheres with
two effective temperatures, namely $\teffm = 5050$\,K, as derived from
photometry, and $\teffm = 4980$\,K, which is close to the result
of the Balmer line analysis. In the statistical equilibrium calculations,
inelastic collisions with hydrogen atoms were accounted for, using the
\citet{hyd} formula with a scaling factor of $S_H = 0.1$ for Ca and $S_H = 1$ for
Fe, as recommended by \citet{ml-nlte-ca,ml-nlte-fe}. 
The NLTE calculations for \ion{Ca}{i-ii} and \ion{Fe}{i-ii} were iterated for various
elemental abundances until agreement between the theoretical and observed
spectra was reached. The gravity was varied between
$\log g = 1.8$ and 2.6 in steps of 0.2\,dex. Microturbulence values were
tested in the range between $\xi = 1.5$ and 2.1\,\kms\ in steps of 0.1\,\kms.
It was found that $\xi = 1.5$, 2.0, and 2.1\,\kms\ led to a steep trend of
abundances found from individual \ion{Fe}{i} lines with equivalent width,
independently of the adopted values of {\tefft} and $\log g$, therefore these
values can be excluded. 

\begin{figure} 
  \resizebox{88mm}{!}{\includegraphics{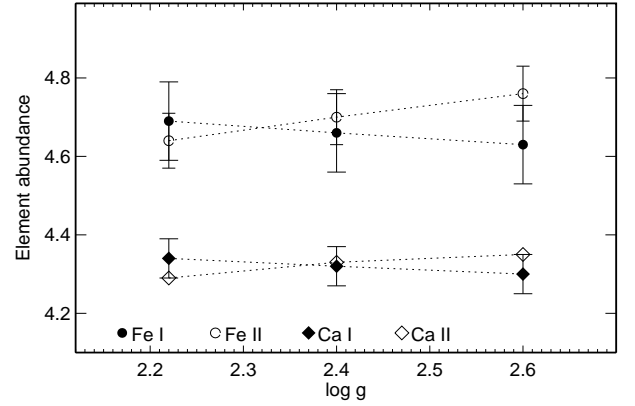}}
  \caption{\label{Fig:logg} NLTE abundances of \ion{Fe}{i} (filled circles)
    and \ion{Fe}{ii} (open circles), \ion{Ca}{i} (filled diamonds) and
    \ion{Ca}{ii} (open diamonds) in {\LyudmilasStar} as a function of surface
    gravity. For better illustration, the symbols for Ca are shifted upwards
    by 0.5\,dex. The calculations are for $\teffm = 5050$\,K, $\mathrm{[Fe/H]}
    = -2.78$, and $\xi = 1.8$\,\kms.}
\end{figure}

\begin{figure} 
  \resizebox{88mm}{!}{\includegraphics{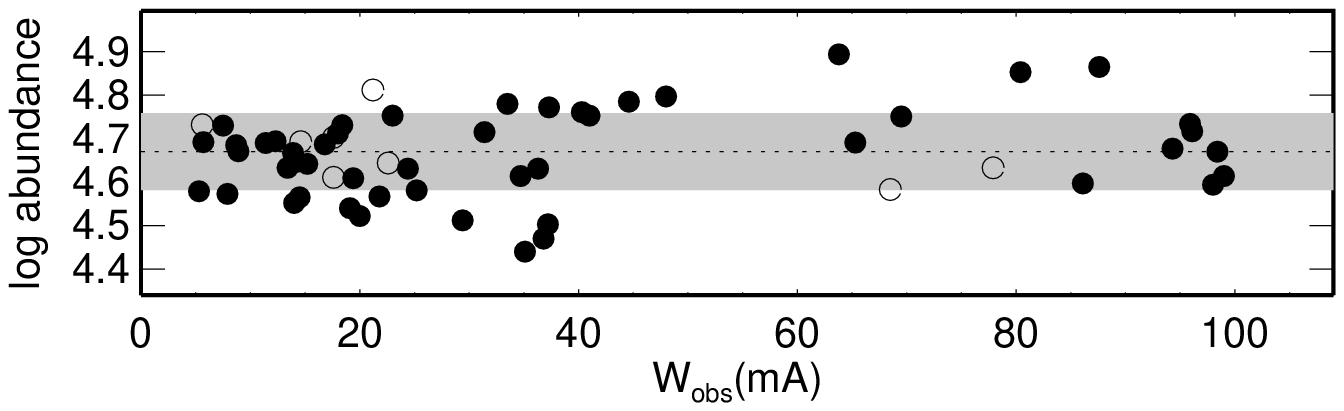}}

  \vspace{-5mm}
  \resizebox{88mm}{!}{\includegraphics{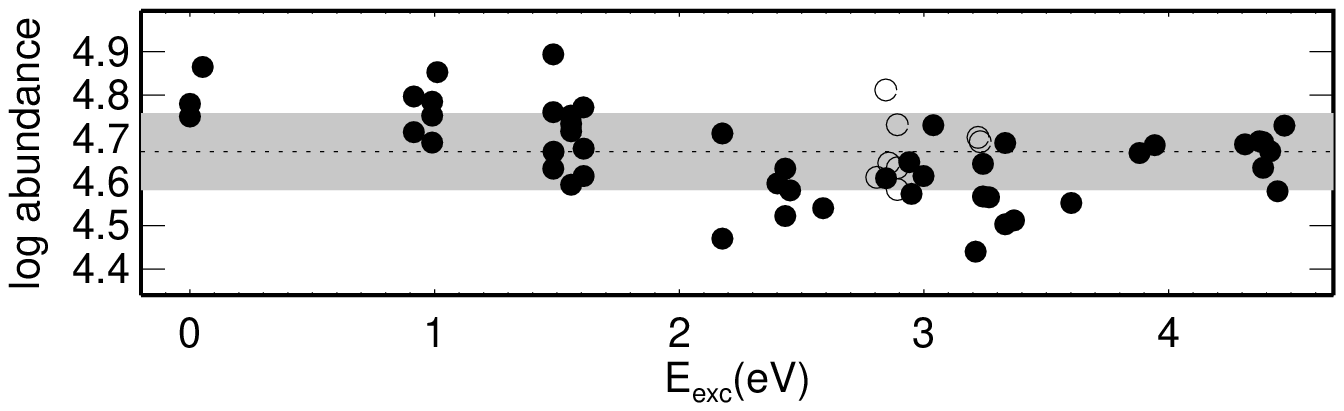}}

  \caption{\label{Fig:fe_w_exc}  Trends of abundances with equivalent width
    and excitation potential, as determined from individual \ion{Fe}{i}
    (filled circles) and \ion{Fe}{ii} (open circles) lines, using our adopted
    stellar parameters. The dotted line indicates the mean Fe abundance from
    two ionization stages and the shaded grey area its statistical error.}
\end{figure}

Adopting $\teffm = 5050$\,K, we obtain consistent iron abundances for the two
ionization stages if $\log g = 2.34$, 2.32, and 2.32, and $\xi$ value, 1.7,
1.8, and 1.9\,\kms, respectively. For Ca, this is achieved for $\log g =
2.28$, 2.37, and 2.47. Fig.~\ref{Fig:logg} illustrates the determination of
the surface gravity from the ionization equilibrium of \ion{Fe}{i/ii} and
\ion{Ca}{i/ii} when the remaining stellar parameters are fixed at $\teffm =
5050$\,K, $\mathrm{[Fe/H]} = -2.78$, and $ \xi = 1.8$\,\kms. When adopting
$\teffm = 4980$\,K, the difference in $\log g$ obtained from Fe and Ca does
not exceed 0.1\,dex if $\xi =$ 1.7\,\kms. Thus, we find two possible
combinations of stellar parameters for {\LyudmilasStar}: (a) 5050/2.34/$-2.78$
with $\xi = 1.7$--$1.8$\,\kms, and (b) 4980/2.23/$-2.85$ with $\xi =
1.7$\,\kms (see Fig.~\ref{Fig:fe_w_exc} for the combination
5050/2.34/$-2.78$). In fact, both sets of the obtained parameters are
consistent with each other within the uncertainties of the stellar parameters.

For consistency reasons, we adopt the effective temperature adopted in
Paper\,II, i.e., $\teffm = 5050\pm70$\,K, and the other stellar parameters as
determined in this study, i.e., $\log g = 2.34\pm0.1$, $\mathrm{[Fe/H]} =
-2.78\pm0.09$, and $ \xi = 1.8\pm0.1$\,\kms\ (Table~\ref{Tab:StellarParameter}). With the obtained effective temperature and surface gravity, the spectroscopic distance of {\LyudmilasStar} is estimated as 4.4 to 4.9 kpc for the star mass of 0.8 to 1 solar mass.

\begin{table} 
 \centering
 \caption{\label{Tab:StellarParameter} Determined stellar parameters of 
   {\LyudmilasStar}.}
  \begin{tabular}{ccl}
   \hline\hline
   Parameter & Value & Uncertainty \\
   \hline
   {\tefft} & 5050 K & $\pm70$K \\
   $\log g$ & 2.34   & $\pm0.1$\\
   $\mathrm{[Fe/H]}$ & $-2.78$ & $\pm0.09$\\
   $ \xi$        & 1.8\,\kms & $ \pm0.1$\kms \\
   \hline
  \end{tabular}
\end{table}

\subsection{Line selection and atomic data}

The lines used in the abundance analysis were selected from the lists
of Paper\,II, \citet{jonsell}, \citet{Lawler_Tb,Lawler_Ho},
\citet{sneden2009}, and \citet{ivans2006}. For atomic lines, we have
endeavored to apply single-source and recent $gf-$values wherever possible, in
order to diminish the uncertainties involved by combining studies that may not
be on the same $gf-$value system. For the selected lines of \ion{Na}{i},
\ion{Mg}{i}, \ion{Al}{i}, \ion{Ca}{i-ii}, \ion{Sr}{ii}, and \ion{Ba}{ii}, we
adopted $gf-$values (mostly from laboratory measurements) and van der Waals
damping constants which were carefully inspected in our previous analyses of
the Solar spectrum (see \citealt{ml-nlte-H} for references).

Fortunately, most neutron-capture element species considered here have been
subjected to extensive laboratory investigations within the past two decades
\citep{Biemont_Yb,Hartog_Nd,Hartog_Gd,Ivarsson_Pr,Ivarsson_Os_Ir,Lawler_La,Lawler_Eu,Lawler_Tb,Lawler_Ho,Lawler_Sm,Lawler_Hf,Lawler_Er,Lawler_Ce,Ljung_Zr,Nilsson_Th,Wickliffe_Tm,Wickliffe_Dy,Xu_Pd}.
We employed $gf-$values determined in these recent laboratory efforts.

Molecular data for two species, CH and NH, were assembled for the abundance
determinations of carbon and nitrogen. For the analysis of the the $A-X$
bands at 4310--4313\,{\AA} and 4362--4367\,{\AA}, we used the CH line list of
Paper\,II, and we use the $^{13}$CH line list described in \citet{hill02}.
The NH molecular line data for the $A-X$ band at 3358--3361\,{\AA} was taken
from \citet{Kur93}. 

 The van der Waals damping for atomic lines was computed following 
the $ABO$ theory, where the data is available, 
using the van der Waals damping
constants $\Gamma_6/N_{\rm H}$ at 10\,000~K as provided by the VALD 
database \citep{vald}. It is worth noting that the correct temperature 
dependence of the ABO theory was accounted for.
 An exception is the selected lines of some elements, 
for which we use the $C_6-$values derived from the 
solar line profile fitting by \citet[][\ion{Na}{i}, \ion{Mg}{i}, 
and \ion{Al}{i}]{mg_c6} and \citet[][\ion{Sr}{ii} and \ion{Ba}{ii}]{ml-nlte-H}.
 If no other data is available, the \citet{Kur94b} $\Gamma_6/N_{\rm H}$ values were employed. 

Many elements considered here are represented in nature by either a single
isotope having an odd number of nucleons (Sc, Mn, Co, Pr, Tb, Ho, and Tm; $
^{139}$La accounts for 99.9\%\ of lanthanum according to \citet{Lodders2003}), or
multiple isotopes with measured wavelength differences ($\Delta\lambda \ge
0.01$\,{\AA} for \ion{Ca}{ii}, \ion{Ba}{ii}, \ion{Nd}{ii}, \ion{Sm}{ii},
\ion{Eu}{ii}, \ion{Yb}{ii}, \ion{Ir}{i}).  Nucleon-electron spin interactions
in odd-$A$ isotopes lead to hyper-fine splitting (HFS) of the energy levels
resulting in absorption lines split into multiple components. Without
accounting properly for HFS and/or isotopic splitting (IS) structure,
abundances determined from the lines sensitive to these effects can be
severely overestimated. For example, in {\LyudmilasStar}, including HFS makes
a difference of $-0.49$\,dex in the Ba abundance derived from the \ion{Ba}{ii}
4554\,{\AA} line, and including IS leads to a 0.13\,dex lower Ca abundance for
\ion{Ca}{ii} 8498\,\AA.

Notes regarding whether HFS/IS were considered in a given feature, and
references to the used HFS/IS data are presented in Table~\ref{linelist}
(Online material). For a number of features, it was helpful to use the data on
wavelengths and relative intensities of the HFS/IS components collected in the
literature by \citet[][\ion{Sc}{ii}, \ion{Mn}{i-ii},\ion{Co}{i}, \ion{La}{ii},
\ion{Tb}{ii}, \ion{Ho}{ii}, and \ion{Yb}{ii}]{jonsell}, \citet[][\ion{Sc}{ii}
and \ion{La}{ii}]{ivans2006}, \citet[][\ion{Nd}{ii} and
\ion{Sm}{ii}]{Roederer_sm_nd}, and \citet[][\ion{Ir}{i}]{cowan2005}.

The selected lines are listed in Table~\ref{linelist} (Online material),
along with the transition information and references to the adopted
$gf-$values.

\section{Abundance results}\label{Sect:Results}

We derived the abundances of 40 elements from Li to Th in {\LyudmilasStar},
and for four elements among them (Ca, Ti, Mn, and Fe), from two ionization
stages.  In Table~\ref{linelist} (Online material) we list the results
obtained from individual lines. For every feature, we provide the obtained LTE
abundance and, for selected species, also the NLTE abundance.  In
Table~\ref{Tab:AbundanceSummary}, we list the mean abundances, dispersion of
the single line measurements around the mean ($\sigma_{\eps{}}$), and the
number of lines used to determine the mean abundances. Also listed are the
Solar photosphere abundances, $\eps{\sun}$, adopted from \citet{Lodders2009}, and the
abundances relative to iron, [X/Fe]. For the computation of [X/Fe],
[Fe/H]$_{\rm NLTE}= -2.78$ has been chosen as the reference, with exception of
the neutral species calculated under the LTE assumption, where the reference
is [Fe~I/H]$_{\rm LTE}= -2.88$.  Below we comment on individual groups of
elements. The sample of cool giants from \citet{cayr04} was chosen as our comparison sample.

\subsection{Li and CNO}\label{Sect:CNO}

With an equivalent width of 15\,m{\AA}, the \ion{Li}{i} 6708\,{\AA} line is
easily detected in this star. The abundance was determined using the spectrum
synthesis approach, to account for the multiple-component structure of the
line caused by the fine structure of the upper energy level and
the existence of two isotopes, $^7$Li and $^6$Li. The calculations of the
synthetic spectra were performed in two ways: (a) without $^6$Li, and (b)
adopting the Solar isotopic ratio, i.e., $^7$Li : $^6$Li = 92.4 : 7.6
\citep{Lodders2003}. In both cases, the result for the Li abundance is
$\eps{LTE}(\mathrm{Li}) = 0.99$. A goodness-of-fit analysis suggests an
asymmetry of the \ion{Li}{i} 6708\,{\AA} line, which could be attributed to a
weak $^6$Li feature in the red wing of the $^7$Li line. Despite that such an asymmetry could also be convection-related \citep{cayrel-li}, we cannot rule out the presence of a significant amount of $^6$Li in {\LyudmilasStar}.
The departures from LTE result in only a
minor increase of the derived lithium abundance; i.e, by 0.04\,dex according
to recent calculations of \citet{lind_li}.

With an abundance of $\eps{ }$(Li) = 0.99, {\LyudmilasStar} is located well
below the lithium plateau for halo stars near the main-sequence turnoff, as
expected for a red giant \citep{iben1967}.

\begin{figure} 
  \resizebox{88mm}{!}{\includegraphics{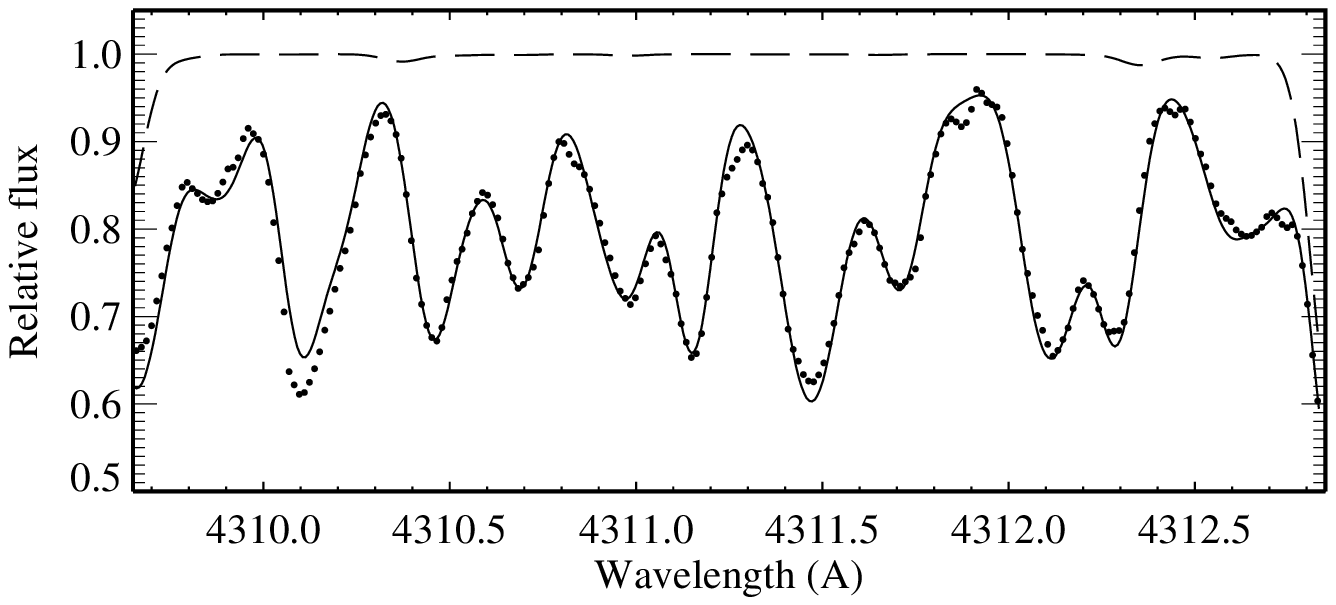}}
  \resizebox{88mm}{!}{\includegraphics{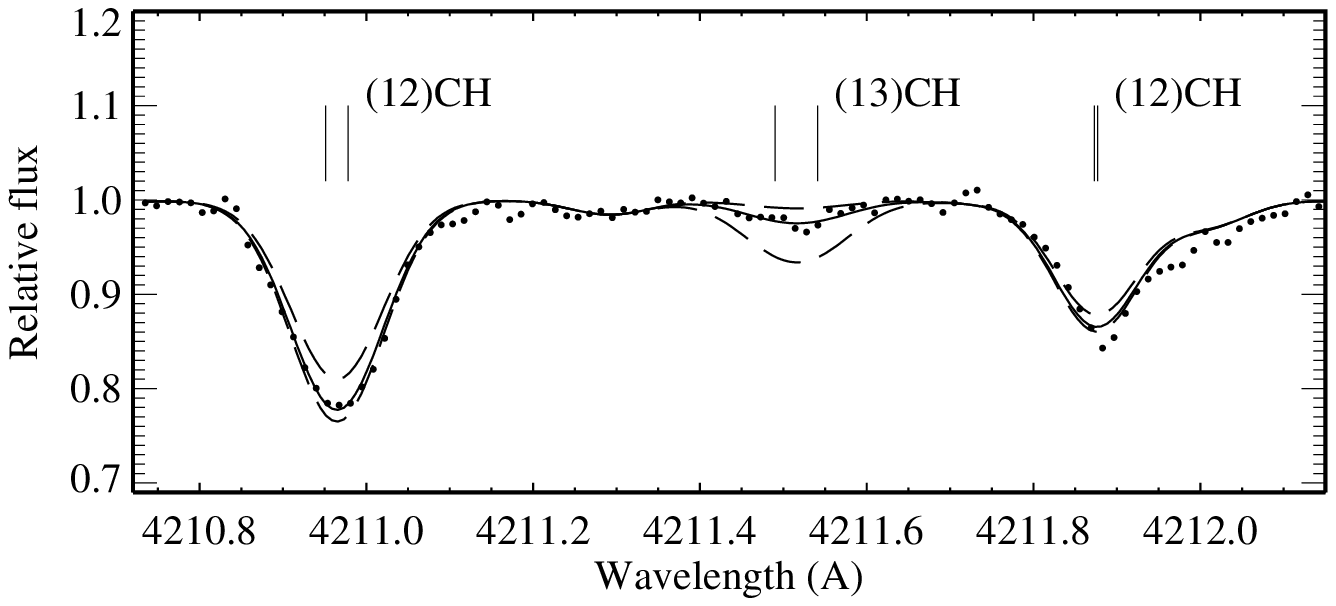}}
  \resizebox{88mm}{!}{\includegraphics{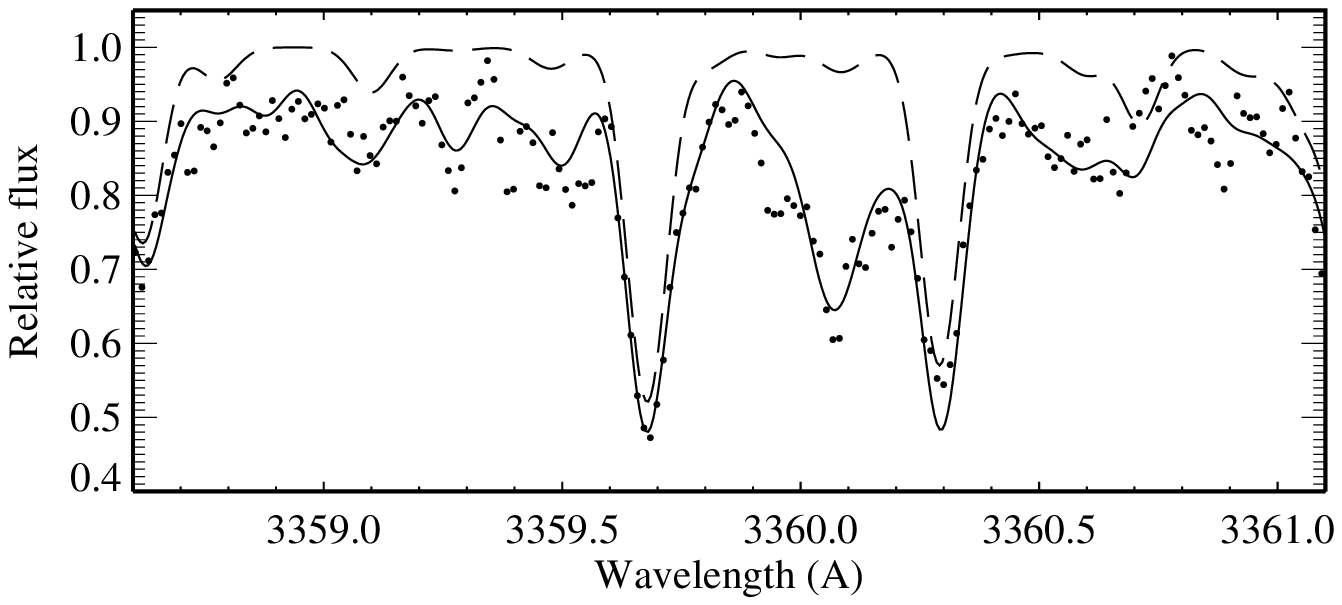}}
  \caption{\label{Fig:ch_nh_lines} Best fits (continuous curve) of the CH 
    features near 4310\,{\AA} (top panel) and 4211\,{\AA} (middle panel), and
    the NH molecular band near 3360\,{\AA} (bottom panel). The observed
    spectrum of {\LyudmilasStar} is shown as bold dots. The dashed curves in
    the top and bottom panels show the synthetic spectra with no carbon and
    nitrogen in the atmosphere.  In the middle panel, the continuous curve
    corresponds to an isotope ratio of $ ^{12}\mathrm{C}/^{13}\mathrm{C} =
    10$, while the dashed curves are synthetic spectra for $
    ^{12}\mathrm{C}/^{13}\mathrm{C} = 3$ and 30.}
\end{figure}

Carbon was measured using CH lines in the regions 4310--4314\,{\AA} and
4362--4367\,{\AA}, which are almost free from intervening atomic lines (see
Fig.~\ref{Fig:ch_nh_lines}, top panel). The C abundances obtained from these
spectral bands are consistent with each other to within 0.03\,dex (see
Table~\ref{linelist}, Online material). The mean abundance is $\mathrm{[C/Fe]}
= 0.13$, which is similar to those of the giants with $\teffm > 4800$\,K
from the sample of \citet{cayr04}.
 
The only detectable $ ^{13}$CH feature near 4211\,{\AA} can be used to
estimate the isotope ratio $ ^{12}$C/$ ^{13}$C. The best fit of the region
4210.7 - 4212.2\,{\AA} including also two $ ^{12}$CH features is achieved with
$ ^{12}$C/$ ^{13}$C = 10 (Fig.~\ref{Fig:ch_nh_lines}, middle panel).
However, with the $S/N \simeq 100$ of the spectrum of
{\LyudmilasStar} around 4211\,\AA, values up to $ ^{12}$C/$ ^{13}$C = 20
can also be accepted.

The abundance of nitrogen can only be determined from the NH band at
3360\,\AA. In the literature, $gf-$values of the NH molecular lines calculated
by \citet{Kur93} were subject to corrections based on analysis of the Solar spectrum around 3360\,{\AA}. \citet{hill02} apply a correction
of --0.807 in $\log gf$ to all of the NH lines, and \citet{hayek09}
--0.4\,dex. We have checked the rather crowded 
spectral region around 3360\,{\AA} in the Solar spectrum
\citep{Atlas} and have fitted it with $gf-$values of the NH
lines scaled down by between --0.3 and --0.4\,dex. With such corrections,
we derived a relative abundance, [N/Fe], between --0.30 and --0.20
(Fig.~\ref{Fig:ch_nh_lines}, bottom panel).

With respect to Li, C, and N abundances, {\LyudmilasStar} is not exceptional.
Unfortunately, the oxygen abundance could not be determined from the available
observed spectrum.

\begin{table} 
 \caption{\label{Tab:AbundanceSummary} Summary of the abundances of 
   {\LyudmilasStar}.}
 \centering
 \begin{tabular}{rllrrcr}\hline\hline
   $Z$ & Species & $\eps{\sun}$ & $N_{lines}$ & $\eps{ }$~~ & $\sigma_{\eps{}}$ & $\mathrm{[X/Fe]}$  \\\hline
   3  & Li I  & 1.10 &  1 & $0.99$~~  & --   & $     $\\
   6  & CH    & 8.39 &  4 & $5.74$~~  & 0.01 & $ 0.13$\\
   7  & NH    & 7.86 &  1 & $4.78$~~  &  --  & $-0.20$\\
   11 & Na I  & 6.30 &  2 & $2.92^N$  & 0.05 & $-0.60$\\
   12 & Mg I  & 7.54 &  3 & $4.95^N$  & 0.04 & $ 0.19$\\
   13 & Al I  & 6.47 &  1 & $3.02^N$  &  --  & $-0.67$\\
   14 & Si I  & 7.52 &  1 & $4.85$~~  &  --  & $ 0.21$\\
   20 & Ca I  & 6.33 &  8 & $3.83^N$  & 0.05 & $ 0.28$\\
   20 & Ca II & 6.33 &  1 & $3.82^N$  & --   & $ 0.27$\\
   21 & Sc II & 3.10 &  4 & $0.10$~~  & 0.02 & $-0.22$\\
   22 & Ti I  & 4.90 & 10 & $2.14$~~  & 0.08 & $ 0.12$\\
   22 & Ti II & 4.90 & 26 & $2.21$~~  & 0.09 & $ 0.09$\\
   23 & V II  & 4.00 &  4 & $0.89$~~  & 0.08 & $-0.33$\\
   24 & Cr I  & 5.64 &  6 & $2.32$~~  & 0.11 & $-0.44$\\
   25 & Mn I  & 5.37 &  1 & $1.89$~~  &  --  & $-0.60$\\
   25 & Mn II & 5.37 &  3 & $1.98$~~  & 0.10 & $-0.61$\\
   26 & Fe I  & 7.45 & 49 & $4.66^N$  & 0.10 & $-0.01$\\
   26 & Fe II & 7.45 &  8 & $4.68^N$  & 0.07 & $ 0.01$\\
   27 & Co I  & 4.92 &  6 & $1.96$~~  & 0.08 & $-0.08$\\   
   28 & Ni I  & 6.23 &  8 & $3.10$~~  & 0.10 & $-0.25$\\    
   30 & Zn I  & 4.62 &  2 & $1.83$~~  & 0.01 & $ 0.09$\\
   38 & Sr II & 2.92 &  2 & $-0.01$~~ & 0.00 & $-0.15$\\
   39 & Y II  & 2.21 &  9 & $-0.69$~~ & 0.06 & $-0.12$\\
   40 & Zr II & 2.58 & 12 & $ 0.04$~~ & 0.05 & $ 0.24$\\
   42 & Mo I  & 1.92 &  1 & $-0.49$~~ &  --  & $ 0.47$\\
   46 & Pd I  & 1.66 &  1 & $-0.72$~~ &  --  & $ 0.50$\\
   56 & Ba II & 2.17 &  3 & $-0.30$~~ & 0.03 & $ 0.31$\\
   57 & La II & 1.14 &  8 & $-1.10$~~ & 0.02 & $ 0.54$\\
   58 & Ce II & 1.61 & 12 & $-0.63$~~ & 0.06 & $ 0.54$\\
   59 & Pr II & 0.76 &  3 & $-1.15$~~ & 0.06 & $ 0.87$\\
   60 & Nd II & 1.45 & 24 & $-0.61$~~ & 0.08 & $ 0.72$\\
   62 & Sm II & 1.00 &  6 & $-0.92$~~ & 0.09 & $ 0.86$\\
   63 & Eu II & 0.52 &  4 & $-1.29$~~ & 0.02 & $ 0.98$\\
   64 & Gd II & 1.11 &  8 & $-0.72$~~ & 0.07 & $ 0.95$\\
   65 & Tb II & 0.28 &  2 & $-1.51$~~ & 0.03 & $ 0.99$\\
   66 & Dy II & 1.13 & 15 & $-0.60$~~ & 0.06 & $ 1.05$\\
   67 & Ho II & 0.51 &  5 & $-1.31$~~ & 0.10 & $ 0.96$\\   
   68 & Er II & 0.96 & 11 & $-0.80$~~ & 0.09 & $ 1.02$\\
   69 & Tm II & 0.14 &  5 & $-1.65$~~ & 0.08 & $ 0.99$\\
   70 & Yb II & 0.86 &  1 & $-1.03$~~ &  --  & $ 0.89$\\
   72 & Hf II & 0.88 &  1 & $-1.32$~~ &  --  & $ 0.58$\\
   76 & Os I  & 1.45 &  1 & $-0.17$~~ &  --  & $ 1.26$\\
   77 & Ir I  & 1.38 &  1 & $-0.05$~~ &  --  & $ 1.45$\\
   90 & Th II & 0.08 &  1 & $-1.67$~~ &  --  & $ 1.03$\\\hline
\multicolumn{7}{l}{  } \\
\multicolumn{7}{l}{$ ^N$ NLTE abundance.} \\
\end{tabular}
\end{table}

\subsection{Sodium to titanium}

In {\LyudmilasStar}, the $\alpha-$process elements Mg, Si, Ca, and Ti are
enhanced relative to iron: $\mathrm{[Mg/Fe]} = 0.19$, $\mathrm{[Si/Fe]} =
0.21$, $\mathrm{[Ca/Fe]} = 0.28$, and $\mathrm{[Ti/Fe]} = 0.10$. This is
consistent with the behavior of other metal-poor halo stars (see, e.g.,
\citealt{cayr04}).

The determination of the abundances of Mg and Ca is
based on NLTE line formation calculations for \ion{Mg}{i} and \ion{Ca}{i-ii},
using the methods described by \citet[][\ion{Mg}{i}]{nlte-mg} and
\citet[][\ion{Ca}{i-ii}]{ml-nlte-ca}. For both elements the same scaling factor, $S_H = 0.1$,
was applied to the \citet{hyd} formula for calculations of the inelastic
collisions with hydrogen atoms. Neutral Mg and Ca are minority species in
the atmosphere of {\LyudmilasStar}, and they are both subject to
overionization caused by super-thermal ultraviolet radiation of non-local
origin, resulting in a weakening of the \ion{Mg}{i} and \ion{Ca}{i} lines
compared to their LTE strengths. The NLTE abundance corrections, $\Delta_{\rm
  NLTE} = \eps{NLTE} - \eps{LTE}$, are in the range 0.08--0.12\,dex for the
\ion{Mg}{i} lines and between 0.17 and 0.29\,dex for the \ion{Ca}{i} lines
(Table~\ref{linelist}, Online material). 

The Si abundance was derived from the only detected line, \ion{Si}{i}
3905\,{\AA}, assuming LTE. Based on the NLTE calculations for \ion{Si}{i}
presented by \citet{nlte-si}, we estimate the NLTE abundance correction for
this line to be positive and on the order of a few hundredths of a dex.

Titanium is observed in {\LyudmilasStar} in two ionization stages, and its
abundance can be reliably determined. We obtained a difference in absolute LTE
abundances of $-0.07$\,dex between \ion{Ti}{i} and \ion{Ti}{ii}. Assuming that
the NLTE effects for \ion{Ti}{ii} are as small, as is the case for
\ion{Fe}{ii}, and that they are of the same order for \ion{Ti}{i} as they are
for \ion{Fe}{i}, we arrive at $\Delta\eps{}(\ion{Ti}{i}-\ion{Ti}{ii}) =
0.03$\,dex.

{\LyudmilasStar} displays an underabundance of the odd$-Z$ elements Na and Al
relative to iron; $\mathrm{[Na/Fe]}=-0.60$ and
$\mathrm{[Al/Fe]}= -0.67$. This is not exceptional
for a metal-poor halo star.
Sodium and aluminium were observed in {\LyudmilasStar} only in the resonance lines of
their neutral species. The abundance determination was based on NLTE line
formation for \ion{Na}{i} and \ion{Al}{i}, using the methods described by
\citet{ml-nlte-na} and \citet{nlte-al}. For both species, $S_H = 0.1$ was
adopted. NLTE leads to $-0.39/-0.28$\,dex smaller abundances derived from the
\ion{Na}{i} 5890/5896\,{\AA} line and a 0.52\,dex larger abundance derived
from the \ion{Al}{i} 3961\,{\AA}, compared to the corresponding LTE values. It
is worth noting that the calculated $\Delta_{\rm NLTE}$ of the Na lines agree
within 0.05/0.02\,dex with the values given by \citet{Andr_na} in their
Table~2 for {\tefft}, $\log g$, and $W_\lambda$ values close to those of
{\LyudmilasStar}, while we found a 0.25\,dex smaller $\Delta_{\rm NLTE}$ for
\ion{Al}{i} compared to that shown by \citet{Andr_al} in their Fig.~2 for
similar stellar parameters. For the relative abundances in {\LyudmilasStar},
we obtained an Al/Na ratio close to Solar ($\mathrm{[Al/Na]} = -0.07$) and
very low odd/even$-Z$ ratios ($\mathrm{[Na/Mg]} = -0.79$; $\mathrm{[Al/Mg]} =
-0.86$).

For determining the abundance of Sc, we employed four lines of the majority
species \ion{Sc}{ii}. For each line, hyperfine structure splitting was taken
into account, using the HFS data of \citet{MPS95}. Neglecting the HFS effect
leads to an overestimation of the Sc abundance of 0.08\,dex for \ion{Sc}{ii}
4246\,{\AA}, the strongest line in the wavelength ranges covered by our
spectra. We obtain $\mathrm{[Sc/Fe]} = -0.22$, which is about 0.2\,dex lower
than the corresponding mean value for the \citet{cayr04} cool halo giants. The
difference can at least partly be explained by the fact that in the cited
study, HFS was not taken into account. NLTE calculations for \ion{Sc}{ii} in
the Sun were performed by \citet{nlte-sc}, with the result that the departures
from LTE are small with negative NLTE abundance corrections of $-0.06$ to $-0.03$~dex.
 
\subsection{Iron group elements and Zn}

We determined the abundance of six elements in this group. For two of them, Mn
and Co, their energy levels are affected by considerable hyper-fine splitting,
and HFS was explicitely taken into account in our spectrum
synthesis calculations where HFS data were available (see
Table~\ref{linelist}, Online material for references).
 
For Mn, we obtained 0.36\,dex lower abundances from the \ion{Mn}{i} resonance
lines at $\sim4030$\,{\AA} compared to the \ion{Mn}{i} subordinate line at
4041\,{\AA} in {\LyudmilasStar}. A similar effect was found for the \ion{Cr}{i} lines:
two lines arising from the ground state, 4254\,{\AA} and 4274\,{\AA}, exhibit
0.26 and 0.32\,dex lower abundances compared to the mean of the other chromium
lines. Our results support the finding of \citet{Johnson2002} and later studies.
 The Mn abundance derived from the \ion{Mn}{i} lines can be
underestimated due to departures from LTE. \citet{nlte-mn} predict
$\Delta_{\rm NLTE} = 0.37-0.41$\,dex for the \ion{Mn}{i} resonance triplet in
the model 5000/4.0/$-3$, and 0.5\,dex for \ion{Mn}{i} 4041\,{\AA}. Usually,
the NLTE effects increase with decreasing $\log g$. However, it is unclear
whether $\Delta_{\rm NLTE}$ will change with surface gravity in similar
proportions for the \ion{Mn}{i} resonance triplet and \ion{Mn}{i} 4041\,{\AA}.
Therefore, the abundances derived from the \ion{Mn}{i} resonance lines were
not taken into account in the mean presented in
Table~\ref{Tab:AbundanceSummary}. 

Fortunately we detected lines of \ion{Mn}{ii}, the majority species of Mn,
which is expected to be hardly affected by departures from LTE, according
to the results of \citet{nlte-mn-sun}.
We note that the relative LTE abundances [Mn\,I (4041\,{\AA})/Fe\,I] and
[Mn\,II/Fe\,II] in {\LyudmilasStar} are consistent with each other to within
0.01\,dex. Though HFS was not taken into account for the
\ion{Mn}{i} 4041\,{\AA} line, its effect on the abundance is expected to be
small, as the line is very weak ($W_\lambda =$ 11\,m{\AA}).  {\LyudmilasStar}
reveals an underabundance of Cr and Mn very similar to that of the comparison
sample \citep{cayr04}.

{\LyudmilasStar} is also deficient in V and Ni relative to iron and the Solar
ratios. The information on V abundances in very metal-poor stars is sparse in
the literature, probably due to difficulties in detecting the vanadium lines.
We used four lines of \ion{V}{ii} located in the blue spectral range, where
severe blending effects are present even in very metal-poor stars. Paper~II
found V/Fe ratios close to Solar for the sample covering a [Fe/H] range from
$-1.5$ to $-3$. However, they noted that the V abundances are based on quite
weak features and hence are susceptible to overestimation due to unresolved
blends. For Ni, we used eight well observed and unblended lines of
\ion{Ni}{i}. The large scatter of the obtained abundances can partly be due to
using four different sources for the $gf-$values (see Table~\ref{linelist}
and Online material for references). For example, the mean abundance derived
from two lines using $gf-$values of \citet{FMW88} is 0.21\,dex larger compared
to the abundances measured from three lines employing $gf-$values taken from
\citet{ni1-BBP89}.

For the cobalt and zinc abundances of {\LyudmilasStar}, we have obtained
values close to the Solar ratios with respect to iron; i.e.,
$\mathrm{[Co\,I/Fe\,I]} = -0.07$ and $\mathrm{[Zn\,I/Fe\,I]} = 0.09$. Using
$\Delta_{\rm NLTE}(\ion{Zn}{i}\,4810\,\mathrm{\AA}) = +0.05$ from the NLTE
calculations of \citet{nlte-zn}, and assuming similar departures from LTE for
\ion{Zn}{i} 4722\,{\AA}, we calculated a NLTE abundance ratio of
$\mathrm{[Zn/Fe]} = 0.04$. \citet{cayr04} found that [Co/Fe] and [Zn/Fe]
increase with decreasing metallicity with $\mathrm{[Co/Fe]} \simeq 0.25$ and
$\mathrm{[Zn/Fe]} \simeq 0.2$ for stars with [Fe/H] close to $-2.8$ (see their
Fig.~12). Note that \citet{cayr04} neglected HFS of the used lines
of \ion{Co}{i}, thus they probably overestimated the Co abundances.
According to our estimate for \ion{Co}{i} 4121\,{\AA} in the atmospheric model
5050/2.34/$-2.78$, ignoring HFS makes a difference in abundance of
$+0.09$\,dex.

\subsection{Heavy elements}\label{Sect:heavy}

In the ``snapshot'' spectra of {\LyudmilasStar}, \citet{HERESpaperII} detected
only six heavy elements beyond strontium. Due to the higher quality and
broader wavelength coverage of the spectra used in this study, we detected 23
elements in the nuclear charge range between $Z = 38$ and 90.  We were not
successful in obtaining abundances for Ru, Rh, and U. The \ion{Ru}{i} 3436,
3728\,{\AA} and \ion{Rh}{i} 3434, 3692\,{\AA} lines are very weak and
therefore could not be detected in our spectra. We have marginally detected
the \ion{U}{ii} 3859.57\,{\AA} line in our spectra of {\LyudmilasStar};
however, the $S/N$ is not high enough for deriving a reliable abundance.

\subsubsection{NLTE effects} 

For six species, \ion{Sr}{ii}, \ion{Zr}{ii}, \ion{Ba}{ii}, \ion{Pr}{ii},
 and \ion{Eu}{ii}, we performed NLTE calculations using the
methods described in our earlier studies
\citep{nlte-sr,nlte-zr,ml-nlte-ba,ml-nlte-pr,ml-nlte-eu} and
determined the NLTE element abundances. They are presented in
Table~\ref{linelist} (Online material).

Our NLTE calculations for {\LyudmilasStar} showed that the \ion{Sr}{ii} and
\ion{Ba}{ii} resonance lines are strengthened compared to the LTE case,
resulting in negative NLTE abundance corrections of $-0.15$\,dex and
$-0.09$\,dex, respectively. The subordinate lines of \ion{Ba}{ii} show a
different behavior: the weakest line at 5853\,{\AA} is weakened, while the two
other lines, at 6141 and 6496\,\AA, are strengthened relative to LTE.

\begin{figure*} 
  \centering
  \resizebox{150mm}{!}{\includegraphics{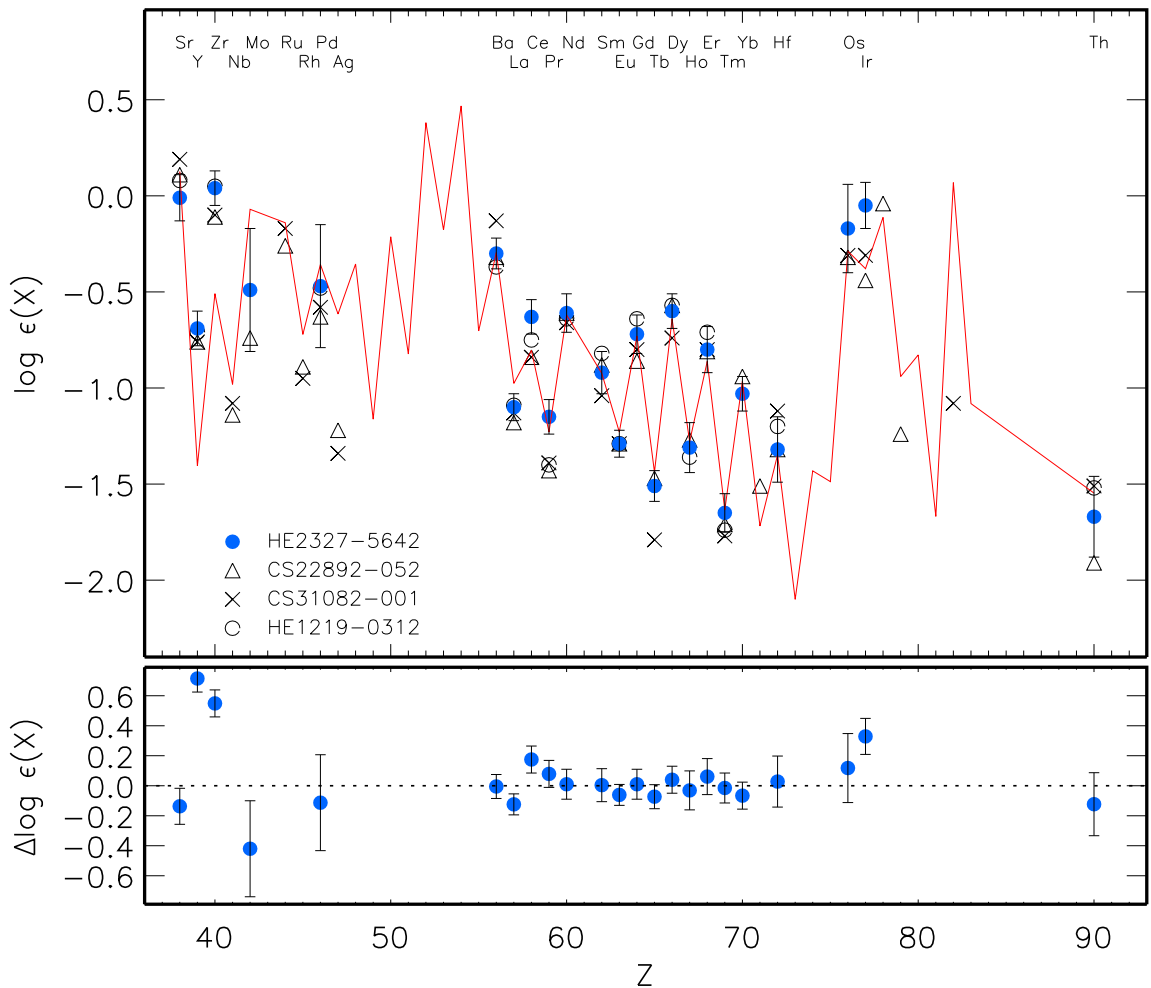}}
  \caption{\label{Fig:AbundancePattern} The heavy-element abundance pattern of 
    {\LyudmilasStar} (filled circles) compared to the Solar system $r$-process
    (SSr) abundance pattern (continuous curve) scaled to match Ba--Hf.
    For comparison, the heavy element abundances of the benchmark r-II stars
    CS\,22892-052 (open triangles), CS\,31082-001 (crosses), and HE\,1219-0312
    (open circles) are shown. They have been normalized to the value derived
    for $\eps{Eu}$ in {\LyudmilasStar}.  The bottom panel displays the
    difference between {\LyudmilasStar} and SSr defined as $\Delta\eps{}(X) =
    \eps{HE\,2327-5642}(X) - \eps{SSr}(X)$. }
\end{figure*}

In contrast to \ion{Sr}{ii} and \ion{Ba}{ii}, the term structure of the other
NLTE species is produced by multiple electronic configurations and consists of
hundreds and thousands of energy levels. For each such species, enhanced
photoexcitation from the ground state leads to overpopulation of the excited
levels in the line formation layers, resulting in a weakening of the lines. We
calculated positive NLTE abundance corrections for the lines of \ion{Zr}{ii},
\ion{Pr}{ii}, \ion{Nd}{ii}, and \ion{Eu}{ii} with values close to $+0.1$\,dex.
All the elements beyond barium are observed in the lines of their majority
species, with term structures as complicated as that for \ion{Eu}{ii}, so
the departures from LTE are expected to be similar to those for \ion{Eu}{ii}.
This is largely true also for osmium and iridium detected in the lines of
their neutrals, \ion{Os}{i} and \ion{Ir}{i}, which have relatively large
ionization energies of 8.44 and 8.97~eV, respectively.  Fortunately, the
abundance \emph{ratios} among heavy elements are probably only weakly affected
by departures from LTE.  For consistency, we use in this study the abundances
of the heavy elements beyond strontium as determined under the LTE assumption.
They are presented in Table~\ref{Tab:AbundanceSummary} and
Figure~\ref{Fig:AbundancePattern}.

Now we explore the abundance patterns of elements in the three $r$-process
peaks.

\subsubsection{The light trans-iron elements} 

Five elements with $38 \le Z \le 46$ were measured in the region of the first
peak. The only molybdenum line in the visible spectrum, \ion{Mo}{i}
3864\,{\AA}, can be used to determine the element abundance of cool stars. In
{\LyudmilasStar}, this line is nearly free of blends, but it is weak: the
central depth of the line is only $\sim 2$\,\% of the continuum. With $S/N
\simeq 50$ of the observed spectrum in this wavelength region, the uncertainty
of the derived Mo abundance was estimated to be $0.3$\,dex.

A similar uncertainty is expected for palladium, which was detected in a
single line, \ion{Pd}{i} 3404\,\AA. In {\LyudmilasStar}, this line is free of
blends and is stronger compared to \ion{Mo}{i} 3864\,{\AA}, but it is located
in a spectral range where the $S/N$ is only 20 to 25.

Strontium is observed in {\LyudmilasStar} in two strong resonance lines,
\ion{Sr}{ii} 4077 and 4215\,{\AA}. Both lines are affected by HFS of the odd isotope \iso{87}{Sr}. 
The synthetic spectrum was calculated with the \iso{87}{Sr} fraction of 0.22 corresponding to a 
pure $r$-process production of strontium  
\citep[][stellar model]{rs99}.

\subsubsection{The second $r$-process peak elements} 

\begin{figure}  
  \resizebox{88mm}{!}{\includegraphics{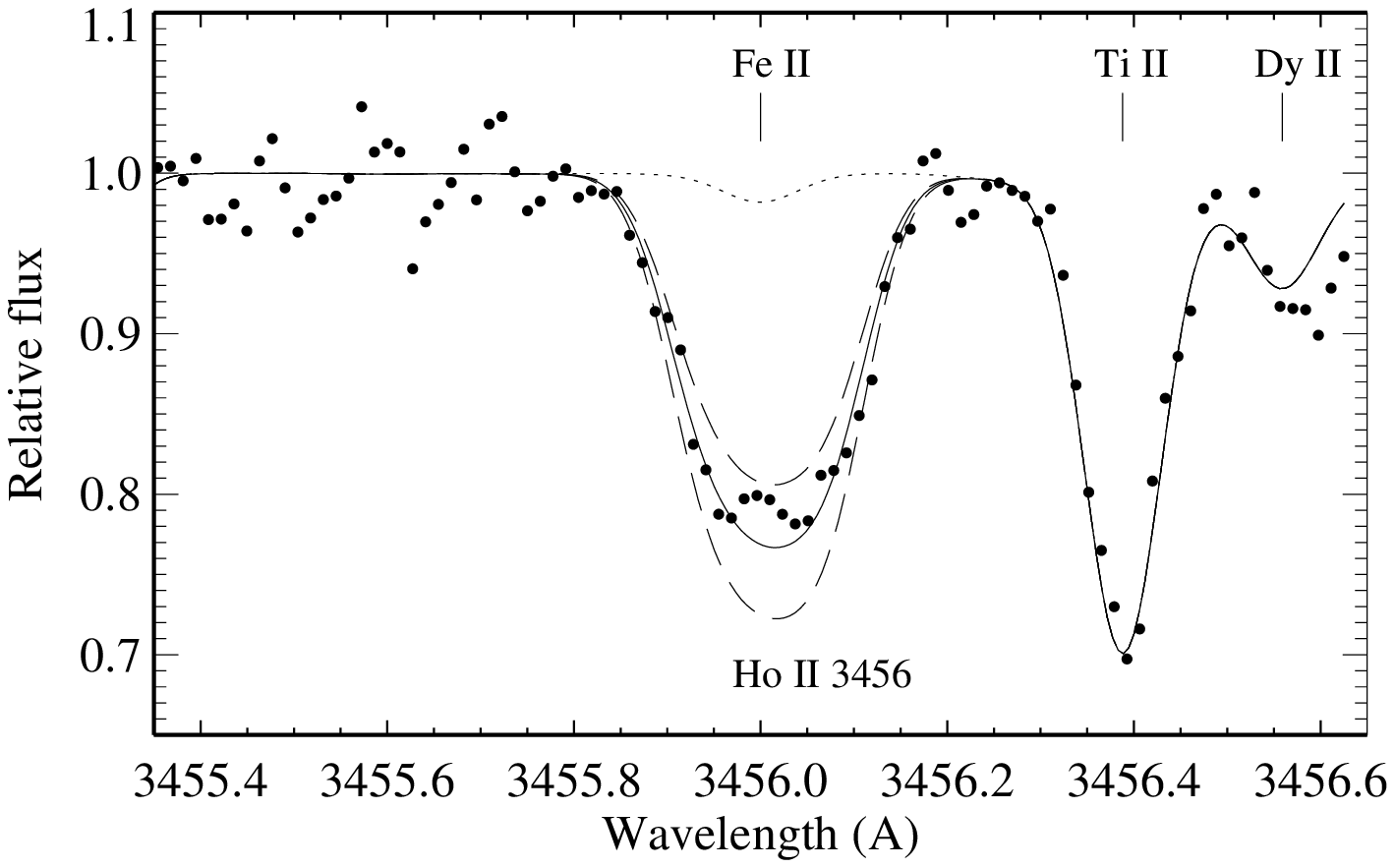}}
  \resizebox{88mm}{!}{\includegraphics{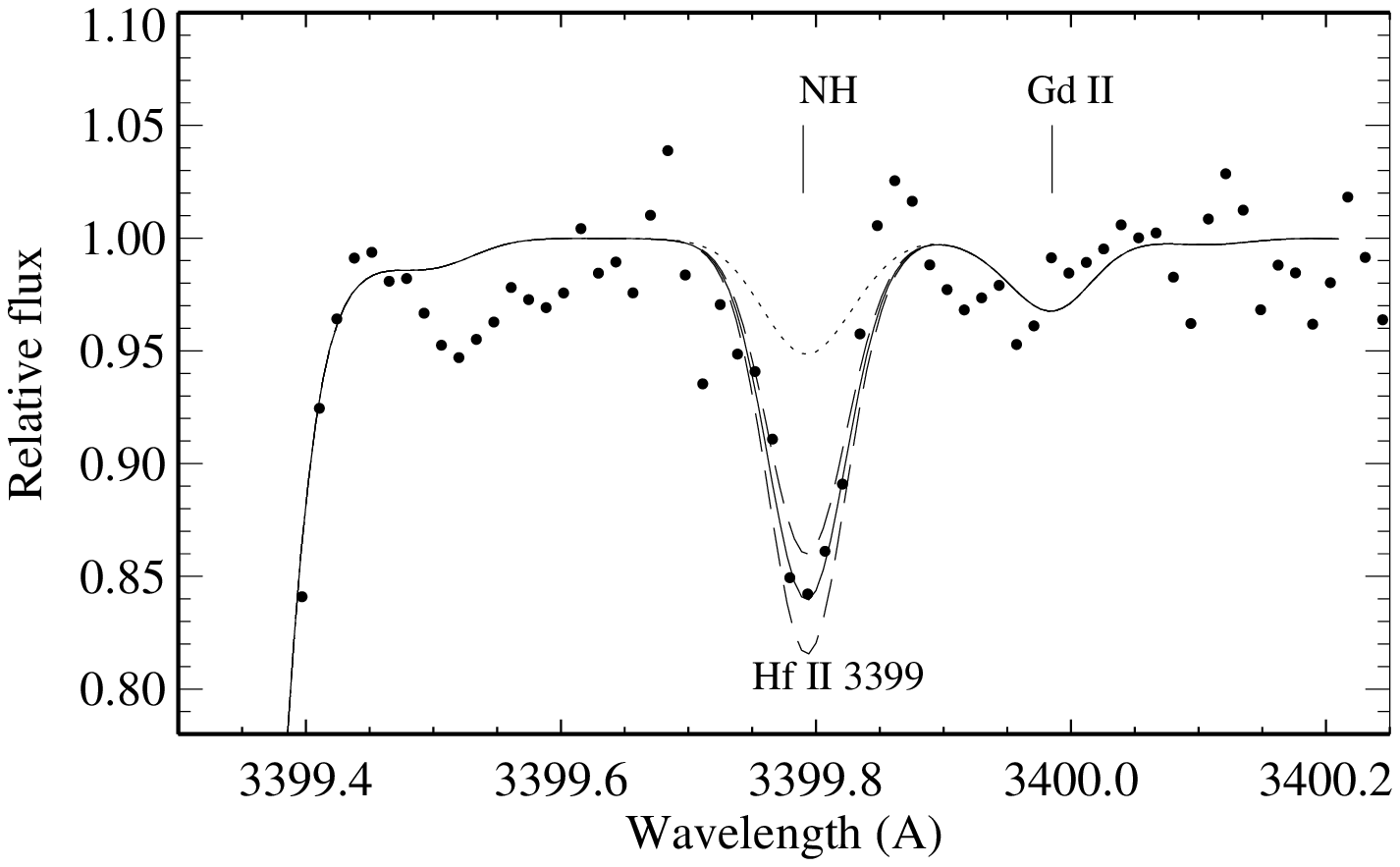}}
  \caption{\label{Fig:hf_ho_lines} The best fits (continuous curve) of
    \ion{Ho}{ii} 3456\,{\AA} (top panel) and \ion{Hf}{ii} 3399\,{\AA} (bottom
    panel) in the observed spectrum of {\LyudmilasStar} (bold dots). The dashed
    curves show the effect of a 0.1\,dex variation of the abundance on the
    synthetic spectrum. The dotted curves show the synthetic
    spectrum with no holmium and hafnium in the atmosphere.}
\end{figure}

With 15 elements measured in the Ba--Hf range (see Fig.~\ref{Fig:hf_ho_lines}
for holmium and hafnium), the second $r$-process peak is the best-constrained
one among the three peaks,

The barium abundance given in Table~\ref{Tab:AbundanceSummary} was determined
from the three subordinate lines, \ion{Ba}{ii} 5853, 6141, and 6497\,{\AA},
which are nearly free of HFS effects. According to our estimate for
\ion{Ba}{ii} 6497\,{\AA}, neglecting HFS makes a difference in abundance of no
more than 0.01\,dex. In contrast, the \ion{Ba}{ii} 4554\,{\AA} resonance line
is strongly affected by HFS. The even isotopes are unaffected by HFS, 
while the odd isotopes show significant HFS, and thus the element abundance 
derived from this line depends on the Ba isotope mixture adopted in the calculations. 
Since the odd isotopes \iso{135}{Ba} and \iso{137}{Ba} have very similar HFS, 
the abundance is essentially dependent on the total fractional abundance of these odd isotopes, 
$f_\mathrm{odd}$.
For example, using the Solar Ba isotope mixture with
$f_\mathrm{odd} = 0.18$, we obtained a $0.34$\,dex larger abundance from \ion{Ba}{ii}
4554\,{\AA} compared to the mean abundance of the subordinate lines. 
In the LTE calculations, the difference was reduced by $0.20$\,dex when we adopted 
a pure $r$-process Ba isotope mixture with $f_\mathrm{odd} = 0.46$, as predicted by
\citet[][stellar model]{rs99}.
The remaining discrepancy between the resonance
and subordinate lines has been largely removed by NLTE calculations. Thus, our
analysis of HFS affecting the \ion{Ba}{ii} 4554\,{\AA} resonance line suggests
a pure $r$-process production of barium in {\LyudmilasStar}.

The lines of \ion{Eu}{ii} and \ion{Yb}{ii} detected in {\LyudmilasStar}
consist of multiple IS and HFS components. To derive the total abundance of the
given element, we adopted in our calculations a pure $r$-process isotope mixture
from the predictions of \citet[][stellar model]{rs99}: \iso{151}{Eu} :
\iso{153}{Eu} = 39 : 61 and \iso{171}{Yb} : \iso{172}{Yb} : \iso{173}{Yb} :
\iso{174}{Yb} : \iso{176}{Yb} = 18.3 : 22.7 : 18.9 : 23.8 : 16.3.  All the
lines of \ion{Nd}{ii} and \ion{Sm}{ii} observed in {\LyudmilasStar} are rather
weak, and they were treated as single lines.

There is only one hafnium line that can be measured in {\LyudmilasStar}. The
observed feature is attributed to a combination of the \ion{Hf}{ii}
3399.79\,{\AA} and NH 3399.79\,{\AA} molecular line. With $\log gf = -1.358$
taken from \citet{Kur93} and scaled down by $-0.3$~dex and the nitrogen abundance $\eps{N} = 4.78$, the molecular line contributes
approximately 45\,\% to the 3399\,{\AA} blend (Fig.~\ref{Fig:hf_ho_lines}).
Ignoring the molecular contaminant completely leads to a $0.17$\,dex larger
hafnium abundance. We therefore estimate the uncertainty of the obtained Hf
abundance to be $0.15$\,dex.

\subsubsection{The heaviest elements} 

The third peak and actinides were probed for three elements, osmium, iridium,
and thorium.  The abundance of osmium was determined from a single line,
\ion{Os}{i} 4260\,\AA. The line is weak, with a center line depth of 2.5\%\ of
the continuum flux, and is nearly free of blends. With a high signal-to-noise
ratio ($S/N\simeq 100$) of the observed spectrum, the uncertainty of the
derived osmium abundance was estimated as 0.2\,dex.

\begin{figure}  
  \resizebox{88mm}{!}{\includegraphics{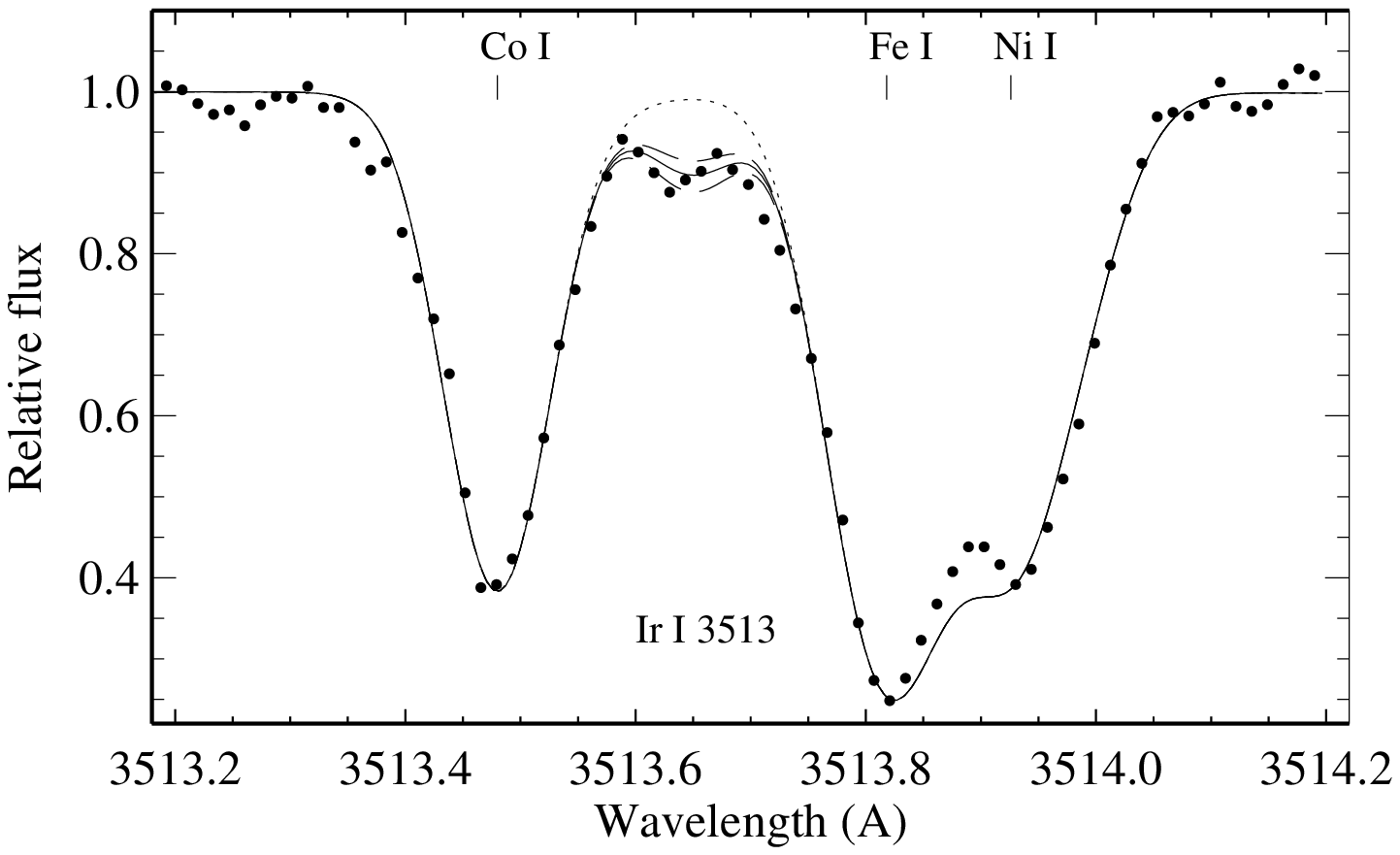}}
  \resizebox{88mm}{!}{\includegraphics{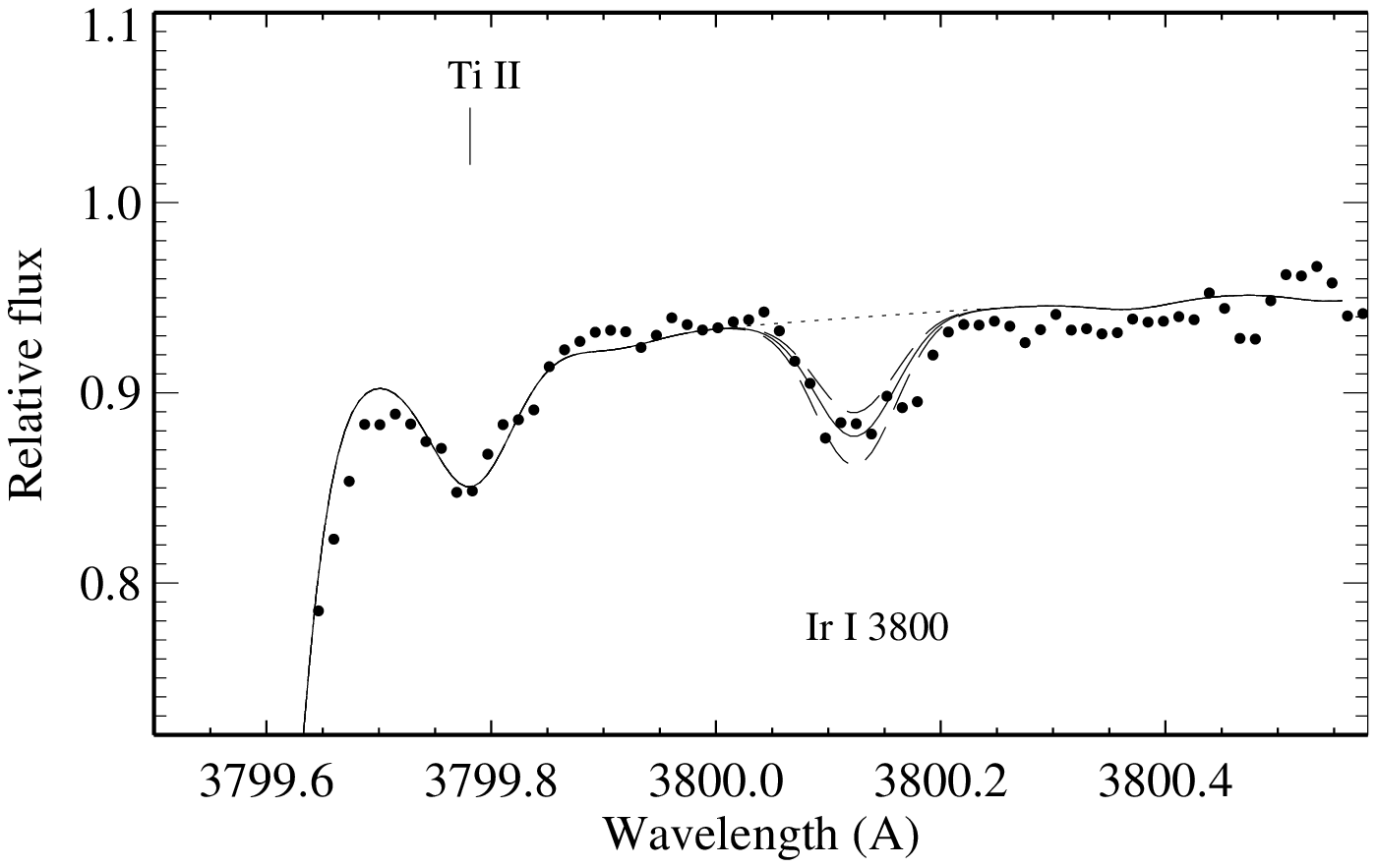}}
  \caption{\label{Fig:ir_lines} The same as in Fig.~\ref{Fig:hf_ho_lines}, but for
    \ion{Ir}{i} 3513 and 3800\,{\AA}.}
\end{figure}

\begin{figure}  
  \resizebox{88mm}{!}{\includegraphics{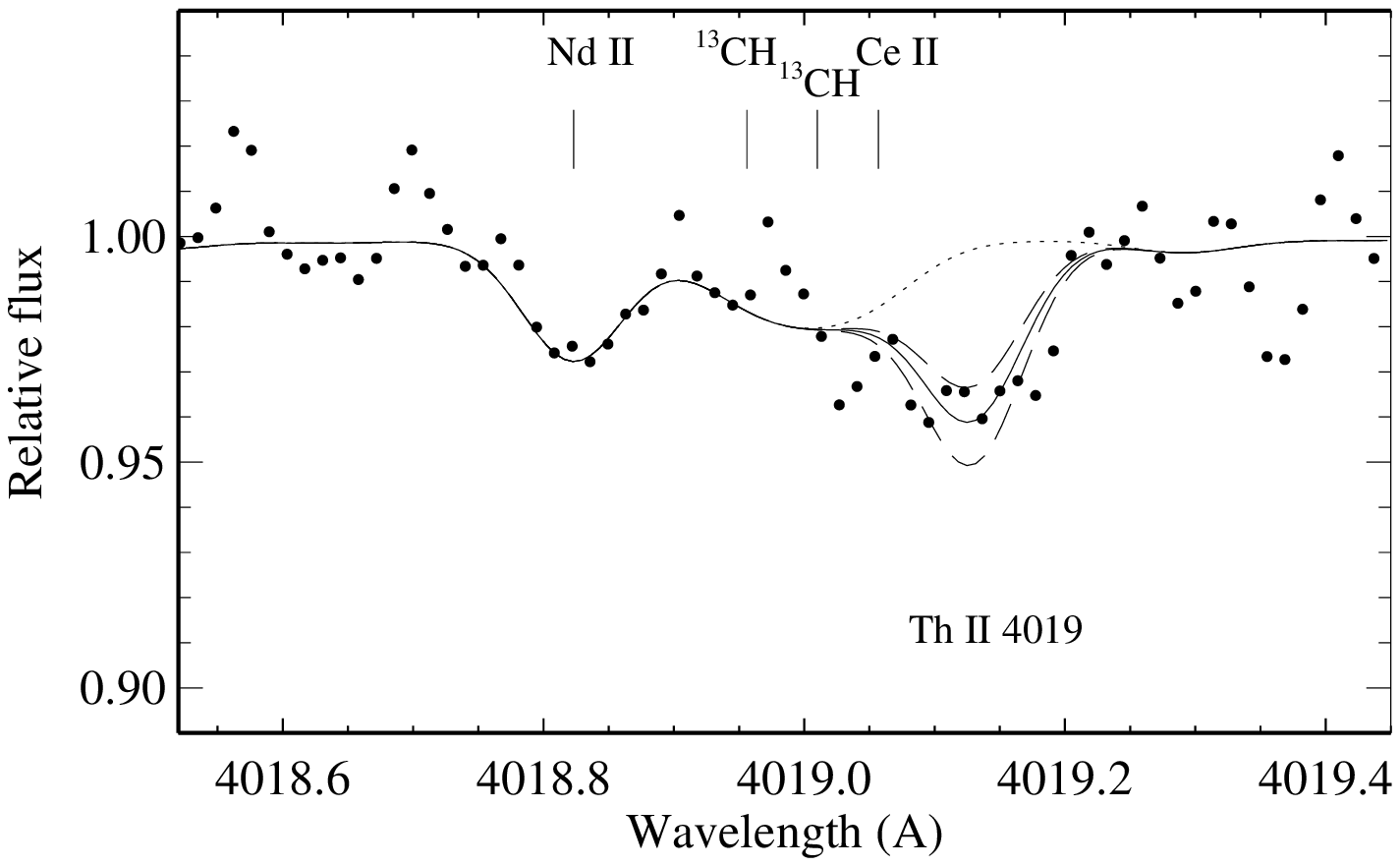}}
  \caption{\label{Fig:th_lines} The same as in Fig.~\ref{Fig:hf_ho_lines}, but for 
    \ion{Th}{ii} 4019\,{\AA}.}
\end{figure}

Two iridium lines, \ion{Ir}{i} 3513 and 3800\,{\AA}, were clearly detected in {\LyudmilasStar} (Fig.~\ref{Fig:ir_lines}). The theoretical profiles were calculated with
taking HFS effect into account and iridium isotope abundance ratio
\iso{191}{Ir}:\iso{193}{Ir} = 37~:~63, which is obtained to be the same for
the Solar system matter \citep{Lodders2003} and the matter produced in the
$r$-process \citep{rs99}. The \ion{Ir}{i} 3800\,{\AA} line was measured in two observed spectra, BLUE390 and 437BLUE, (Table~\ref{Tab:Observations}) and seems reasonably reliable. With its equivalent width of 5.8\,m\AA\ and $S/N\simeq 100$ of the observed spectrum, the uncertainty of the
derived iridium abundance was estimated as 0.08\,dex. The \ion{Ir}{i} 3513\,{\AA} line served as a verification of that it agrees with the other line. The blend at 3513.6\,{\AA} is well fitted with $\eps{Ir} = -0.05$ found from \ion{Ir}{i} 3800\,{\AA}, as shown in  Fig.~\ref{Fig:ir_lines}.

The radioactive element thorium was clearly detected in {\LyudmilasStar} in the \ion{Th}{ii} 4019\,{\AA} line (Fig.~\ref{Fig:th_lines}), but proved rather challenging for the
determination of stellar age of {\LyudmilasStar}. Unhappily, the observed spectrum around \ion{Th}{ii} 4019\,{\AA} has low signal-to-noise ratio ($S/N\simeq 50$), and the uncertainty of the
derived thorium abundance was estimated as 0.2\,dex.


\subsection{Error budget} 

We performed a detailed error analysis on {\LyudmilasStar}, in order to
estimate the uncertainties of the abundance measurements for the heavy
elements beyond the iron group. Stochastic errors ($\sigma_{obs}$) arising
from random uncertainties of the continuum placement, line profile fitting,
and $gf$-values, are represented by a dispersion of the measurements of
multiple lines around the mean ($\sigma_{\eps{}}$), as given in
Table~\ref{Tab:AbundanceSummary} when $N \ge 2$ lines of an element are
observed. Observational errors of the species with a single line used in
abundance analysis were discussed in Sect.\,\ref{Sect:heavy}. Systematic
uncertainties include those which exist in the adopted stellar parameters, in
the used hydrostatic model atmospheres, and in the LTE line formation
calculations. As argued in Sect.\,\ref{Sect:heavy}, the latter is not expected
to influence the abundance pattern of the elements in the range from La to Th.

It is difficult to estimate the uncertainty introduced by using the 1D model
atmosphere. The elements in the La--Th range, for example, are observed in 
the lines of their majority species. The detected lines arise from either the
ground or low-excitation levels, and most of them are relatively weak; i.e.,
$W_\lambda < 30$\,m{\AA}. This means that they all are formed in the same
atmospheric layers. It would therefore be rather unexpected if there were significantly different 3D
effects for individual elements in the La--Th range. This is different from 
the case of the elements in the Sr--Pd range, Ba, and Yb, which are observed
in either the strong lines (e.g. \ion{Sr}{ii} 4215\AA, \ion{Ba}{ii} 6141\AA,
\ion{Yb}{ii} 3694\AA) or the lines of the minority species \ion{Mo}{i} and
\ion{Pd}{i}.  Both NLTE and 3D effects may have a strong influence on their
derived abundance. Hence we examined here only those uncertainties linked to
our choice of stellar parameters. These were estimated by varying {\tefft} by
$-70$~K, $\log g$ by $-0.1$\,dex, and $\xi$ by $-0.1$~\kms\ in the stellar
atmosphere model.

\begin{table} 
 \caption{\label{Tab:AbundanceErrors} Error budget for neutron-capture
   elements in {\LyudmilasStar}.}                   
 \centering                                                                                                             
  \begin{tabular}{lrrcclc}\hline\hline                                                                                  
    ~El. & \multicolumn{1}{c}{$\Delta T$} & $\Delta \log g$ &$\Delta \xi$ & 
    $\Delta$ & $\sigma_{obs}$ &  $\sigma_{tot}$ \\           
    & $-70$\,K & $-0.1$\,dex & $-0.1$\,\kms & ($T,g,\xi$) & & \\          
    \multicolumn{1}{c}{\scriptsize (1)} & \multicolumn{1}{c}{\scriptsize (2)} 
    & \multicolumn{1}{c}{\scriptsize (3)} & {\scriptsize (4)} & {\scriptsize (5)} 
    & \multicolumn{1}{c}{\scriptsize (6)} & {\scriptsize (7)} \\\hline
    Sr II & $-0.08$ & $-0.01$ & 0.09   & 0.12 & 0.00 & 0.12 \\
    Y II  & $-0.05$ & $-0.03$ & 0.02   & 0.06 & 0.06 & 0.09 \\
    Zr II & $-0.06$ & $-0.03$ & 0.02   & 0.07 & 0.05 & 0.09 \\
    Mo I  & $-0.10$ & $<0.01$ & $<0.01$& 0.10 & 0.3  & 0.32 \\
    Pd I  & $-0.10$ & $<0.01$ & $<0.01$& 0.10 & 0.3  & 0.32 \\
    Ba II & $-0.06$ & $-0.03$ & 0.03   & 0.07 & 0.03 & 0.08 \\
    La II & $-0.06$ & $-0.03$ & $<0.01$& 0.07 & 0.02 & 0.07 \\
    Ce II & $-0.06$ & $-0.03$ & $<0.01$& 0.07 & 0.06 & 0.09 \\
    Pr II & $-0.06$ & $-0.03$ & $<0.01$& 0.07 & 0.06 & 0.09 \\
    Nd II & $-0.06$ & $-0.03$ & 0.01   & 0.07 & 0.08 & 0.10 \\
    Sm II & $-0.06$ & $-0.03$ & $<0.01$& 0.07 & 0.09 & 0.11 \\
    Eu II & $-0.06$ & $-0.03$ & 0.01   & 0.07 & 0.02 & 0.07 \\
    Gd II & $-0.06$ & $-0.03$ & 0.01   & 0.07 & 0.07 & 0.10 \\
    Tb II & $-0.07$ & $-0.03$ & 0.01   & 0.08 & 0.03 & 0.08 \\
    Dy II & $-0.06$ & $-0.03$ & 0.03   & 0.07 & 0.06 & 0.09 \\
    Ho II & $-0.07$ & $-0.03$ & $<0.01$& 0.08 & 0.10 & 0.13 \\
    Er II & $-0.07$ & $-0.03$ & $<0.01$& 0.08 & 0.09 & 0.12 \\
    Tm II & $-0.06$ & $-0.03$ & 0.01   & 0.07 & 0.08 & 0.10 \\
    Yb II & $-0.04$ & $-0.03$ & 0.07   & 0.09 & 0.02 & 0.09 \\
    Hf II & $-0.07$ & $-0.03$ & 0.01   & 0.08 & 0.15 & 0.17 \\
    Os I  & $-0.11$ & $<0.01$ & $<0.01$& 0.11 & 0.2  & 0.23 \\
    Ir I  & $-0.09$ & $-0.01$ & $<0.01$& 0.09 & 0.08 & 0.12 \\
    Th II & $-0.07$ & $-0.03$ & $<0.01$& 0.08 & 0.2  & 0.21 \\
    \hline
    \multicolumn{7}{l}{  } \\
  \end{tabular}
\end{table}

Table~\ref{Tab:AbundanceErrors} summarizes the various sources of
uncertainties. The quantity $\Delta(T,g,\xi)$ listed in Col.~5 is the total
impact of varying each of the three parameters, computed as the quadratic sum
of Cols.~2, 3, and 4. The total uncertainty $\sigma_{tot}$ (Col.~7) of the
absolute abundance of each element is computed by the quadratic sum of the
stochastic ($\sigma_{obs}$) and systematic ($\Delta(T,g,\xi)$) errors.

\section{Comparison to other r-II stars and Solar r-process abundances}\label{Sect:ssr}

The abundance pattern of the neutron-capture elements in the range from Sr to
Os in {\LyudmilasStar} is very similar to that of other well-studied r-II
stars. Figure~\ref{Fig:AbundancePattern} shows comparisons with CS\,22892-052
\citep{sne03}, CS\,31082-001 \citep{hill02, plez04}, and HE\,1219-0312
\citep{hayek09}. For example, the dispersion around the mean of the quantities
$(\eps{HE\,2327-5642} - \eps{CS\,22892-052})$ amounts to $0.10$\,dex, which is
at the level of $1\sigma$ error bars in the abundance determinations of the
individual elements. 

Iridium in {\LyudmilasStar} seems to be overabundant with respect to
CS\,22892-052 and CS\,31082-001. \citet{Roederer2009} compiled and/or revised the
neutron-capture element abundances for a sample of $r$-process rich stars
including CS\,22892-052 and CS\,31082-001, and they obtained a mean iridium to
europium ratio of $\log(\mathrm{Ir}/\mathrm{Eu}) = 0.90\pm 0.09$ for 15 stars.
For {\LyudmilasStar}, we derived $\log(\mathrm{Ir}/\mathrm{Eu}) = 1.24$. We
investigated possible sources of the difference between our analysis and that
of \citet{Roederer2009}, and found one which could at least partly explain it.
The iridium abundances in the \citet{Roederer2009} study were underestimated
by approximately 0.2\,dex due to using the partition functions of
\ion{Ir}{ii} as implemented in the MOOG code (C. Sneden, private
communication), which are about a factor of two lower compared to the ones we
used in our study. We have compared the partition functions thoroughly 
and are satisfied that our new data, which are based on the best available 
energy level data, are  more accurate.
 In all other relevant respects (i.e., $gf$-values, HFS
data, ionization potential of \ion{Ir}{i}), our data and methods and those of
\citet{Roederer2009} are identical. The difference in the codes used for the
abundance determination cannot play a role only for iridium, leaving all the
other determinations unaffected. Hence, we are left with a discrepancy of
0.14\,dex in $\log(\mathrm{Ir}/\mathrm{Eu})$ between {\LyudmilasStar} and the
stellar sample of \citet{Roederer2009}. However, in order to draw a firm
conclusion on this point, the abundance determinations have to confirmed from
better measurements and from new detections of the third $r$-process peak
elements.

In Fig.~\ref{Fig:AbundancePattern}, we also plot the Solar
$r$-process residuals. The decomposition of the $s$- and $r$-process contributions is
based on the meteoritic abundances of \citet{Lodders2009} and the $s$-process
abundances of \citet[][stellar model]{rs99}. The absolute $s$-process
abundances were obtained by \citet{rs99} by normalization to the Solar
abundance of the pure $s$-process isotope \iso{150}{Sm} taken from \citet{AG89}.
The difference in the Sm abundance between \citet{AG89} and
\citet{Lodders2009} was taken into account. Hereafter, the Solar $r$-residuals
are referred to as the Solar system $r$-process (SSr) abundance pattern.

The elements in the range from Ba to Hf in {\LyudmilasStar} were found to match
the scaled Solar $r$-process pattern very well, with a dispersion of 0.07\,dex
around the mean of the differences $(\eps{HE\,2327-5642} - \eps{SSr})$.  This
is in line with earlier results obtained for other $r$-process rich stars,
e.g. CS\,22892-052 \citep{sne03}, CS\,31082-001 \citep{hill02}, HD\,221170
\citep{ivans2006}, CS\,29491-069 and HE\,1219-0312 \citep{hayek09}, and it
gives further evidence for an universal production ratio of these elements during
the Galaxy history. It is worth noting that the use of the partition
function of \ion{Ho}{ii} from \citet{ho2-pf} improves the comparison to the scaled Solar $r$-process for holmium.

For the lighter elements in {\LyudmilasStar}, the difference $\Delta\eps{}(X)
= \eps{obs}(X) - \eps{SSr}(X)$ reveals a large spread of the data, between
$-0.4$\,dex (Mo) and $+0.7$\,dex (Y). In the following we will show that at
least a major fraction of the departures from the Solar $r$-process found for
the light trans-iron elements is likely to be due to inaccurate Solar
$r$-residuals.

\begin{figure*} 
 \centering
  \includegraphics[clip=true,bb=54 420 530 724,width=\textwidth]{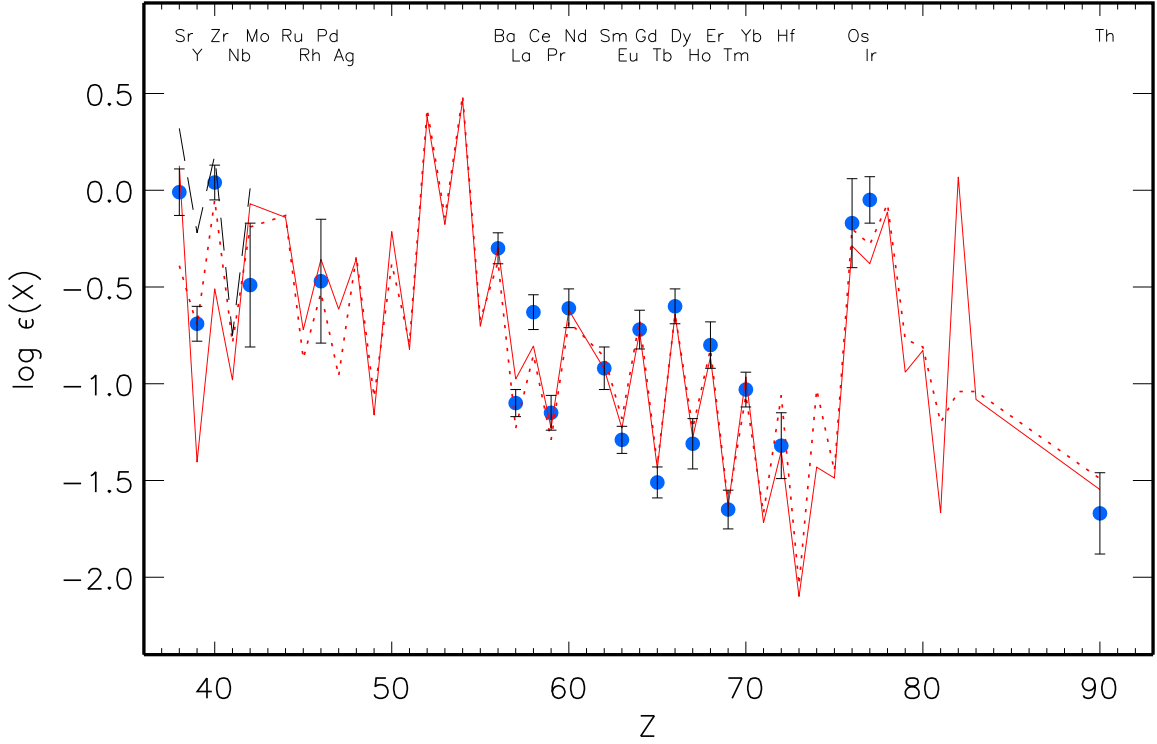}
  \caption{\label{Fig:SSr} The heavy-element abundance pattern of 
    {\LyudmilasStar} (filled circles) compared to the $r$-residuals calculated
    with various Solar total abundances and $s$-process abundances. Continuous
    and dotted curves correspond to the predicted $s$-process abundances of
    \citet[][stellar model]{rs99} and the Solar total abundances from
    \citet{Lodders2009} and \citet{aspl09}, respectively. The dashed curve
    (only Sr--Mo) corresponds to the $s$-process abundances of
    \citet{Travaglio04} and the Solar total abundances of \citet{Lodders2009}.
    Each $r$-process abundance pattern was scaled to match the Ba--Hf in
    {\LyudmilasStar}. }
\end{figure*}

For a given element, the $r$-residual is obtained by subtracting theoretical
$s$-process yields from the observed total Solar abundance.  Consider, e.g.,
the $s$-process abundances from \citet[][stellar model]{rs99} and use the
Solar total abundances from two different sources,
\citet[][meteoritic]{Lodders2009} and \citet[][photospheric]{aspl09}.  The
0.02\,dex increase of the yttrium abundance as one changes from
\citet{Lodders2009} to the \citet{aspl09} data leads to a 0.65\,dex increase
of the $r$-residual. A notable difference between two sets of the Solar
$r$-process abundances was found for all elements with dominant $s$-process
contribution to their Solar abundances; for example, $-0.58$\,dex for Sr,
$-0.40$\,dex for Zr, $-0.31$\,dex for La, etc. (see Fig.~\ref{Fig:SSr}). This
is because the calculation of the $r$-residuals involves the subtraction of a
large number from another large number, so that any small variations of one of
them leads to a dramatic change of the difference.

Significant uncertainties in the $r$-residual arise also from differences
between $s$-process calculations.  For example, \citet{rs99} obtained the $s$-process
abundance distribution as a best-fit to the Solar \emph{main} s-component
using stellar AGB models of 1.5 $M_\odot$ and 3 $M_\odot$ with half-Solar
metallicity. \citet{Travaglio99,Travaglio04} calculated the $s$-process
contribution to the Solar abundances by integration of the $s$-process yields
of different generations of AGB stars, i.e. considering the whole range of
Galactic metallicities. In both studies, very similar results were found for
Ba and Eu; however, \citet{Travaglio04} predicted lower $s$-process abundances
for the elements in the Sr--Mo range. Consequently, their Solar $r$-residuals
were significantly increased, as shown in Fig.~\ref{Fig:SSr}. From this
discussion, it is clear that no solid conclusion can be drawn with respect to
the existence of departures from the scaled Solar $r$-process pattern in the
Sr--Pd region in r-II stars.

Observations of metal-poor halo stars give evidence for the existence of a
distinct production mechanism for the light trans-iron (Sr--Zr) and heavy
elements beyond Ba in the early Galaxy \citep{aoki05,francois2007,sr_y_zr}. We
have chosen strontium and europium as representative elements of the first and
second neutron-capture element peaks and inspected the Sr/Eu abundance ratios
in a pre-selected sample of halo stars with a dominant contribution of the
$r$-process to the production of heavy elements beyond Ba, i.e. with
$\mathrm{[Ba/Eu]} \le -0.4$. The stars are separable into three groups,
depending on the observed europium enhancement. Nine r-II stars
($\mathrm{[Eu/Fe]} > 1.0$) were taken from
\citet{hill02,sne03,HERESpaperI,honda2004,HERESpaperII,francois2007,Lai2008,hayek09},
32 r-I stars ($0.2 \ge \mathrm{[Eu/Fe]} < 1.0$) from
\citet{cowan2002,honda2004,HERESpaperII,ivans2006,francois2007,sr_y_zr,Lai2008,hayek09},
and 12 stars with $\mathrm{[Eu/Fe]} \le 0.06$ (hereafter Eu-poor stars) from
\citet{honda2004,honda2006,honda2007,HERESpaperII,francois2007}.

\begin{figure}  
  \resizebox{88mm}{!}{\includegraphics{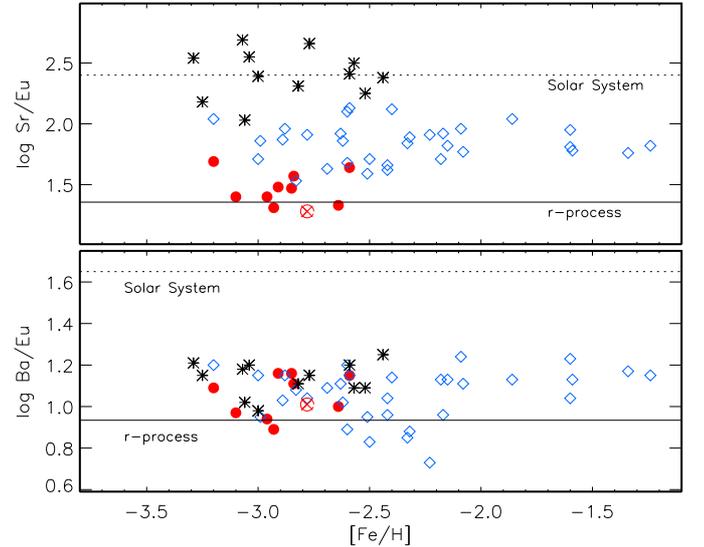}}
  \caption{\label{Fig:lepp} The Sr/Eu (top panel) and Ba/Eu (bottom panel) 
    abundance ratios in r-II (filled circles), r-I (open rombs), and
    Eu-poor (asterisks) stars (for the sources of the data, see text). 
{\LyudmilasStar} is shown by a crossed open circle. The solid and dotted 
    lines indicate the pure $r$-process and Solar system ratios, respectively.}
\end{figure}

As expected, each group of stars reveals very similar Ba/Eu abundance ratios,
as shown in the bottom panel of Fig.~\ref{Fig:lepp}, with mean Ba/Eu ratios of
$\log(\mathrm{Ba}/\mathrm{Eu}) = 1.05 \pm 0.10$ (r-II stars), $1.06 \pm 0.13$
(r-I stars), and $1.14 \pm 0.08$ (Eu-poor stars). 
Note that $\log(\mathrm{Ba}/\mathrm{Eu})_r = 0.93$ for a pure $r-$process production of heavy elements \citep{rs99}.
This suggests that only a
small number of $s$-nuclei (including those of strontium) existed in the
matter out of which these stars formed. The Sr/Eu abundance ratios reveal a
clear separation between each of these three groups (Fig.~\ref{Fig:lepp}).
Note that the Zr/Eu ratios exhibit very similar behavior. The mean Zr/Eu
abundance ratios are $\log(\mathrm{Zr}/\mathrm{Eu}) = 1.42 \pm 0.19$ (r-II
stars), $1.75 \pm 0.15$ (r-I stars), and $2.30 \pm 0.20$ (Eu-poor stars).
{\LyudmilasStar}, having $\log(\mathrm{Ba}/\mathrm{Eu}) = 1.01$,
$\log(\mathrm{Sr}/\mathrm{Eu}) = 1.28$ (crossed circle in
Fig.~\ref{Fig:lepp}), and $\log(\mathrm{Zr}/\mathrm{Eu}) = 1.33$, clearly
belongs to the group of r-II stars. 

From Figs.~\ref{Fig:AbundancePattern} and \ref{Fig:lepp}, it is clear that the
first and second $r$-process peak elements in the r-II stars are of common
origin. However, the origin of the first neutron-capture peak elements in the
r-I and Eu-poor stars is still unclear, despite of a number of studies
\citep{truran02,Travaglio04,far09}.

\section{Age determination}\label{Sect:ages}

The detection of thorium permits a nucleo-chronometric age
estimation of {\LyudmilasStar} by means of a comparison of the observed
Th-to-stable neutron capture element abundance ratios with $\log$(Th/$X$)$_0$,
the corresponding initial values at the time when the star was born:
\begin{displaymath}
\tau = 46.7\,\mathrm{Gyr}\,\left[\log({\rm Th}/X)_0 - \log({\rm Th}/X)_{obs}\right].
\end{displaymath}
Considering the uncertainty of the thorium abundance of 0.2\,dex, which
translates into an age uncertainty of 9.3\,Gyr, a precise age estimation is
not possible. Nevertheless, we investigated how the results depend on the
adopted values of $\log$(Th/$X$)$_0$, and which Th/$X$ pairs are possibly
reliable chronometers. We assume that thorium was produced together with the
lighter elements in the range between Ba and Hf.

We first determined the age of the star using the initial abundance ratios
from the dynamical network calculations of \citet{far08a}. They considered a
core-collapse supernova (SN\,II) with an adiabatically expanding high-entropy
wind (HEW) as the astrophysical environment for the $r$-process. In the HEW
scenario, the total nucleosynthetic yield is the sum of SN ejecta with
multiple components in different entropy ranges. \citet{far08a} found that
heavy elements beyond $Z \simeq 52$ are produced in the highest entropy ($S >
150$) zones in the so-called ``main'' $r$-process. The HEW model production
ratios, $\mathrm{PR} = \log(\mathrm{Th}/X)_0$, as given by \citet{hayek09},
and the calculated ages for multiple Th/$X$ pairs (where $X$ is one of the
elements in the Ba--Ir range for which we determined an abundance) are listed
in Table~\ref{Tab:age}.

The age uncertainties introduced by the measurement uncertainties are listed
in the column ``Error''. They were calculated as $\sigma_\tau =
46.7\,\sqrt{\sigma^2_{obs}(X) + \sigma^2_{obs}({\rm Th})}$\,Gyr with $\sigma_{obs}$
from Table~\ref{Tab:AbundanceErrors}. Variations of the stellar parameters
{\tefft} and $\log g$ yielded an uncertainty of 1.5\,Gyr in the final age. It
can be seen that individual pairs reveal a large spread of stellar ages. The
mean value is $\tau = 15.1\pm7.4$\,Gyr. As noted by \citet{hayek09}, the
estimate for hafnium is bracketed for the HEW model due to problems with the
nuclear data.  Neglecting hafnium and also osmium in the view of its less
reliable stellar abundance yields $\tau = 13.3\pm 6.2$\,Gyr, which agrees well with the expectation of the age of an extremely metal-poor star which formed in the early Galaxy. Note that the cosmic age derived from the results of the {\em Wilkinson Microwave Anisotropy Probe (WMAP)} experiment is 
$13.7\pm 0.2$\,Gyr \citep{spergel2003}.

For an additional estimate of the age of {\LyudmilasStar}, we employed the
Solar $r$-residual ratios $(\mathrm{Th}/X)_0$ for the elements for which the
$r$-process fraction exceeds 70\,\% (columns SSr in
Table~\ref{Tab:age}). Since these are measured values, they
depend only weakly on theoretial predictions and their associated nuclear
physics uncertainties. Since the Sun is approximately 4.5\,Gyr old, the
corresponding correction accounting for the thorium radiative decay was
introduced to the Solar current thorium abundance. The resulting mean age of
{\LyudmilasStar}, $\tau = 7.4\pm 5.5$\,Gyr, calculated using all stable elements from Sm to Ir seems low for a halo star. If we neglect the estimate based on Th/Ir, which is clearly an outlier (i.e.,
$\tau = 21.0$\,Gyr), we come to even lower stellar age of $\tau = 5.9\pm 2.8$\,Gyr.

Note that the stochastic error of the stellar age based on the HEW model predictions is large compared to that for the Solar $r$-residual ratios. This is most likely due to the
uncertainty of the theoretical yields for individual elements. 

\begin{table}[htbp]
\caption{\label{Tab:age} Logarithmic production ratios (PR)
for the HEW model and Solar system $r$-process (SSr) and 
  corresponding radioactive decay ages in {\LyudmilasStar}. 
}
\centering
 \begin{tabular}{cccrccr}\hline\hline
   Element & Stellar & \multicolumn{2}{c}{HEW $S=10\dots 300$} & & 
   \multicolumn{2}{c}{SSr} \\
   \cline{3-4}
   \cline{6-7}
   pair  & ratio   &  PR      & \multicolumn{1}{c}{age,} &  Error, 
         &  PR   & \multicolumn{1}{c}{age,} \\
         &         &          & \multicolumn{1}{c}{Gyr} & Gyr &  & Gyr \\
\hline
   Th/Ba & $-1.37$ &  $-0.96$ & 19.1~~~~  &  9.4 & $     $ &	\\
   Th/La & $-0.57$ &  $-0.22$ & 16.3~~~~  &  9.4 & $     $ &	\\
   Th/Ce & $-1.04$ &  $-0.62$ & 19.6~~~~  &  9.8 & $     $ &	\\
   Th/Pr & $-0.52$ &  $-0.20$ & 14.9~~~~  &  9.8 & $     $ &	\\
   Th/Nd & $-1.06$ &  $-0.87$ &  8.9~~~~  & 10.0 & $     $ &	\\
   Th/Sm & $-0.75$ &  $-0.71$ &  1.9~~~~  & 10.2 & $-0.62$ &  6.1 \\
   Th/Eu & $-0.38$ &  $-0.32$ &  2.8~~~~  &  9.4 & $-0.32$ &  2.8 \\
   Th/Gd & $-0.95$ &  $-0.73$ & 10.3~~~~  &  9.9 & $-0.82$ &  6.1 \\
   Th/Tb & $-0.16$ &  $     $ &       &      & $-0.11$ &  2.3 \\
   Th/Dy & $-1.07$ &  $-0.69$ & 17.7~~~~  &  9.8 & $-0.91$ &  7.5 \\
   Th/Ho & $-0.36$ &  $ 0.04$ & 18.7~~~~  & 10.4 & $-0.27$ &  4.2 \\
   Th/Er & $-0.87$ &  $-0.48$ & 18.2~~~~  & 10.2 & $-0.69$ &  8.4 \\
   Th/Tm & $-0.02$ &  $ 0.22$ & 11.2~~~~  & 10.0 & $ 0.08$ &  4.7 \\
   Th/Hf & $-0.35$ &  $ 0.20$ & 25.7~~~~  & 11.7 & $     $ &      \\
   Th/Os & $-1.50$ &  $-0.93$ & 26.6~~~~  & 13.2 & $-1.26$ & 11.2 \\
   Th/Ir & $-1.62$ &  $     $ &       & 10.1 & $-1.17$ & 21.0 \\
   \hline
   \multicolumn{3}{l}{~~~mean(Ba--Ir)}  & \multicolumn{2}{l}{~~15.1$\pm$7.4} & \multicolumn{2}{r}{7.4$\pm$5.5}  \\
   \hline
   \multicolumn{3}{l}{~~~final} & \multicolumn{2}{l}{~~13.3$\pm$6.2} & \multicolumn{2}{r}{5.9$\pm$2.8}  \\
   \multicolumn{3}{l}{} & \multicolumn{2}{l}{~~(Ba--Tm)} & \multicolumn{2}{r}{(Sm--Os)}  \\
   \hline
 \end{tabular}
\end{table}

\section{Conclusions}\label{Sect:DiscussionConclusions}

The high-quality VLT/UVES spectra of {\LyudmilasStar} enabled the
determination of accurate abundances for 40 elements, including 23 elements
in the nuclear charge range between $Z = 38$--$90$. We confirmed that
{\LyudmilasStar} is strongly $r$-process enhanced, having $[r/\mathrm{Fe}] =
0.99\pm0.03$ where $r$ denotes the average of the abundances of seven
elements (i.e., Eu, Gd, Tb, Dy, Ho, Er, and Tm), with an $r$-process contribution to
the Solar system matter of more than 85\,\% according to the $r$-residuals of
\citet{rs99}. {\LyudmilasStar} and three benchmark r-II stars,
CS\,22892-052 \citep{sne03}, CS\,31082-001 \citep{hill02}, and HE\,1219-0312
\citep{hayek09}, have very similar abundance patterns of the elements in the range
from Sr to Os. Hence, {\LyudmilasStar} is a member of the small sample of
currently- known r-II stars.

The elements in the range from Ba to Hf in {\LyudmilasStar} match the scaled
Solar $r$-process pattern very well. We showed that the Solar $r$-residuals
for the first $r$-process peak elements are rather uncertain. They may vary as
much as 0.5\,dex or even more, depending on the adopted Solar total abundances
and $s$-process fractions. Therefore, no firm conclusion can be drawn with
respect to a relation of the light trans-iron elements in {\LyudmilasStar} and
other r-II stars to the Solar $r$-process.

We found a clear distinction in Sr/Eu abundance ratios between the halo stars
with different europium enhancement and suggest using the [Sr/Eu] ratio in
addition to [Eu/Fe] to separate the strongly $r$-process enhanced (r-II) stars
from the other halo stars with dominant contribution of the $r$-process to
heavy element production. The r-II stars, whose stellar matter presumably has
experienced a single nucleosynthesis event, have $\mathrm{[Eu/Fe]} > 1\pm
0.1$, $\mathrm{[Ba/Eu]} = -0.60\pm 0.10$, and a low Sr/Eu abundance ratio of
$\mathrm{[Sr/Eu]} = -0.92\pm 0.13$. Stars with very similar Ba/Eu ratios have
two times (0.36\,dex) larger Sr/Eu ratios if their Eu/Fe ratio is in the range
$1.0 < \mathrm{[Eu/Fe]} < 0.2$ (i.e., r-I stars), and nearly an order of
magnitude (0.93\,dex) higher Sr/Eu ratios if $\mathrm{[Eu/Fe]} < 0$ (Eu-poor
stars). The origin of the first neutron-capture peak elements in the r-I stars
and Eu-poor stars is still unclear. Further theoretical studies will be needed
to elucidate this problem.

Only two elements, Os and Ir, of the third $r$-process peak were detected in
{\LyudmilasStar}. Iridium appears to be overabundant compared to the Ir
abundance determined in other $r$-process enhanced stars. However, due to the
uncertainty of the Ir abundance we cannot yet draw a firm conclusion on this
point.

The detection of thorium permitted an estimate of the radioactive decay age of
{\LyudmilasStar}, although the age uncertainty of 9.3\,Gyr introduced by the uncertainty of the thorium abundance is rather large. Employing multiple Th/$X$ chronometers and initial
production ratios based on the Solar $r$-residuals, an age of $5.9\pm
2.8$\,Gyr was obtained from nine Th/$X$ pairs, involving elements in the
Sm--Os range. Using the predictions of the HEW $r$-process model, as given by
\citet{hayek09}, we obtained $\tau = 13.3\pm 6.2$\,Gyr from 12 Th/$X$ pairs.

Based on our high-resolution spectra, covering $\sim 4.3$ years, we suspect that {\LyudmilasStar} is radial-velocity variable with a highly elliptical orbit of the system. Determination of the orbital period would provide the unique opportunity to determine a lower limit for the mass of the secondary in this system. Scenarios for the site of the $r-$process include a high-entropy wind from a type-II supernova \citep[e.g.][]{Woosley1994,takahashi1994}, ejecta from neutron star mergers \citep[e.g.][]{freiburg1999}, or the neutrino-driven wind of a neutron star newly-formed in an accretion induced collapse (AIC) event \citep[e.g.][]{woosley1992,qian2003}. According to these scenarios, it is expected that the secondary is a neutron-star. With a lower limit for the mass of the secondary it might be possible to constrain a scenario, because in the AIC case the neutron star is expected to have a mass just slightly above the Chandrasekhar mass, while core-collapse supernovae or neutron star mergers would result in remants of significantly higher mass. 

\begin{acknowledgements}
  
  The authors thank Thomas Gehren for the NLTE calculations for \ion{Al}{i} and Tatyana Ryabchikova for help with collecting the atomic data. 
L.M. and A.V. are supported by the Russian Foundation for Basic
  Research (grant 08-02-92203-GFEN-a), the Russian Federal Agency on Science and Innovation (No.  02.740.11.0247), and the Swiss National Science Foundation (SCOPES project No.~IZ73Z0-128180/1).
  N.C. is supported by the Knut and Alice Wallenberg Foundation, and by
  Deutsche Forschungsgemeinschaft through grants Ch~214/3 and Re~353/44.
P.S.B is a Royal Swedish Academy of Sciences Research Fellow supported by a grant from the Knut and Alice Wallenberg Foundation. P.S.B also acknowledges additional support from the Swedish Research Council. 
T.C.B. acknowledges partial
funding of this work from grants PHY 02-16783 and PHY 08- 22648:
Physics Frontier Center/Joint Institute for Nuclear Astrophysics
(JINA), awarded by the U.S. National Science Foundation.
We made use of model atmosphere from the MARCS library, and
  the NIST and VALD databases.

\end{acknowledgements}

\Online

\longtab{7}{
\begin{longtable}[]{rcccrlcrrrrcl}   
\caption{\label{linelist} Line data and abundances from an analysis of {\LyudmilasStar}. $\Gamma_6$ corresponds to 10\,000~K. Column 6 gives references to the adopted $gf-$values. Column 13 gives references to the sources of the used IS and HFS data and adopted $\Gamma_6-$values.}  \\
\hline \\ 
 Z & Atom/ & $\lambda$ & $E_{exc}$ & $\log gf$ & \multicolumn{1}{c}{Ref} & \multicolumn{3}{c}{$\eps{}$} & [X/Fe] & $W_\lambda$ &  $\log \Gamma_6/N_{\rm H}$ & Note, Ref. \\
\cline{7-9}
   &  mol  & (\AA)     & (eV)     &   &   & Solar & LTE  & NLTE & & (m\AA) & (rad/s$\cdot$cm$^3$)& \\
\hline 
 1 & 2 & 3 & 4 & \multicolumn{1}{c}{5} & \multicolumn{1}{c}{6} & 7 & \multicolumn{1}{c}{8} & \multicolumn{1}{c}{9} & \multicolumn{1}{c}{10} & \multicolumn{1}{c}{11} & 12 & \multicolumn{1}{c}{13} \\
\hline \\
\endfirsthead
\caption{continued.}\\
\hline \\
 Z & Atom/ & $\lambda$ & $E_{exc}$ & $\log gf$ & Ref & \multicolumn{3}{c}{$\eps{}$} & [X/Fe] & $W_\lambda$ &  $\log \Gamma_6/N_{\rm H}$ & Note, Ref. \\
\cline{7-9}
   &  mol  & (\AA)     & (eV)     &   &   & Solar & LTE  & NLTE & & (m\AA) & (rad/s$\cdot$cm$^3$) & \\
\hline 
 1 & 2 & 3 & 4 & \multicolumn{1}{c}{5} & \multicolumn{1}{c}{6} & 7 & \multicolumn{1}{c}{8} & \multicolumn{1}{c}{9} & \multicolumn{1}{c}{10} & \multicolumn{1}{c}{11} & 12 & \multicolumn{1}{c}{13} \\
\hline  \\   
\endhead
\hline
\endfoot
\hline
\endlastfoot
  3 & Li 1 & 6707.80 & 0.00 &  0.167 & NIST8 &  1.10 &  0.99 & &  2.67 & Syn &  -7.574 & IS:SRE95 \\
    &      &         &      &        &       &      &       &      &      &      &        & $\Gamma_6$:ABO \\
\multicolumn{13}{c}{ } \\
  6 & CH   &  \multicolumn{3}{l}{4310.0 - 4312.5} & BCB05 & 8.39 & 5.74 & & 0.13 & Syn & \\
  6 & CH   &  \multicolumn{3}{l}{4313.4 - 4313.8} & BCB05 & 8.39 & 5.74 & & 0.13 & Syn & \\
  6 & CH   &  \multicolumn{3}{l}{4362.4 - 4364.6} & BCB05 & 8.39 & 5.76 & & 0.15 & Syn & \\
  6 & CH   &  \multicolumn{3}{l}{4366.0 - 4367.0} & BCB05 & 8.39 & 5.73 & & 0.12 & Syn & \\
\multicolumn{13}{c}{ } \\
  7 & NH   &  \multicolumn{3}{l}{3358 - 3361} & K94  & 7.86 & 4.78 & & -0.30 & Syn & \\
\multicolumn{13}{c}{ } \\
 11 & Na 1 & 5889.96 & 0.00 &  0.110 & NIST8 &  6.30 &  3.28 & 2.89 & -0.63 & Syn &  -7.712 & $\Gamma_6$:GLS04\\
 11 & Na 1 & 5895.93 & 0.00 & -0.190 & NIST8 &  6.30 &  3.24 & 2.96 & -0.56 & Syn &  -7.712 & the same \\
\multicolumn{13}{c}{ } \\
 12 & Mg 1 & 4571.10 & 0.00 & -5.620 & NIST8 &  7.54 &  4.86 & 4.98 &  0.22 & Syn &  -7.856 & $\Gamma_6$:GLS04\\
 12 & Mg 1 & 4702.99 & 4.33 & -0.440 & NIST8 &  7.54 &  4.79 & 4.90 &  0.14 & Syn &  -6.956 & the same \\
 12 & Mg 1 & 5528.41 & 4.33 & -0.498 & NIST8 &  7.54 &  4.84 & 4.96 &  0.20 & Syn &  -7.152 & the same \\
\multicolumn{13}{c}{ } \\
 13 & Al 1 & 3961.52 & 0.01 & -0.340 & NIST8 &  6.47 &  2.50 & 3.02 & -0.67 & Syn &  -7.552 & $\Gamma_6$:GLS04\\
\multicolumn{13}{c}{ } \\
 14 & Si 1 & 3905.53 & 1.91 & -1.041 & BL91  &  7.52 &  4.85 &      &  0.11 & Syn &  -7.440 & $\Gamma_6$:ABO\\
\multicolumn{13}{c}{ } \\
 20 & Ca 1 & 4425.44 & 1.88 & -0.358 & SON75 & 6.33 &  3.52 & 3.76 & 0.20 & 23.7 & -7.432 & $\Gamma_6$:MKP07 \\
 20 & Ca 1 & 5349.46 & 2.71 & -0.310 & SR81  & 6.33 &  3.52 & 3.78 & 0.21 &  4.6 & -7.652 & $\Gamma_6$:S81 \\
 20 & Ca 1 & 5588.75 & 2.53 &  0.358 & SR81  & 6.33 &  3.57 & 3.86 & 0.26 & 22.8 & -7.628 & $\Gamma_6$:S81 \\
 20 & Ca 1 & 5857.45 & 2.93 &  0.240 & SR81  & 6.33 &  3.67 & 3.86 & 0.36 & 11.2 & -7.316 & $\Gamma_6$:S81 \\
 20 & Ca 1 & 6122.22 & 1.89 & -0.315 & SON75 & 6.33 &  3.68 & 3.90 & 0.36 & 29.5 & -7.189 & $\Gamma_6$:ABO \\
 20 & Ca 1 & 6162.17 & 1.90 & -0.089 & SON75 & 6.33 &  3.63 & 3.84 & 0.32 & 38.6 & -7.189 & $\Gamma_6$:ABO \\
 20 & Ca 1 & 6169.56 & 2.53 & -0.478 & SR81  & 6.33 &  3.52 & 3.81 & 0.21 &  4.7 & -7.264 & $\Gamma_6$:S81 \\
 20 & Ca 1 & 6439.08 & 2.53 &  0.390 & SR81  & 6.33 &  3.63 & 3.80 & 0.31 & 27.4 & -7.704 & $\Gamma_6$:S81 \\
 20 & Ca 2 & 8498.02 & 1.69 & -1.416 & Th89  & 6.33 &  -    & 3.82 & 0.27 & Syn~ & -7.676 & IS:NBI98 \\
    &      &         &      &        &       &      &       &      &      &      &        & $\Gamma_6$:ABO \\
\multicolumn{13}{c}{ } \\
 21 & Sc 2 & 4246.82 & 0.32 &  0.240 & LD89 &  3.10 &  0.08 & & -0.24 & Syn &         & HFS:MPS95 \\
 21 & Sc 2 & 4400.39 & 0.60 & -0.540 & LD89 &  3.10 &  0.08 & & -0.24 & Syn &         & the same  \\
 21 & Sc 2 & 4415.56 & 0.60 & -0.670 & LD89 &  3.10 &  0.13 & & -0.19 & Syn &         & the same \\
 21 & Sc 2 & 5031.02 & 1.36 & -0.400 & LD89 &  3.10 &  0.09 & & -0.23 & Syn &         & the same \\
\multicolumn{13}{c}{ } \\
 22 & Ti 1 & 3998.64 & 0.05 & -0.056 & WF75 & 4.90 &  2.14 &  &   0.02 &  37.3 & -7.654 & $\Gamma_6$:ABO \\
 22 & Ti 1 & 4533.25 & 0.85 &  0.476 & WF75 & 4.90 &  2.06 &  &  -0.06 &  19.7 & -7.593 & the same      \\
 22 & Ti 1 & 4534.78 & 0.84 &  0.280 & WF75 & 4.90 &  2.01 &  &  -0.11 &  12.5 & -7.596 & the same      \\
 22 & Ti 1 & 4981.73 & 0.85 &  0.504 & WF75 & 4.90 &  2.15 &  &   0.03 &  25.6 & -7.626 & the same      \\
 22 & Ti 1 & 4991.06 & 0.84 &  0.380 & WF75 & 4.90 &  2.15 &  &   0.03 &  21.0 & -7.629 & the same      \\
 22 & Ti 1 & 5022.87 & 0.83 & -0.434 & WF75 & 4.90 &  2.14 &  &   0.02 &   3.9 & -7.633 & the same      \\
 22 & Ti 1 & 5024.85 & 0.82 & -0.602 & WF75 & 4.90 &  2.32 &  &   0.20 &   4.1 & -7.635 & the same      \\
 22 & Ti 1 & 5039.96 & 0.02 & -1.130 & WF75 & 4.90 &  2.09 &  &  -0.03 &   5.7 & -7.720 & the same      \\
 22 & Ti 1 & 5064.66 & 0.05 & -0.991 & WF75 & 4.90 &  2.18 &  &   0.06 &   8.8 & -7.719 & the same      \\
 22 & Ti 1 & 5210.39 & 0.05 & -0.884 & WF75 & 4.90 &  2.21 &  &   0.09 &  11.8 & -7.724 & the same      \\
 22 & Ti 2 & 3913.47 & 1.12 & -0.420 & PTP01 & 4.90 &  2.29 &  &   0.17 &  87.4 &  &	  \\
 22 & Ti 2 & 4028.34 & 1.89 & -0.960 & PTP01 & 4.90 &  2.25 &  &   0.13 &  21.8 &  &	  \\
 22 & Ti 2 & 4053.83 & 1.89 & -1.130 & PTP01 & 4.90 &  2.11 &  &  -0.01 &  11.9 &  &	  \\
 22 & Ti 2 & 4290.22 & 1.16 & -0.850 & PTP01 & 4.90 &  2.16 &  &   0.04 &  66.5 &  &	  \\
 22 & Ti 2 & 4300.05 & 1.18 & -0.440 & PTP01 & 4.90 &  2.41 &  &   0.29 &  93.3 &  &	  \\
 22 & Ti 2 & 4394.05 & 1.22 & -1.780 & PTP01 & 4.90 &  2.24 &  &   0.12 &  20.5 &  &	  \\
 22 & Ti 2 & 4395.03 & 1.08 & -0.540 & PTP01 & 4.90 &  2.22 &  &   0.10 &  88.0 &  &	  \\
 22 & Ti 2 & 4395.85 & 1.24 & -1.930 & PTP01 & 4.90 &  2.29 &  &   0.17 &  16.2 &  &	  \\
 22 & Ti 2 & 4399.77 & 1.24 & -1.190 & PTP01 & 4.90 &  2.29 &  &   0.17 &  51.4 &  &	   \\
 22 & Ti 2 & 4417.72 & 1.16 & -1.190 & PTP01 & 4.90 &  2.23 &  &   0.11 &  53.0 &  &	  \\
 22 & Ti 2 & 4418.33 & 1.24 & -1.970 & PTP01 & 4.90 &  2.23 &  &   0.11 &  13.3 &  &	  \\
 22 & Ti 2 & 4443.79 & 1.08 & -0.720 & PTP01 & 4.90 &  2.22 &  &   0.10 &  79.9 &  &	  \\
 22 & Ti 2 & 4444.56 & 1.12 & -2.240 & PTP01 & 4.90 &  2.29 &  &   0.17 &  11.7 &  &	  \\
 22 & Ti 2 & 4450.48 & 1.08 & -1.520 & PTP01 & 4.90 &  2.24 &  &   0.12 &  40.2 &  &	  \\
 22 & Ti 2 & 4464.45 & 1.16 & -1.810 & PTP01 & 4.90 &  2.33 &  &   0.21 &  26.0 &  &	  \\
 22 & Ti 2 & 4468.51 & 1.13 & -0.600 & BHN93 & 4.90 &  2.20 &  &   0.08 &  81.8 &  &	  \\
 22 & Ti 2 & 4470.86 & 1.16 & -2.020 & PTP01 & 4.90 &  2.07 &  &  -0.05 &  10.6 &  &	  \\
 22 & Ti 2 & 4501.27 & 1.12 & -0.770 & PTP01 & 4.90 &  2.23 &  &   0.11 &  77.5 &  &	  \\
 22 & Ti 2 & 4533.97 & 1.24 & -0.530 & PTP01 & 4.90 &  2.10 &  &  -0.02 &  76.0 &  &	  \\
 22 & Ti 2 & 4563.76 & 1.22 & -0.690 & PTP01 & 4.90 &  2.05 &  &  -0.07 &  67.7 &  &	  \\
 22 & Ti 2 & 4571.97 & 1.57 & -0.320 & PTP01 & 4.90 &  2.08 &  &  -0.04 &  67.7 &  &	  \\
 22 & Ti 2 & 5185.91 & 1.89 & -1.490 & PTP01 & 4.90 &  2.14 &  &   0.02 &   7.3 &  &	  \\
 22 & Ti 2 & 5188.70 & 1.58 & -1.050 & PTP01 & 4.90 &  2.20 &  &   0.08 &  36.4 &  &	  \\
 22 & Ti 2 & 5226.55 & 1.57 & -1.260 & PTP01 & 4.90 &  2.21 &  &   0.09 &  26.9 &  &	  \\
 22 & Ti 2 & 5336.79 & 1.58 & -1.590 & PTP01 & 4.90 &  2.20 &  &   0.08 &  14.3 &  &	  \\
 22 & Ti 2 & 5381.01 & 1.57 & -1.920 & PTP01 & 4.90 &  2.15 &  &   0.03 &   6.6 &  &	  \\
\multicolumn{13}{c}{ } \\
 23 & V  2 & 3530.76 & 1.07 & -0.600 & FW96 &  4.00 &  0.84 & & -0.38 & Syn &        & \\
 23 & V  2 & 3545.19 & 1.10 & -0.390 & FW96 &  4.00 &  0.84 & & -0.38 & Syn &         & \\
 23 & V  2 & 3592.03 & 1.10 & -0.370 & FW96 &  4.00 &  0.88 & & -0.34 & Syn &         & \\
 23 & V  2 & 4005.71 & 1.82 & -0.460 & FW96 &  4.00 &  1.01 & & -0.21 & Syn &         & \\
\multicolumn{13}{c}{ } \\
 24 & Cr 1 & 4254.33 & 0.00 & -0.114 & MFW88 & 5.64 &  2.22 &  &  -0.64 &  80.4 & -7.675 & $\Gamma_6$:ABO      \\
 24 & Cr 1 & 4274.80 & 0.00 & -0.230 & MFW88 & 5.64 &  2.16 &  &  -0.70 &  73.9 & -7.675 & the same     \\
 24 & Cr 1 & 5206.04 & 0.94 &  0.019 & MFW88 & 5.64 &  2.31 &  &  -0.55 &  45.1 & -7.597 & the same     \\
 24 & Cr 1 & 5345.81 & 1.00 & -0.980 & MFW88 & 5.64 &  2.43 &  &  -0.43 &   8.7 & -7.620 & the same     \\
 24 & Cr 1 & 5348.33 & 1.00 & -1.290 & FW96  & 5.64 &  2.34 &  &  -0.52 &   3.6 & -7.620 & the same     \\
 24 & Cr 1 & 5409.80 & 1.03 & -0.720 & MFW88 & 5.64 &  2.44 &  &  -0.42 &  14.1 & -7.620 & the same     \\
\multicolumn{13}{c}{ } \\
 25 & Mn 1 & 4030.75 & 0.00 & -0.470 & BBP84 &  5.37 &  1.53 & & -1.06 & Syn &  -7.552 & HFS:LGB03\\
    &      &         &      &        &       &       &       & &       &     &         & $\Gamma_6$:ABO \\
 25 & Mn 1 & 4033.06 & 0.00 & -0.620 & BBP84 &  5.37 &  1.53 & & -1.06 & Syn &  -7.552 & the same \\
 25 & Mn 1 & 4034.48 & 0.00 & -0.810 & BBP84 &  5.37 &  1.53 & & -1.06 & Syn &  -7.552 & the same \\
 25 & Mn 1 & 4041.35 & 2.11 &  0.280 & BBP84 &  5.37 &  1.89 & & -0.70 & Syn &  -7.701 & $\Gamma_6$:ABO\\
 25 & Mn 2 & 3460.31 & 1.81 & -0.640 & KG00  &  5.37 &  1.91 & & -0.68 & Syn &         & HFS:HSR99\\
 25 & Mn 2 & 3482.90 & 1.83 & -0.840 & KG00  &  5.37 &  1.94 & & -0.65 & Syn &         & HFS:HSR99\\
 25 & Mn 2 & 3488.67 & 1.85 & -0.950 & KG00  &  5.37 &  2.10 & & -0.49 & Syn &         & HFS:HSR99\\
\multicolumn{13}{c}{ } \\
 26 & Fe 1 & 3917.18 & 0.99 & -2.155 & OWL91 & 7.45 & 4.61 & 4.69 &  0.02 & 65.3 & -7.695 & $\Gamma_6$:ABO  \\
 26 & Fe 1 & 3949.95 & 2.18 & -1.251 & OWL91 & 7.45 & 4.39 & 4.47 & -0.19 & 36.8 & -7.820 & the same  \\
 26 & Fe 1 & 4005.24 & 1.56 & -0.580 & OWL91 & 7.45 & 4.66 & 4.72 &  0.07 & 96.1 & -7.620 & the same  \\
 26 & Fe 1 & 4132.06 & 1.61 & -0.675 & OWL91 & 7.45 & 4.61 & 4.68 &  0.02 & 94.3 & -7.620 & the same  \\
 26 & Fe 1 & 4132.90 & 2.85 & -1.006 & OWL91 & 7.45 & 4.53 & 4.61 & -0.06 & 19.4 & -7.659 & the same  \\
 26 & Fe 1 & 4143.87 & 1.56 & -0.511 & OWL91 & 7.45 & 4.53 & 4.59 & -0.06 & 98.0 & -7.636 & the same  \\
 26 & Fe 1 & 4147.67 & 1.49 & -2.074 & OWL91 & 7.45 & 4.68 & 4.76 &  0.10 & 40.3 & -7.648 & the same  \\
 26 & Fe 1 & 4202.03 & 1.49 & -0.708 & FMW88 & 7.45 & 4.64 & 4.67 &  0.05 & 98.4 & -7.648 & the same  \\
 26 & Fe 1 & 4216.18 & 0.00 & -3.356 & FMW88 & 7.45 & 4.66 & 4.75 &  0.08 & 69.5 & -7.797 & the same  \\
 26 & Fe 1 & 4250.79 & 1.56 & -0.714 & OWL91 & 7.45 & 4.69 & 4.73 &  0.10 & 95.9 & -7.648 & the same  \\
 26 & Fe 1 & 4260.47 & 2.40 &  0.080 & OWL91 & 7.45 & 4.45 & 4.51 & -0.14 & 86.1 & -7.274 & the same  \\
 26 & Fe 1 & 4415.12 & 1.61 & -0.615 & FMW88 & 7.45 & 4.56 & 4.61 & -0.02 & 99.0 & -7.652 & the same  \\
 26 & Fe 1 & 4427.31 & 0.05 & -2.924 & OWL91 & 7.45 & 4.77 & 4.86 &  0.18 & 87.6 & -7.813 & the same  \\
 26 & Fe 1 & 4602.94 & 1.49 & -2.209 & OWL91 & 7.45 & 4.55 & 4.63 & -0.04 & 36.3 & -7.790 & the same  \\
 26 & Fe 1 & 4647.43 & 2.95 & -1.351 & OWL91 & 7.45 & 4.49 & 4.57 & -0.10 &  7.9 & -7.685 & the same  \\
 26 & Fe 1 & 4966.09 & 3.33 & -0.871 & OWL91 & 7.45 & 4.59 & 4.69 &  0.00 & 11.4 & -7.218 & the same  \\
 26 & Fe 1 & 4994.13 & 0.92 & -2.970 & OWL91 & 7.45 & 4.63 & 4.72 &  0.04 & 31.4 & -7.744 & the same  \\
 26 & Fe 1 & 5001.86 & 3.88 &  0.010 & FMW88 & 7.45 & 4.57 & 4.67 & -0.02 & 13.9 & -7.273 & the same  \\
 26 & Fe 1 & 5014.94 & 3.94 & -0.303 & OWL91 & 7.45 & 4.59 & 4.68 &  0.00 &  8.7 & -7.268 & $\Gamma_6$:ABO  \\
 26 & Fe 1 & 5051.63 & 0.92 & -2.765 & OWL91 & 7.45 & 4.71 & 4.80 &  0.12 & 48.0 & -7.746 & the same  \\
 26 & Fe 1 & 5068.77 & 2.94 & -1.042 & OWL91 & 7.45 & 4.56 & 4.65 & -0.03 & 14.1 & -7.265 & the same  \\
 26 & Fe 1 & 5150.84 & 0.99 & -3.040 & OWL91 & 7.45 & 4.67 & 4.75 &  0.08 & 23.0 & -7.742 & the same  \\
 26 & Fe 1 & 5166.28 & 0.00 & -4.200 & OWL91 & 7.45 & 4.69 & 4.78 &  0.10 & 33.5 & -7.826 & the same  \\
 26 & Fe 1 & 5171.60 & 1.49 & -1.720 & OWL91 & 7.45 & 4.81 & 4.89 &  0.22 & 63.8 & -7.687 & the same  \\
 26 & Fe 1 & 5192.34 & 3.00 & -0.421 & OWL91 & 7.45 & 4.53 & 4.61 & -0.06 & 34.7 & -7.266 & the same  \\
 26 & Fe 1 & 5194.94 & 1.56 & -2.020 & OWL91 & 7.45 & 4.67 & 4.75 &  0.08 & 41.0 & -7.680 & the same  \\
 26 & Fe 1 & 5216.27 & 1.61 & -2.080 & OWL91 & 7.45 & 4.69 & 4.77 &  0.10 & 37.3 & -7.674 & the same  \\
 26 & Fe 1 & 5281.79 & 3.04 & -0.830 & OWL91 & 7.45 & 4.65 & 4.73 &  0.06 & 18.4 & -7.266 & the same  \\
 26 & Fe 1 & 5283.62 & 3.24 & -0.520 & OWL91 & 7.45 & 4.48 & 4.57 & -0.11 & 21.8 & -7.221 & the same  \\
 26 & Fe 1 & 5324.18 & 3.21 & -0.100 & BKK91 & 7.45 & 4.35 & 4.44 & -0.24 & 35.1 & -7.235 & the same  \\
 26 & Fe 1 & 5339.93 & 3.27 & -0.720 & OWL91 & 7.45 & 4.48 & 4.57 & -0.11 & 14.5 & -7.221 & the same  \\
 26 & Fe 1 & 5364.87 & 4.45 &  0.228 & OWL91 & 7.45 & 4.47 & 4.58 & -0.12 &  5.3 & -7.136 & the same  \\
 26 & Fe 1 & 5367.47 & 4.41 &  0.440 & OWL91 & 7.45 & 4.56 & 4.67 & -0.03 &  8.9 & -7.153 & the same  \\
 26 & Fe 1 & 5369.96 & 4.37 &  0.536 & OWL91 & 7.45 & 4.58 & 4.69 & -0.01 & 12.3 & -7.179 & the same  \\
 26 & Fe 1 & 5383.37 & 4.31 &  0.645 & OWL91 & 7.45 & 4.57 & 4.69 & -0.02 & 16.8 & -7.219 & the same  \\
 26 & Fe 1 & 5393.17 & 3.24 & -0.720 & BKK91 & 7.45 & 4.56 & 4.64 & -0.03 & 15.2 & -7.235 & the same  \\
 26 & Fe 1 & 5410.91 & 4.47 &  0.398 & OWL91 & 7.45 & 4.62 & 4.73 &  0.03 &  7.5 & -7.132 & the same  \\
 26 & Fe 1 & 5415.20 & 4.39 &  0.642 & OWL91 & 7.45 & 4.52 & 4.63 & -0.07 & 13.4 & -7.182 & the same  \\
 26 & Fe 1 & 5434.52 & 1.01 & -2.130 & OWL91 & 7.45 & 4.77 & 4.85 &  0.18 & 80.4 & -7.749 & the same  \\
 26 & Fe 1 & 5445.04 & 4.39 & -0.020 & FMW88 & 7.45 & 4.58 & 4.69 & -0.01 &  5.7 & -7.189 & the same  \\
 26 & Fe 1 & 5506.78 & 0.99 & -2.797 & OWL91 & 7.45 & 4.70 & 4.79 &  0.11 & 44.6 & -7.753 & the same  \\
 26 & Fe 1 & 5586.76 & 3.37 & -0.100 & BKK91 & 7.45 & 4.42 & 4.51 & -0.17 & 29.4 & -7.221 & the same  \\
 26 & Fe 1 & 5615.64 & 3.33 &  0.050 & BKK91 & 7.45 & 4.41 & 4.50 & -0.18 & 37.2 & -7.234 & the same  \\
 26 & Fe 1 & 6136.62 & 2.45 & -1.410 & OWL91 & 7.45 & 4.50 & 4.58 & -0.09 & 25.2 & -7.609 & the same  \\
 26 & Fe 1 & 6137.69 & 2.59 & -1.350 & OWL91 & 7.45 & 4.46 & 4.54 & -0.13 & 19.1 & -7.589 & the same  \\
 26 & Fe 1 & 6191.56 & 2.43 & -1.417 & OWL91 & 7.45 & 4.55 & 4.63 & -0.04 & 24.4 & -7.615 & the same  \\
 26 & Fe 1 & 6393.60 & 2.43 & -1.570 & OWL91 & 7.45 & 4.44 & 4.52 & -0.14 & 20.0 & -7.622 & the same  \\
 26 & Fe 1 & 6400.00 & 3.60 & -0.290 & BKK91 & 7.45 & 4.46 & 4.55 & -0.13 & 14.0 & -7.232 & the same  \\
 26 & Fe 1 & 6430.85 & 2.18 & -1.950 & OWL91 & 7.45 & 4.63 & 4.71 &  0.04 & 18.0 & -7.704 & the same  \\
 26 & Fe 2 & 4508.29 & 2.86 & -2.440 & MB09  & 7.45 & 4.64 & 4.64 & -0.03 & 22.6 & -7.870 & the same  \\
 26 & Fe 2 & 4515.34 & 2.84 & -2.600 & MB09  & 7.45 & 4.81 & 4.81 &  0.14 & 21.2 & -7.880 & the same  \\
 26 & Fe 2 & 4520.22 & 2.81 & -2.650 & MB09  & 7.45 & 4.61 & 4.61 & -0.06 & 17.6 & -7.880 & the same  \\
 26 & Fe 2 & 4923.93 & 2.89 & -1.260 & MB09  & 7.45 & 4.59 & 4.58 & -0.07 & 68.5 & -7.890 & the same  \\
 26 & Fe 2 & 5018.44 & 2.89 & -1.100 & MB09  & 7.45 & 4.64 & 4.63 & -0.02 & 77.9 & -7.890 & the same  \\
 26 & Fe 2 & 5197.58 & 3.23 & -2.220 & MB09  & 7.45 & 4.69 & 4.69 &  0.03 & 14.6 & -7.880 & the same  \\
 26 & Fe 2 & 5234.62 & 3.22 & -2.180 & MB09  & 7.45 & 4.70 & 4.70 &  0.03 & 17.6 & -7.880 & the same \\
 26 & Fe 2 & 5284.11 & 2.89 & -3.110 & MB09  & 7.45 & 4.73 & 4.73 &  0.06 &  5.6 & -7.890 & the same \\
\multicolumn{13}{c}{ } \\
 27 & Co 1 & 3412.33 & 0.51 &  0.030 & CSS82 &  4.92 &  1.87 & & -0.27 & Syn &  -7.666 & HFS:P96\\
    &      &         &      &        &       &       &       & &       &     &         & $\Gamma_6$:ABO \\
 27 & Co 1 & 3417.16 & 0.58 & -0.470 & CSS82 &  4.92 &  1.97 & & -0.17 & Syn &  -7.657 & $\Gamma_6$:ABO \\
 27 & Co 1 & 3489.40 & 0.92 &  0.150 & CSS82 &  4.92 &  1.86 & & -0.28 & Syn &  -7.610 & HFS:P96\\
    &      &         &      &        &       &       &       & &       &     &         & $\Gamma_6$:ABO \\
 27 & Co 1 & 3894.07 & 1.05 &  0.090 & NKW99 &  4.92 &  1.98 & & -0.16 & Syn &  -7.648 & HFS:P96\\
    &      &         &      &        &       &       &       & &       &     &         & $\Gamma_6$:ABO \\
 27 & Co 1 & 4118.77 & 1.05 & -0.470 & NKW99 &  4.92 &  2.05 & & -0.09 & Syn &  -7.710 & $\Gamma_6$:ABO \\
 27 & Co 1 & 4121.31 & 0.92 & -0.300 & NKW99 &  4.92 &  2.05 & & -0.09 & Syn &  -7.724 & HFS:P96\\
    &      &         &      &        &       &       &       & &       &     &         & $\Gamma_6$:ABO \\
\multicolumn{13}{c}{ } \\
 28 & Ni 1 & 3413.47 & 0.16 & -1.480 & HS80  &  6.23 &  3.18 & & -0.27 & Syn &  -7.690 & $\Gamma_6$:ABO \\
 28 & Ni 1 & 3413.93 & 0.11 & -1.720 & FMW88 &  6.23 &  3.26 & & -0.19 & Syn &  -7.785 & the same \\
 28 & Ni 1 & 3519.76 & 0.28 & -1.420 & BBP89 &  6.23 &  3.08 & & -0.37 & Syn &  -7.689 & the same \\
 28 & Ni 1 & 3783.52 & 0.42 & -1.310 & HS80  &  6.23 &  3.00 & & -0.45 & Syn &  -7.780 & the same \\
 28 & Ni 1 & 3807.14 & 0.42 & -1.220 & BBP89 &  6.23 &  3.00 & & -0.45 & Syn &  -7.694 & the same \\
 28 & Ni 1 & 3858.29 & 0.42 & -0.950 & BBP89 &  6.23 &  2.98 & & -0.47 & Syn &  -7.700 & the same \\
 28 & Ni 1 & 5035.37 & 3.63 &  0.290 & WL97a &  6.23 &  3.09 & & -0.36 & Syn &  -7.231 & $\Gamma_6$:ABO \\
 28 & Ni 1 & 5476.90 & 1.83 & -0.890 & FMW88 &  6.23 &  3.20 & & -0.25 & Syn &         & \\
\multicolumn{13}{c}{ } \\
 30 & Zn 1 & 4722.16 & 4.01 & -0.390 & BG80 &  4.62 &  1.82 & & -0.02 & Syn &     & \\
 30 & Zn 1 & 4810.54 & 4.06 & -0.170 & BG80 &  4.62 &  1.84 & &  0.00 & Syn &     & \\
\multicolumn{13}{c}{ } \\
 38 & Sr 2 & 4077.72 & 0.00 &  0.150 & RCW80 &  2.92 & -0.01 & -0.16 & -0.30 & Syn &  -7.792 & HFS:BBH83\\
    &      &         &      &        &       &       &       & &       &     &         & $\Gamma_6$:MZG08 \\
 38 & Sr 2 & 4215.54 & 0.00 & -0.170 & RCW80 &  2.92 & -0.01 & -0.16 & -0.30 & Syn &  -7.792 & the same\\
\multicolumn{13}{c}{ } \\
 39 & Y  2 & 3549.01 & 0.13 & -0.280 & HLG82 &  2.21 & -0.74 & & -0.17 & Syn &     & \\
 39 & Y  2 & 3600.74 & 0.18 &  0.280 & HLG82 &  2.21 & -0.67 & & -0.10 & Syn &     & \\
 39 & Y  2 & 3611.04 & 0.13 &  0.010 & HLG82 &  2.21 & -0.73 & & -0.16 & Syn &     & \\
 39 & Y  2 & 3774.33 & 0.13 &  0.210 & HLG82 &  2.21 & -0.67 & & -0.10 & Syn &     & \\
 39 & Y  2 & 3788.69 & 0.10 & -0.070 & HLG82 &  2.21 & -0.66 & & -0.09 & Syn &     & \\
 39 & Y  2 & 3950.35 & 0.10 & -0.490 & HLG82 &  2.21 & -0.56 & &  0.01 & Syn &     & \\
 39 & Y  2 & 4883.68 & 1.08 &  0.070 & HLG82 &  2.21 & -0.66 & & -0.09 & Syn &     & \\
 39 & Y  2 & 5087.43 & 1.08 & -0.170 & HLG82 &  2.21 & -0.75 & & -0.18 & Syn &     & \\
 39 & Y  2 & 5205.73 & 1.03 & -0.340 & HLG82 &  2.21 & -0.75 & & -0.18 & Syn &     & \\
\multicolumn{13}{c}{ } \\
 40 & Zr 2 & 3404.83 & 0.36 & -0.490 & LNA06 &  2.58 & -0.01 & 0.07 &  0.26 & Syn &     & \\
 40 & Zr 2 & 3430.53 & 0.47 & -0.160 & LNA06 &  2.58 & -0.03 & 0.08 &  0.28 & Syn &     & \\
 40 & Zr 2 & 3457.56 & 0.56 & -0.470 & MBM06 &  2.58 &  0.06 & 0.14 &  0.34 & Syn &     & \\
 40 & Zr 2 & 3479.39 & 0.71 & -0.180 & LNA06 &  2.58 &  0.02 & 0.16 &  0.36 & Syn &     & \\
 40 & Zr 2 & 3499.57 & 0.41 & -1.060 & LNA06 &  2.58 & -0.02 & 0.02 &  0.22 & Syn &     & \\
 40 & Zr 2 & 3505.67 & 0.16 & -0.390 & LNA06 &  2.58 & -0.02 & 0.03 &  0.23 & Syn &     & \\
 40 & Zr 2 & 3551.95 & 0.09 & -0.360 & LNA06 &  2.58 &  0.01 & 0.07 &  0.27 & Syn &     & \\
 40 & Zr 2 & 3573.08 & 0.32 & -0.910 & MBM06 &  2.58 &  0.09 & 0.18 &  0.38 & Syn &     & \\
 40 & Zr 2 & 3766.82 & 0.41 & -0.830 & LNA06 &  2.58 &  0.12 & 0.16 &  0.36 & Syn &     & \\
 40 & Zr 2 & 3998.97 & 0.56 & -0.520 & LNA06 &  2.58 &  0.07 & 0.15 &  0.35 & Syn &     & \\
 40 & Zr 2 & 4161.21 & 0.71 & -0.590 & LNA06 &  2.58 &  0.08 & 0.16 &  0.36 & Syn &     & \\
 40 & Zr 2 & 4208.98 & 0.71 & -0.510 & LNA06 &  2.58 &  0.07 & 0.15 &  0.35 & Syn &     & \\
\multicolumn{13}{c}{ } \\
 42 & Mo 1 & 3864.11 & 0.00 & -0.010 & FW96  &  1.92 & -0.49 & &  0.37 & Syn &     & \\
\multicolumn{13}{c}{ } \\
 46 & Pd 1 & 3404.58 & 0.81 &  0.330 & XSD06 &  1.66 & -0.72 & &  0.40 & Syn &     & \\
\multicolumn{13}{c}{ } \\
 56 & Ba 2 & 4554.03 & 0.00 &  0.167 & RCW80 &  2.17 & -0.16 & -0.25 &  0.36 & Syn &  -7.732 & HFS:M98\\
    &      &         &      &        &       &       &       & &       &     &         & $\Gamma_6$:MZG08 \\
 56 & Ba 2 & 5853.67 & 0.60 & -1.000 & RCW80 &  2.17 & -0.29 & -0.23 &  0.38 & Syn &  -7.584 & $\Gamma_6$:ABO \\
 56 & Ba 2 & 6141.71 & 0.70 & -0.076 & RCW80 &  2.17 & -0.34 & -0.36 &  0.25 & Syn &  -7.584 & $\Gamma_6$:ABO \\
 56 & Ba 2 & 6496.90 & 0.60 & -0.377 & RCW80 &  2.17 & -0.28 & -0.32 &  0.29 & Syn &  -7.584 & $\Gamma_6$:ABO \\  
\multicolumn{13}{c}{ } \\
 57 & La 2 & 3849.01 & 0.00 & -0.450 & LBS01 &  1.14 & -1.11 & &  0.53 & Syn &     & HFS:LBS01\\
 57 & La 2 & 3949.10 & 0.40 &  0.490 & LBS01 &  1.14 & -1.13 & &  0.51 & Syn &     & HFS:LBS01\\
 57 & La 2 & 3988.51 & 0.40 &  0.210 & LBS01 &  1.14 & -1.13 & &  0.51 & Syn &     & HFS:LBS01\\
 57 & La 2 & 3995.74 & 0.17 & -0.060 & LBS01 &  1.14 & -1.09 & &  0.55 & Syn &     & HFS:LBS01\\
 57 & La 2 & 4077.34 & 0.24 & -0.060 & LBS01 &  1.14 & -1.11 & &  0.53 & Syn &     & HFS:LBS01\\
 57 & La 2 & 4086.71 & 0.00 & -0.070 & LBS01 &  1.14 & -1.06 & &  0.58 & Syn &     & HFS:LBS01\\
 57 & La 2 & 4196.55 & 0.32 & -0.300 & LBS01 &  1.14 & -1.11 & &  0.53 & Syn &     & HFS:LBS01\\
 57 & La 2 & 4920.98 & 0.13 & -0.580 & LBS01 &  1.14 & -1.08 & &  0.56 & Syn &     & HFS:LBS01\\
\multicolumn{13}{c}{ } \\
 58 & Ce 2 & 3942.15 & 0.00 & -0.220 & LSC09 &  1.61 & -0.60 & &  0.57 & Syn &     & \\
 58 & Ce 2 & 3992.38 & 0.45 & -0.220 & LSC09 &  1.61 & -0.57 & &  0.60 & Syn &     & \\
 58 & Ce 2 & 3993.82 & 0.91 &  0.290 & LSC09 &  1.61 & -0.59 & &  0.58 & Syn &     & \\
 58 & Ce 2 & 3999.24 & 0.30 &  0.060 & LSC09 &  1.61 & -0.67 & &  0.50 & Syn &     & \\
 58 & Ce 2 & 4073.47 & 0.48 &  0.210 & LSC09 &  1.61 & -0.69 & &  0.48 & Syn &     & \\
 58 & Ce 2 & 4075.70 & 0.70 &  0.230 & LSC09 &  1.61 & -0.68 & &  0.49 & Syn &     & \\
 58 & Ce 2 & 4083.22 & 0.70 &  0.270 & LSC09 &  1.61 & -0.58 & &  0.59 & Syn &     & \\
 58 & Ce 2 & 4127.36 & 0.68 &  0.310 & LSC09 &  1.61 & -0.70 & &  0.47 & Syn &     & \\
 58 & Ce 2 & 4137.64 & 0.52 &  0.400 & LSC09 &  1.61 & -0.55 & &  0.62 & Syn &     & \\
 58 & Ce 2 & 4222.60 & 0.12 & -0.150 & LSC09 &  1.61 & -0.59 & &  0.58 & Syn &     & \\
 58 & Ce 2 & 4562.36 & 0.48 &  0.210 & LSC09 &  1.61 & -0.68 & &  0.49 & Syn &     & \\
 58 & Ce 2 & 4628.16 & 0.52 &  0.140 & LSC09 &  1.61 & -0.63 & &  0.54 & Syn &     & \\
\multicolumn{13}{c}{ } \\
 59 & Pr 2 & 4143.12 & 0.37 &  0.610 & ILW01 &  0.76 & -1.21 & -1.13 &  0.89 & Syn &     & HFS:G89\\
 59 & Pr 2 & 4179.40 & 0.20 &  0.480 & ILW01 &  0.76 & -1.09 & -1.01 &  1.01 & Syn &     & HFS:G89\\
 59 & Pr 2 & 4408.82 & 0.00 &  0.180 & ILW01 &  0.76 & -1.14 & -1.07 &  0.95 & Syn &     & HFS:G89\\
\multicolumn{13}{c}{ } \\
 60 & Nd 2 & 3784.25 & 0.38 &  0.150 & DLS03 &  1.45 & -0.78 & &  0.55 & Syn &     & \\
 60 & Nd 2 & 3838.98 & 0.00 & -0.240 & DLS03 &  1.45 & -0.68 & &  0.65 & Syn &     & \\
 60 & Nd 2 & 3900.22 & 0.47 &  0.100 & DLS03 &  1.45 & -0.70 & &  0.63 & Syn &     & \\
 60 & Nd 2 & 3941.51 & 0.06 & -0.020 & DLS03 &  1.45 & -0.64 & &  0.69 & Syn &     & \\
 60 & Nd 2 & 3990.10 & 0.47 &  0.130 & DLS03 &  1.45 & -0.68 & &  0.65 & Syn &     & \\
 60 & Nd 2 & 3991.74 & 0.00 & -0.260 & DLS03 &  1.45 & -0.63 & &  0.70 & Syn &     & \\
 60 & Nd 2 & 4012.70 & 0.00 & -0.600 & DLS03 &  1.45 & -0.58 & &  0.75 & Syn &     & \\
 60 & Nd 2 & 4018.82 & 0.06 & -0.850 & DLS03 &  1.45 & -0.58 & &  0.75 & Syn &     & \\
 60 & Nd 2 & 4021.33 & 0.32 & -0.100 & DLS03 &  1.45 & -0.53 & &  0.80 & Syn &     & \\
 60 & Nd 2 & 4023.00 & 0.56 &  0.040 & DLS03 &  1.45 & -0.48 & &  0.85 & Syn &     & \\
 60 & Nd 2 & 4059.95 & 0.20 & -0.520 & DLS03 &  1.45 & -0.58 & &  0.75 & Syn &     & \\
 60 & Nd 2 & 4061.08 & 0.47 &  0.550 & DLS03 &  1.45 & -0.61 & &  0.72 & Syn &     & \\
 60 & Nd 2 & 4069.26 & 0.06 & -0.570 & DLS03 &  1.45 & -0.61 & &  0.72 & Syn &     & \\
 60 & Nd 2 & 4109.07 & 0.06 & -0.160 & DLS03 &  1.45 & -0.63 & &  0.70 & Syn &     & \\
 60 & Nd 2 & 4109.45 & 0.32 &  0.350 & DLS03 &  1.45 & -0.63 & &  0.70 & Syn &     & \\
 60 & Nd 2 & 4135.32 & 0.63 & -0.070 & DLS03 &  1.45 & -0.53 & &  0.80 & Syn &     & \\
 60 & Nd 2 & 4156.08 & 0.18 &  0.160 & DLS03 &  1.45 & -0.63 & &  0.70 & Syn &     & \\
 60 & Nd 2 & 4177.33 & 0.06 & -0.100 & DLS03 &  1.45 & -0.62 & &  0.71 & Syn &     & \\
 60 & Nd 2 & 4385.66 & 0.20 & -0.300 & DLS03 &  1.45 & -0.62 & &  0.71 & Syn &     & \\
 60 & Nd 2 & 4462.98 & 0.56 &  0.040 & DLS03 &  1.45 & -0.48 & &  0.85 & Syn &     & \\
 60 & Nd 2 & 4706.54 & 0.00 & -0.710 & DLS03 &  1.45 & -0.56 & &  0.77 & Syn &     & \\
 60 & Nd 2 & 5212.35 & 0.20 & -0.960 & DLS03 &  1.45 & -0.65 & &  0.68 & Syn &     & \\
 60 & Nd 2 & 5255.51 & 0.20 & -0.670 & DLS03 &  1.45 & -0.53 & &  0.80 & Syn &     & \\
 60 & Nd 2 & 5319.82 & 0.55 & -0.140 & DLS03 &  1.45 & -0.67 & &  0.66 & Syn &     & \\
\multicolumn{13}{c}{ } \\
 62 & Sm 2 & 3568.26 & 0.48 &  0.290 & LDS06 &  1.00 & -1.09 & &  0.69 & Syn &     & \\
 62 & Sm 2 & 3760.71 & 0.19 & -0.400 & LDS06 &  1.00 & -0.89 & &  0.89 & Syn &     & \\
 62 & Sm 2 & 4318.94 & 0.28 & -0.250 & LDS06 &  1.00 & -0.86 & &  0.92 & Syn &     & \\
 62 & Sm 2 & 4434.32 & 0.38 & -0.070 & LDS06 &  1.00 & -0.96 & &  0.82 & Syn &     & \\
 62 & Sm 2 & 4467.34 & 0.66 &  0.150 & LDS06 &  1.00 & -0.84 & &  0.94 & Syn &     & \\
 62 & Sm 2 & 4519.63 & 0.54 & -0.350 & LDS06 &  1.00 & -0.86 & &  0.92 & Syn &     & \\
\multicolumn{13}{c}{ } \\
 63 & Eu 2 & 3819.67 & 0.00 &  0.510 & LWD01 &  0.52 & -1.32 & -1.17 &  1.09 & Syn &     & HFS:LWD01\\
 63 & Eu 2 & 3907.11 & 0.21 &  0.170 & LWD01 &  0.52 & -1.27 & -1.07 &  1.19 & Syn &     & HFS:LWD01\\
 63 & Eu 2 & 4129.72 & 0.00 &  0.220 & LWD01 &  0.52 & -1.27 & -1.12 &  1.14 & Syn &     & HFS:LWD01\\
 63 & Eu 2 & 4205.02 & 0.00 &  0.210 & LWD01 &  0.52 & -1.29 & -1.14 &  1.12 & Syn &     & HFS:LWD01\\
\multicolumn{13}{c}{ } \\
 64 & Gd 2 & 3358.62 & 0.03 &  0.250 & DLS06 &  1.11 & -0.79 & &  0.88 & Syn &     & \\
 64 & Gd 2 & 3422.46 & 0.24 &  0.710 & DLS06 &  1.11 & -0.63 & &  1.04 & Syn &     & \\
 64 & Gd 2 & 3481.80 & 0.49 &  0.110 & DLS06 &  1.11 & -0.83 & &  0.84 & Syn &     & \\
 64 & Gd 2 & 3768.40 & 0.08 &  0.210 & DLS06 &  1.11 & -0.69 & &  0.98 & Syn &     & \\
 64 & Gd 2 & 3796.38 & 0.03 &  0.020 & DLS06 &  1.11 & -0.66 & &  1.01 & Syn &     & \\
 64 & Gd 2 & 4085.57 & 0.73 & -0.010 & DLS06 &  1.11 & -0.77 & &  0.90 & Syn &     & \\
 64 & Gd 2 & 4130.37 & 0.73 &  0.140 & DLS06 &  1.11 & -0.73 & &  0.94 & Syn &     & \\
 64 & Gd 2 & 4215.02 & 0.43 & -0.440 & DLS06 &  1.11 & -0.66 & &  1.01 & Syn &     & \\
\multicolumn{13}{c}{ } \\
 65 & Tb 2 & 3509.14 & 0.00 &  0.700 & LWC01 &  0.28 & -1.53 & &  0.97 & Syn &     & HFS:LWB01\\
 65 & Tb 2 & 3848.73 & 0.00 &  0.280 & LWC01 &  0.28 & -1.49 & &  1.01 & Syn &     & \\
\multicolumn{13}{c}{ } \\
 66 & Dy 2 & 3407.80 & 0.00 &  0.180 & WLN00 &  1.13 & -0.68 & &  0.97 & Syn &     & \\
 66 & Dy 2 & 3445.57 & 0.00 & -0.150 & WLN00 &  1.13 & -0.56 & &  1.09 & Syn &     & \\
 66 & Dy 2 & 3454.32 & 0.10 & -0.140 & WLN00 &  1.13 & -0.61 & &  1.04 & Syn &     & \\
 66 & Dy 2 & 3460.97 & 0.00 & -0.070 & WLN00 &  1.13 & -0.56 & &  1.09 & Syn &     & \\
 66 & Dy 2 & 3506.81 & 0.10 & -0.600 & WLN00 &  1.13 & -0.71 & &  0.94 & Syn &     & \\
 66 & Dy 2 & 3531.71 & 0.00 &  0.770 & WLN00 &  1.13 & -0.63 & &  1.02 & Syn &     & \\
 66 & Dy 2 & 3534.96 & 0.10 & -0.040 & WLN00 &  1.13 & -0.51 & &  1.14 & Syn &     & \\
 66 & Dy 2 & 3536.02 & 0.54 &  0.530 & WLN00 &  1.13 & -0.64 & &  1.01 & Syn &     & \\
 66 & Dy 2 & 3538.52 & 0.00 & -0.020 & WLN00 &  1.13 & -0.64 & &  1.01 & Syn &     & \\
 66 & Dy 2 & 3550.22 & 0.59 &  0.270 & WLN00 &  1.13 & -0.54 & &  1.11 & Syn &     & \\
 66 & Dy 2 & 3563.15 & 0.10 & -0.360 & WLN00 &  1.13 & -0.59 & &  1.06 & Syn &     & \\
 66 & Dy 2 & 3694.81 & 0.10 & -0.110 & WLN00 &  1.13 & -0.66 & &  0.99 & Syn &     & \\
 66 & Dy 2 & 3944.68 & 0.00 &  0.110 & WLN00 &  1.13 & -0.57 & &  1.08 & Syn &     & \\
 66 & Dy 2 & 3996.69 & 0.59 & -0.260 & WLN00 &  1.13 & -0.57 & &  1.08 & Syn &     & \\
 66 & Dy 2 & 4077.96 & 0.10 & -0.040 & WLN00 &  1.13 & -0.51 & &  1.14 & Syn &     & \\
\multicolumn{13}{c}{ } \\
 67 & Ho 2 & 3398.94 & 0.00 &  0.410 & LSC04 &  0.51 & -1.37 & &  0.90 & Syn &     & HFS:LSC04\\
 67 & Ho 2 & 3416.44 & 0.08 &  0.260 & LSC04 &  0.51 & -1.45 & &  0.82 & Syn &     & HFS:LSC04\\
 67 & Ho 2 & 3456.00 & 0.00 &  0.760 & LSC04 &  0.51 & -1.30 & &  0.97 & Syn &     & HFS:LSC04\\
 67 & Ho 2 & 3796.67 & 0.00 &  0.160 & LSC04 &  0.51 & -1.19 & &  1.08 & Syn &     & HFS:LSC04\\
 67 & Ho 2 & 3810.00 & 0.00 &  0.190 & LSC04 &  0.51 & -1.22 & &  1.05 & Syn &     & HFS:LSC04\\
\multicolumn{13}{c}{ } \\
 68 & Er 2 & 3499.10 & 0.06 &  0.290 & LSC08 &  0.96 & -0.89 & &  0.93 & Syn &     & \\
 68 & Er 2 & 3559.89 & 0.00 & -0.690 & LSC08 &  0.96 & -0.90 & &  0.92 & Syn &     & \\
 68 & Er 2 & 3616.57 & 0.00 & -0.310 & LSC08 &  0.96 & -0.73 & &  1.09 & Syn &     & \\
 68 & Er 2 & 3633.54 & 0.00 & -0.530 & LSC08 &  0.96 & -0.91 & &  0.91 & Syn &     & \\
 68 & Er 2 & 3692.65 & 0.06 &  0.280 & LSC08 &  0.96 & -0.79 & &  1.03 & Syn &     & \\
 68 & Er 2 & 3729.52 & 0.00 & -0.590 & LSC08 &  0.96 & -0.66 & &  1.16 & Syn &     & \\
 68 & Er 2 & 3786.84 & 0.00 & -0.520 & LSC08 &  0.96 & -0.73 & &  1.09 & Syn &     & \\
 68 & Er 2 & 3830.48 & 0.00 & -0.220 & LSC08 &  0.96 & -0.84 & &  0.98 & Syn &     & \\
 68 & Er 2 & 3896.23 & 0.06 & -0.120 & LSC08 &  0.96 & -0.89 & &  0.93 & Syn &     & \\
 68 & Er 2 & 3906.31 & 0.00 &  0.120 & LSC08 &  0.96 & -0.74 & &  1.08 & Syn &     & \\
 68 & Er 2 & 3938.63 & 0.00 & -0.520 & KB95  &  0.96 & -0.76 & &  1.06 & Syn &     & \\
\multicolumn{13}{c}{ } \\
 69 & Tm 2 & 3462.20 & 0.00 &  0.030 & WL97b &  0.14 & -1.68 & &  0.96 & Syn &     & \\
 69 & Tm 2 & 3700.26 & 0.03 & -0.380 & WL97b &  0.14 & -1.61 & &  1.03 & Syn &     & \\
 69 & Tm 2 & 3701.36 & 0.00 & -0.540 & WL97b &  0.14 & -1.63 & &  1.01 & Syn &     & \\
 69 & Tm 2 & 3795.76 & 0.03 & -0.230 & WL97b &  0.14 & -1.55 & &  1.09 & Syn &     & \\
 69 & Tm 2 & 3848.02 & 0.00 & -0.140 & WL97b &  0.14 & -1.76 & &  0.88 & Syn &     & \\
\multicolumn{13}{c}{ } \\
 70 & Yb 2 & 3694.19 & 0.00 & -0.300 & BDM98 &  0.86 & -1.03 & &  0.89 & Syn &     & HFS:MGH94\\
\multicolumn{13}{c}{ } \\
 72 & Hf 2 & 3399.79 & 0.00 & -0.570 & LDL07 &  0.88 & -1.32 & &  0.58 & Syn &     & \\
\multicolumn{13}{c}{ } \\
 76 & Os 1 & 4260.85 & 0.00 & -1.430 & IAN03 &  1.45 & -0.17 & &  1.16 & Syn &     & \\
\multicolumn{13}{c}{ } \\
 77 & Ir 1 & 3513.65 & 0.00 & -1.210 & IAN03 &  1.38 & -0.04 & &  1.36 & Syn &     & HFS:CSB05\\
 77 & Ir 1 & 3800.12 & 0.00 & -1.440 & IAN03 &  1.38 & -0.05 & &  1.35 & Syn &     & HFS:CSB05\\
\multicolumn{13}{c}{ } \\
 90 & Th 2 & 4019.13 & 0.00 & -0.228 & NZL02 &  0.08 & -1.67 & &  1.03 & Syn &     & \\
\hline 
\end{longtable}
\begin{table}
\begin{tabular}{ll}
\multicolumn{1}{l}{ABO} & \multicolumn{1}{l}{\citet{omara_sp, omara_pd, omara_df};} \\
\multicolumn{1}{l}{ } & \multicolumn{1}{l}{\citet{omara_ion, omara-fe2};} \\
\multicolumn{1}{l}{BBP89} & \multicolumn{1}{l}{\citet{ni1-BBP89};} \\
\multicolumn{1}{l}{BBH83} & \multicolumn{1}{l}{\citet{borghs1983};} \\
\multicolumn{1}{l}{BCB05} & \multicolumn{1}{l}{\citet{HERESpaperII};} \\
\multicolumn{1}{l}{BDM98} & \multicolumn{1}{l}{\citet{Biemont_Yb};} \\
\multicolumn{1}{l}{BG80} & \multicolumn{1}{l}{ \citet{zn1-BG80};} \\
\multicolumn{1}{l}{BHN93} & \multicolumn{1}{l}{\citet{BHN93};} \\
\multicolumn{1}{l}{BKK91} & \multicolumn{1}{l}{\citet{fe-BKK91};}\\
\multicolumn{1}{l}{BL91} & \multicolumn{1}{l}{ \citet{BL91};}\\
\multicolumn{1}{l}{CSB05} & \multicolumn{1}{l}{ \citet{cowan2005};}\\
\multicolumn{1}{l}{CSS82} & \multicolumn{1}{l}{ \citet{Cardon_co1};}\\
\multicolumn{1}{l}{DLS03} & \multicolumn{1}{l}{ \citet{Hartog_Nd};}\\
\multicolumn{1}{l}{DLS06} & \multicolumn{1}{l}{ \citet{Hartog_Gd};}\\
\multicolumn{1}{l}{FMW88} & \multicolumn{1}{l}{ \citet{FMW88};}\\
\multicolumn{1}{l}{FW96} & \multicolumn{1}{l}{ \citet{Fuhr_Mo};}\\
\multicolumn{1}{l}{G89} & \multicolumn{1}{l}{ \citet{ginibre-pr2};}\\
\multicolumn{1}{l}{GLS04} & \multicolumn{1}{l}{\citet{mg_c6};} \\
\multicolumn{1}{l}{HLG82} & \multicolumn{1}{l}{\citet{Hannaford_Y};}\\
\multicolumn{1}{l}{HS80} & \multicolumn{1}{l}{ \citet{ni1-HS80};}\\
\multicolumn{1}{l}{HSR99} & \multicolumn{1}{l}{\citet{hfs-mn2};}\\
\multicolumn{1}{l}{IAN03} & \multicolumn{1}{l}{\citet{Ivarsson_Os_Ir};}\\
\multicolumn{1}{l}{ILW01} & \multicolumn{1}{l}{\citet{Ivarsson_Pr};}\\
\multicolumn{1}{l}{K94 } & \multicolumn{1}{l}{ \citet{Kur94a};}\\
\multicolumn{1}{l}{KB95} & \multicolumn{1}{l}{ \citet{Kur94b};}\\
\multicolumn{1}{l}{KG00} & \multicolumn{1}{l}{  \citet{KG-mn2};}\\
\multicolumn{1}{l}{LBS01} & \multicolumn{1}{l}{\citet{Lawler_La};}\\
\multicolumn{1}{l}{LD89} & \multicolumn{1}{l}{ \citet{LD89};}\\
\multicolumn{1}{l}{LDL07} & \multicolumn{1}{l}{ \citet{Lawler_Hf};}\\
\multicolumn{1}{l}{LDS06} & \multicolumn{1}{l}{ \citet{Lawler_Sm};}\\
\multicolumn{1}{l}{LGB03} & \multicolumn{1}{l}{ \citet{LGB03};}\\
\multicolumn{1}{l}{LSC04} & \multicolumn{1}{l}{ \citet{Lawler_Ho};}\\
\multicolumn{1}{l}{LSC08} & \multicolumn{1}{l}{ \citet{Lawler_Er};}\\
\multicolumn{1}{l}{LSC09} & \multicolumn{1}{l}{ \citet{Lawler_Ce};}\\
\multicolumn{1}{l}{LNA06} & \multicolumn{1}{l}{ \citet{Ljung_Zr};}\\
\multicolumn{1}{l}{LWB01} & \multicolumn{1}{l}{ \citet{Lawler_Tb_hfs};}\\
\multicolumn{1}{l}{LWC01} & \multicolumn{1}{l}{ \citet{Lawler_Tb};}\\ 
\multicolumn{1}{l}{LWD01} & \multicolumn{1}{l}{ \citet{Lawler_Eu};}\\
\multicolumn{1}{l}{M98} & \multicolumn{1}{l}{ \citet{M98};}\\
\multicolumn{1}{l}{MB09} & \multicolumn{1}{l}{ \citet{fe-MB09};}\\ 
\multicolumn{1}{l}{MBM06} & \multicolumn{1}{l}{ \citet{Malcheva_Zr};}\\ 
\multicolumn{1}{l}{MFW88} & \multicolumn{1}{l}{ \citet{MFW88};}\\
\multicolumn{1}{l}{MGH94} & \multicolumn{1}{l}{ \citet{hfs-yb2};}\\
\multicolumn{1}{l}{MKP07} & \multicolumn{1}{l}{ \citet{ml-nlte-ca};}\\
\multicolumn{1}{l}{MPS95} & \multicolumn{1}{l}{ \citet{MPS95};}\\
\multicolumn{1}{l}{MZG08} & \multicolumn{1}{l}{ \citet{ml-nlte-H};}\\
\multicolumn{1}{l}{NBI98} & \multicolumn{1}{l}{ \citet{ca_is};}\\ 
\multicolumn{1}{l}{NIST8} & \multicolumn{1}{l}{ \citet{NIST08};}\\ 
\multicolumn{1}{l}{NKW99} & \multicolumn{1}{l}{ \citet{NKW99-co1};}\\ 
\multicolumn{1}{l}{NZL02} & \multicolumn{1}{l}{ \citet{Nilsson_Th};}\\
\multicolumn{1}{l}{OWL91} & \multicolumn{1}{l}{ \citet{fe-OWL91};}\\ 
\multicolumn{1}{l}{P96} & \multicolumn{1}{l}{ \citet{hfs-co1};}\\
\multicolumn{1}{l}{PTP01} & \multicolumn{1}{l}{ \citet{ti2-PTP01};}\\ 
\multicolumn{1}{l}{RCW80} & \multicolumn{1}{l}{ \citet{RCW80};}\\ 
\multicolumn{1}{l}{S81} & \multicolumn{1}{l}{ \citet{smith81};}\\
\multicolumn{1}{l}{SON75} & \multicolumn{1}{l}{\citet{SON75};}\\ 
\multicolumn{1}{l}{SR81} & \multicolumn{1}{l}{ \citet{SR81};}\\
\multicolumn{1}{l}{SRE95} & \multicolumn{1}{l}{\citet{SRE95};}\\
\multicolumn{1}{l}{Th89} & \multicolumn{1}{l}{ \citet{gf-ca8498};}\\
\multicolumn{1}{l}{WF75} & \multicolumn{1}{l}{  \citet{NBS-ti1};}\\
\multicolumn{1}{l}{WL97a} & \multicolumn{1}{l}{ \citet{ni1-WL97a};}\\
\multicolumn{1}{l}{WL97b} & \multicolumn{1}{l}{ \citet{Wickliffe_Tm};}\\
\multicolumn{1}{l}{WLN00} & \multicolumn{1}{l}{ \citet{Wickliffe_Dy};}\\
\multicolumn{1}{l}{XSD06} & \multicolumn{1}{l}{ \citet{Xu_Pd};}\\
\end{tabular}
\end{table}
}

\end{document}